\documentclass[preprint,12pt,authoryear]{elsarticle}

\usepackage{graphicx}
\usepackage{amsmath}
\usepackage{algorithm}
\usepackage{algorithmic}
\usepackage{array}
\usepackage{amssymb}
\usepackage{multirow}
\usepackage{float}
\usepackage{longtable}
\usepackage{hyperref}
\usepackage{cleveref}
\usepackage{graphicx}
\usepackage{caption}
\usepackage{subcaption}
\usepackage{tikz}
\usetikzlibrary{shapes.geometric, arrows.meta, positioning}
\usepackage[most]{tcolorbox}
\tcbset{
  colback=gray!5!white,
  colframe=black!75!black,
  boxrule=0.8pt,
  arc=4pt,
  boxsep=5pt,
  left=5pt,
  right=5pt,
  top=5pt,
  bottom=5pt
}

\journal{Journal of Air Transport Management} 
\begin{document}
\begin{frontmatter}

\title{Modeling the Impact of Communication and Human Uncertainties on Runway Capacity in Terminal Airspace}

\affiliation[inst1]{organization={Department of Aerospace Engineering and Engineering Mechanics, \\ The University of Texas at Austin}, 
city={Austin},
postcode={78712}, 
state={TX},
country={USA}}

\author[inst1]{Yutian Pang\corref{mycorrespondingauthor}}
\cortext[mycorrespondingauthor]{Corresponding author.}
\ead{yutian.pang@austin.utexas.edu}
\author[inst1]{Andrew Kendall}
\author[inst1]{John-Paul Clarke}

\begin{highlights}
\item A comprehensive literature review on runway capacity estimation is provided, encompassing simulation-based and analytical approaches, empirical investigations, and policy perspectives.
\item We investigate and model aeronautical communication and human-induced uncertainties in the merging scenario during final approach of terminal arrivals.
\item A computer simulation based final approach merging scenario is developed, with theoretical validation and sensitivity study on various uncertainties. 
\item Runway capacity metrics such as the throughput and downwind immediate turn rates are adopted to quantify the arrival performance. 
\item As a more realistic scenario, an inverse optimal planning based automated final approach planning algorithm is evaluated in a similar way. 
\item The reliability contour of both studies are generated, providing better runway capacity estimation for optimal decision-making. 
\end{highlights}

\begin{abstract}
Runway capacity is a major constraint in airport and terminal area operations. Thus, improving runway capacity is critical to reducing delays while maintaining safety and efficiency. More so, because operations are expected to become more complex and more susceptible to disruptions with the integration of highly automated and autonomous aircraft into the existing airspace. To this end, we investigate the potential impact of communication and human performance uncertainties on runway operations. Specifically, we consider these impacts within the context of an arrival scenario with two converging flows: a straight-in approach stream and a downwind stream merging into it. Both arrival stream are modeled using a modified Possion distribution that incorporate the separation minima as well as the runway occupancy time. Various system level uncertainties are addressed in this process, including communication link- and human-related uncertainties. In this research, we first build a Monte Carlo-based discrete-time simulation, where aircraft arrivals are generated by modified Possion processes subject to minimum separation constraints, simulating various traffic operations. The merging logic incorporates standard bank angle continuous turn-to-final, pilot response delays, and dynamic gap availability in real time. Then, we investigate an automated final approach vectoring model (i.e., Auto-ATC), in which inverse optimal control is used to learn decision advisories from human expert records. By augmenting trajectories and incorporating the aforementioned uncertainties into the planning scenario, we create a setup analogous to the discrete event simulation. For both studies, runway capacity is measured by runway throughput, the fraction of downwind arrivals that merge immediately without holding, and the average delay (i.e., holding time/distance) experienced on the downwind leg. This research provides a method for runway capacity estimation in merging scenarios, and demonstrates that aeronautical communication link uncertainties significantly affect runway capacity in current voice-based operations, whereas the impact can be mitigated in autonomous operational settings. This work emphasizes implications and guidance for future vectoring procedures in congested terminal areas, highlighting the uncertainty impacts to runway capacities, providing better decision support to ensure efficiency and safety of near-terminal operation operations. The code used in this research can be found from this \href{https://github.com/YutianPangASU/arrival-approach-sim}{LINK}.
\end{abstract}

\begin{keyword}
Runway Capacity, Near-Terminal Operations, High-Density Airspace, Radar Vectoring, Air Traffic Management
\end{keyword}

\end{frontmatter}

\section{Introduction \label{sec: introduction}}
Runways are a major bottleneck in the air transportation system, and the lack of adequate runway capacity is a leading cause of delays \citep{ng2022impact}. The persistent and substantial growth in global air traffic has intensified the challenge of ensuring the efficiency and reliability of airside operations. Flight delays are increasingly observed in both arrivals and departures, and the finite capacity of both airspace and the airport infrastructure are viewed as the primary bottleneck \citep{cook2011european, tee2018modelling, eurocontrol2019forecast, ribeiro2025delay}. Runway capacity generally refers to the maximum number of aircraft that can be safely handled by the controller, while maintaining acceptable levels of delay \citep{horonjeff1962planning, tascon2021air}. Effective management of airport operations relies heavily on aligning the capacity of critical components with the actual and projected demand for air transport services. Runway capacity thus becomes the key criterion for assessing the feasibility and efficiency of both current operational practices and future design solutions \citep{jurczyk2023simulation}. Expanding runway capacity can address delays, but infrastructure improvements require long lead times and are difficult to align with uncertain future operational demand. Ineffective planning may result in either underused capacity or elevated congestion, as demand forecasts often deviate significantly from actual traffic \citep{xiao2013demand}. Accurate runway capacity estimation enables the airport planner to make better decisions on infrastructure, configuration, and operational procedures \citep{horonjeff1962planning}. Moreover, the declaration of runway capacity by the managing authority is inherently complex, involving factors that are stochastic and hard to predict (i.e., aircraft performance, human factors, communication infrastructures, weather conditions) \citep{putra2017review}. Capacity studies are therefore critical not only for day-to-day traffic management and delay minimization, but also for long-term airport planning, airspace modernization, and resilience assessment under uncertain or adverse conditions \citep{icao9971}. 

With the anticipated rise of high-density airspace operations, the study of runway and airspace capacity will become even more critical and complex. As current air traffic management systems already operate near capacity during peak periods, the integration of novel aircraft types, including Advanced Air Mobility (AAM) vehicles with short takeoff and landing (STOL) capabilities, along with the vertical takeoff and landing (VTOL) designs, will introduce new operational dynamics and constraints. These diverse vehicle characteristics, combined with potentially different performance envelopes and separation requirements, will complicate traffic flows and challenge traditional capacity modeling approaches. Ensuring the safe and efficient accommodation of such heterogeneous traffic will require rigorous investigation into new capacity metrics, dynamic scheduling algorithms, and adaptive operational procedures. Accordingly, the development of robust methods for capacity estimation and management has been identified as a key research priority for the future integration of AAM into the national airspace system \citep{patterson2021advanced, ellis2021defining}. As these technologies progress toward large-scale deployment, capacity analysis will serve as a foundation for both operational safety and system scalability. Understanding the behavior and properly forecast runway conditions in future high density scenarios are critical to enhance the safety and efficiency of aviation operations. Given the transformative potential and challenges posed by integrating AAM operations into urban and near-terminal environments, understanding and effectively managing demand-capacity interactions becomes particularly urgent. The complexity introduced by heterogeneous aircraft performance, operational diversity, and infrastructure limitations demands robust modeling frameworks capable of simulating high-density scenarios under realistic operational uncertainties \citep{vascik2018analysis, alvarez2021demand}.

\begin{figure}
    \centering
    \includegraphics[width=0.85\textwidth]
    {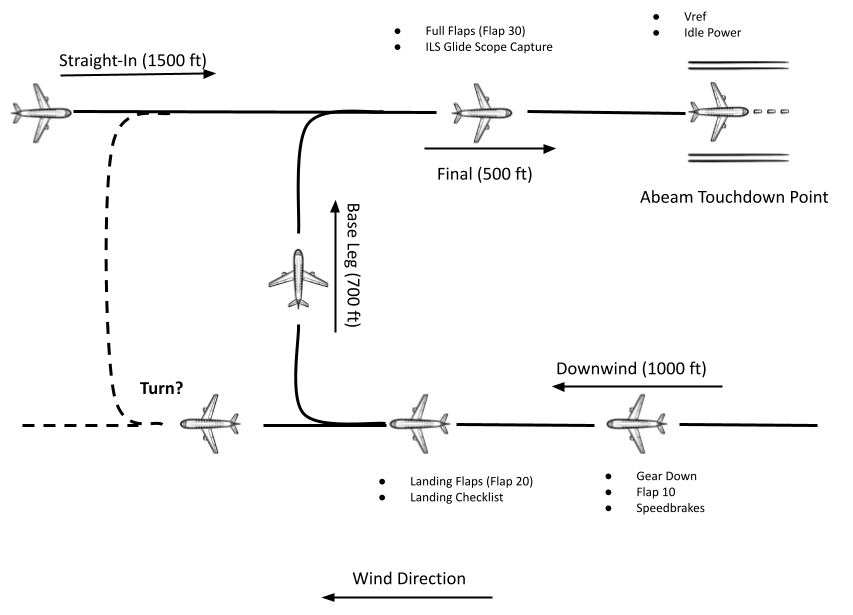}
    \caption{Illustration of the simulated merging scenario in this study. Two arrival streams are included. In the scenario, the Straight-In arrivals have aligned with the runway centerline, while the Downwind arrivals have to determine if the gap between two Straight-Ins are sufficient enough to make a complete 180$^\circ$ continuous turn-to-final to merge into the straight-in flow, as well as maintaining proper separation and adjusted with runway occupancy time. To further complicate the simulation, the Downwind arrivals also suffer from the communication loss and pilot response delay uncertainties while making the turning decision.}
    \label{fig: flowchart}
\end{figure}

Runway capacity is increasingly becoming more sensitive to uncertainties from multiple sources. For example, intermittent aeronautical communication links or latency in datalinks can hinder in-time controller-pilot situational awareness \citep{trsek2007last, dave2022cyber, ukwandu2022cyber}, and studies show that degraded communication (such as communication latency or communication loss) can drastically erode an aircraft’s performance and stability margins \citep{thirtyacre2021effects, thirtyacre2022remotely, bulusu2022impact, pang2026reliability}. Likewise, human-in-the-loop factors such as pilot response delays or missed clearances become more problematic at higher traffic densities, with large delays or non-responses significantly degrading safety in crowded airspace \citep{consiglio2008impact}. Compounding these issues is the growing presence of highly automated and autonomous aircraft. Advanced Air Mobility (AAM) vehicles and other autonomous platforms are poised to operate in high-density traffic environments \citep{cohen2024planning}, creating a new integration challenge for traditional Air Traffic Management (ATM) systems. Ensuring that these novel aircraft can safely merge into conventional ATM procedures is critical. As evidenced by the need for new standards and protocols, the integration of remotely piloted or autonomous aircraft into conventional airspace has been identified as a major challenge by regulators and researchers \citep{patterson2021advanced, ellis2021defining, gao2023developing}. This context motivates the present research into how high-density arrival operations perform under varied uncertainty conditions. 

Many studies have employed fast-time simulations to estimate runway capacity for throughput improvements. Monte Carlo simulations, for example, have been utilized to assess the benefits of newly proposed arrival procedures and operational strategies across various airports \citep{levy2004opportunities}. Discrete-event simulations similarly enable detailed analysis of complex runway operations, capturing factors like fleet mix, separation rules, and queuing phenomena \citep{bubalo2011airport, tee2018modelling, jurczyk2023simulation}. Recent studies have emphasized integrating uncertainty into runway capacity estimation. Stochastic and robust scheduling models consider variability in demand, operational timing, and traffic flow dynamics, highlighting the significantly influence of unpredictable factors to real-world runway capacity \citep{yin2021delay, ng2022impact}. Such uncertainties, arising from weather, human factors, and communication infrastructures, can disrupt planned inter-arrival spacing and compromise efficiency \citep{tascon2021air, ng2022impact}. The development of automated computer-aided systems has been long regarded as a key objectives of air traffic control researchers \citep{erzberger1992ctas, davis1995final, mcnally2015dynamic, pang2024machine}. Considering the cost of conducting real-world flight tests on runway capacity study, computer simulations are a natural choice. However, few runway capacity estimation tools that are open-sourced and user-friendly \citep{di2020critical, mascio2020analytical, jurczyk2023simulation}. The use of simulation models allows for the flexibility of analyzing multiple stochastic factors with different arrival streams configurations and separation rules. This can save time, money, and help decision makers optimize their operating systems. Concurrently, research has explored incorporating automation tools to enhance runway operations, recognizing the potential of advanced air traffic management technologies in managing high-density terminal scenarios \citep{mueller2017enabling}. 

Despite these advancements, several limitations persist in the literature. Past studies, such as those examining departure and arrival constraints at busy airports, underscore the critical role of accurately managing demand and capacity to mitigate congestion, reduce delays, and minimize operational costs and environmental impacts \citep{pujet1999input, idris2001observation}. However, these studies also reveal structural limitations in existing capacity assessment methodologies. Firstly, many simulation tools employed for runway capacity analysis are either proprietary or project-specific (i.e., targeting a specific airport), resulting in a notable scarcity of open-source platforms accessible to the broader community \citep{jurczyk2023simulation, bubalo2011airport}. Additionally, prior studies frequently address uncertainties in an oversimplified manner or under idealized conditions, inadequately capturing human-induced availabilities and aeronautical communication disruptions prevalent in real-world operations \citep{putra2017review}, let alone providing high-fidelity impact analyses on varying uncertainty levels to runway capacities. Furthermore, realistic merging scenarios with converging traffic flows remain underrepresented, with existing capacity evaluations often oversimplifying merging dynamics and neglecting critical operational intricacies \citep{ng2022impact}. Lastly, the integration of automated or autonomous systems into capacity modeling has been limited, with many analyses primarily assuming traditional human-in-the-loop operations, thereby overlooking the transformational potential of automated decision-support systems \citep{mueller2017enabling}.

To address these limitations, this research introduces two detailed case studies. First, an open-source discrete event simulation framework is developed explicitly for runway capacity analysis in complex merging scenarios under uncertainty. This model simulates converging traffic flows consisting of a straight-in stream and a downwind stream merging via a continuous turn-to-final, seeking appropriate gaps in the straight-in queue once turn advisories are issued (see \Cref{fig: flowchart}). Aircraft arrivals are modeled through modified Possion processes, incorporating human-induced delays, aeronautical communication uncertainties (e.g., latency, availability, continuity) \citep{pang2026reliability}, voice communication transaction times, and separation standards, while allowing varying arrival rates from both arrival streams. Consequently, the proposed model provides a more realistic and comprehensive runway capacity assessment compared to existing approaches. The second case study leverages an automated final approach benchmark model \citep{tolstaya2019inverse}, modified to include multiple layers of uncertainty. This model functions as an automated air traffic management tool, offering precise vectoring recommendations through planned flight paths. Both case studies are evaluated through robust runway capacity metrics, quantifying throughput trends and downwind delays as arrival rates increase. Specifically, metrics utilized include runway throughput, average delay for downwind aircraft, and the fraction of downwind arrivals able to merge without entering holding patterns. Reliability contour lines for these metrics under varying downwind and straight-in arrival rates, and different uncertainty levels, are provided. These contours enable the accurate estimation of expected times of arrival (ETAs) and time-to-lose (TTL), facilitating proactive decision-making by arrival managers to mitigate congestion and significant delays during peak and high-density operations \citep{de2014supporting, jun2022towards}.

In summary, the primary contributions of this research include,
\begin{itemize}
\item Development of an discrete event simulation tailored to terminal-area merging scenarios, incorporating variable arrival rates and realistic operational conditions.
\item Comprehensive incorporation and evaluation of aeronautical communication uncertainties (e.g., latency, availability, continuity) and human-induced factors (e.g., pilot response delays, communication message transaction time) through high-fidelity Monte Carlo simulations, systematically quantifying their impacts on runway capacity metrics.
\item Visualization of reliability surface for runway capacity metrics across varying uncertainty levels, providing actionable insights to air traffic controllers and airport planners for effective decision-making, especially for future automated scenarios under various traffic density.
\end{itemize}

The rest of the paper is structured as follows. The related literature of this work and a thorough of runway capacity estimation are discussed in \Cref{sec: literature review}. \Cref{sec: methodologies} introduces our approaches on modeling the uncertainties and performance evaluation metrics. \Cref{sec: case1} explains the detailed setup of the discrete event simulation, where the mathematical modeling of generating arrival flows and merging logic is emphasized, followed by the figures showing the reliability contour lines of the discrete event simulation for performance metrics in \Cref{subsec: case1-mc}. Notably, an analytical study is also given to valid the simulation results. Similarly, \Cref{sec: case2} provides a brief summary of the mechanism of inverse optimal planning, the basic evaluation of the planner, and the trajectory augmentation approach to simulate high-density arrivals. The results of the second case study is provided in \Cref{subsec: case2-mc}. \Cref{sec: conclusion} concludes this research. 

\section{Related Studies \label{sec: literature review}}
FAA Advisory Circular 150/5060-5 defines capacity as the peak number of operations that can be handled in one hour by the airport’s runways under specific conditions. The practical capacity of a runway capacity is the maximum number of takeoff and/or landing operations that can be safely handled with less than an acceptable delay (i.e., 4 minutes) \citep{kuzminski2013improved, de2020airport, jurczyk2023simulation}. These definitions underscore that runway capacity is not necessarily an absolute constant, but rather depends on the chosen level of service (i.e., acceptable delay), safety requirements (i.e., wake vortex separations), and operating conditions (weather, fleet mix, etc.). Accurate estimation of runway capacity is critical to both tactical and strategic air traffic management that balance the trade-off between minimizing delays and maximizing throughput, which also quantify the safety margin when unpredictable operational events arise \citep{ikli2021aircraft}. 

The literature on runway capacity studies falls into four categories: (i) simulation-based approaches; (ii) analytical models; (iii) empirical studies; and (iv) the insights to policy/planning to improve runway capacity. In the following subsections, we discuss these categories with a modeling techniques applied, the made assumptions (e.g. regarding aircraft separations and delays), the handling of uncertainty, and their applicability to real-world airport operations.

\subsection{Simulation-based Approach}
Simulation is a widely used approach to estimate runway capacity. High-fidelity fast-time simulation tools (e.g. FAA’s SIMMOD/ADSIM \citep{faaSimmodManual, monk1984adsim} or commercial models like TAAM \citep{taam2005manual} and AirTOP \citep{airtop2025}) have informed hundreds of capacity studies by modeling gate-to-gate or surface operations with realistic aircraft trajectories and controller rules. Such models typically incorporate fine-grained details (e.g., runway entry/exit taxiways, approach routes, wake vortex separation rules, fleet mix, and ATC procedures) \citep{ratner1970methodology, harris1973models, swedish1981upgraded, kuzminski2013improved}. By simulating hours of operations under varying demand, they can estimate the maximum sustainable throughput (i.e., capacity) and even generate capacity envelopes (i.e., trade-off curves of arrivals v.s. departures). For example, MITRE’s runwaySimulator \citep{faa2024runwaysimulator} is a fast-time model developed to estimate airport runway system capacity under NextGen scenarios, which produces capacity curves (i.e., hourly arrival/departure combinations) and identifies bottlenecks, while being more lightweight than full ATC simulations \citep{kuzminski2013improved}. 

Simulation allows flexibility to test \textit{what-if} scenarios such as usage of new runways or procedures. \cite{mascio2020analytical, mascio2020hourly} demonstrated this by comparing a baseline airport layout to a scenario with an added runway using AirTOP simulator, showing how capacity would increase under the new configuration. Likewise, an agent-based simulation modeled parallel runway operations and provided capacity estimates under different control strategies \citep{peng2013capacity}. These studies highlight that simulation-based methods can capture interactions and nonlinear effects that simpler models often miss. For instance, they can pinpoint specific delay causes or runway occupancy patterns that limit arrival throughput \citep{mascio2020hourly}.

Discrete-event and Monte Carlo simulation has also been applied to runway capacity analysis, as general-purpose tools. \cite{Jurczyk2023} build an airport operations model in FlexSim \citep{nordgren2003flexsim} to calculate runway and taxiway capacity. Their study confirmed that discrete-event simulation is a viable tool for determining maximum hourly runway throughput, supporting investment decisions on infrastructure modernization. The model, tailored to a specific airport, was used to evaluate the \textit{basic} vs \textit{practical} capacity by simulating numerous operational scenarios. Such simulations provide valuable insights for airport managers, albeit at the cost of requiring detailed data and modeling effort. Monte Carlo simulations are used to evaluate new procedures by randomly sampling inter-arrival times and aircraft performance. \cite{Levy2004} identified pitfalls in using simplistic assumptions in Monte Carlo models for arrival capacity. They found that assuming uniform distributions for aircraft landing speeds and inter-arrival spacings can underestimate arrival rates. By analyzing over 100,000 real arrivals at Memphis (KMEM), they showed the true spacing distributions are skewed. Using empirical spacing distributions instead of uniform ones yielded up to 7 more landings per hour in capacity estimates. This underscores the importance of correct stochastic assumptions, their improved simulation (with weight-class-conditioned speed/spacing generation) reduced error to less than one arrival/hour difference from observed rates. In practice, simulation models increasingly integrate such empirical data. 

Some recent studies use machine learning within simulation. \cite{Herrema2019} embedded an ML model to predict runway exit times at Vienna airport, allowing more accurate simulation of runway occupancy and thus capacity. Similarly, \cite{de2018data} applied ML to enhance throughput predictions at a large European hub, identifying how adjusting sequencing could improve arrival rates. Surface congestion also impacts runway capacity. Delays from congested surfaces propagate to runway operations, causing additional airborne holding. This airborne holding, in turn, worsens surface conditions by creating arrival bunching, exacerbating congestion and further diminishing runway capacity \citep{khadilkar2014network, raphael2022control}. Runway occupancy time (ROT) is an important quantity for runway capacity estimation, which can be fit into an Normal distribution of $\mathcal{N}(48, 11)$ seconds \citep{kumar2009runway}. However, in extreme scenarios, the ROT can be very significant. (i.e., up to 120 seconds for super heavy wake category \citep{meijers2019data}).

Simulation approaches are indispensable for capacity estimation under complex or future scenarios, ensuring that planning decisions account for real-world variability and interactions that simpler methods might overlook. Fast-time simulations can output not only capacity values but also detailed performance metrics (delay distributions, controller workload, fuel burn, etc.), aiding holistic airport management. On the downside, these simulations require extensive input data and can be time-consuming to set up and validate. In fact, FAA notes that high-resolution simulation may be impractical for early-stage studies due to the resource and time requirements. However, improved tools like runwaySimulator aim to strike a balance by providing medium-resolution fast-time modeling that is faster to configure, while still capturing the essential dynamics of runway queues and separations.

\subsection{Analytical Approach}
Analytical approaches to arrival capacity range from classic theoretical models to modern optimization formulations. Queuing theory provides a foundation for many analytical models of runway capacity. In a simple view, a single runway can be modeled as a server with aircraft arrivals as a stochastic process and service times related to runway occupancy and wake separations \citep{Blumstein1960}. A large body of work applies M/M/1 or M/G/1 queue models to runways \citep{bauerle2007waiting, itoh2022modeling}, which require Possion arrival streams or specific separation distributions. While such models can provide closed-form estimates and insights (e.g. how capacity degrades as variance in inter-arrival times increases), they are limited to a single runway and idealized conditions, and are not suitable for complex layouts and cannot easily incorporate non-runway constraints or dynamic controller strategies \citep{gilbo2002airport}. The output of such studies can be sensitive to input assumptions and often serve as upper-bound capacity estimates (i.e., neglecting real-world inefficiencies). For instance, FAA Advisory Circular (AC) 150/5060-5 \citep{faa15050605} provides charts derived from analytical/empirical formulas (the Airfield Capacity Model) to estimate hourly capacity given runway configuration and fleet mix. The AC distinguishes multiple levels of analysis, (i) Level 1 uses simple geometric analogies and base charts; (ii) Level 2 uses tabulated curves and spreadsheets (refining for mix and touch-and-goes); (iii) Level 3 uses queuing formulas for arrivals and departures. These methods require progressively more data (from just runway layout at Level 1 up to detailed separation times at Level 3) and provide correspondingly more accuracy. Importantly, even the FAA acknowledges that such analytical tools might not capture everything (e.g. they treat runways, taxiways, gates separately without feedback loops), but they are fast and easy to apply, which is considered a crucial advantage in early planning stages.

Optimization models have been developed to directly maximize throughput or minimize delays subject to separation constraints. One stream of research formulates the runway scheduling problem as an optimization problem. Given a set of arriving flights with timings, find the sequence (and possibly runway assignment if multiple runways) that maximizes the number of landings in a time window or minimizes total delay \citep{pang2024machine}. \cite{gilbo2002airport} introduced a formulation of airport capacity as a constrained optimization, and defined the concept of a capacity envelope (the trade-off curve between arrival and departure rates) which can be computed by optimizing different objective weights. Modern approaches often use mixed-integer programming or dynamic programming to schedule aircraft landings. \cite{lieder2016scheduling} modeled take-off and landing scheduling on multiple interdependent runways as a sequencing problem, capturing interactions between runways and achieving optimized throughput relative to heuristic ATC procedures. 

To handle uncertainty in operations, stochastic and robust optimization methods have been applied. \cite{solveling2011scheduling} developed a two-stage stochastic runway scheduling model for departures and arrivals under uncertain taxi and pushback times. In their model, the first stage determines an ideal weight-class sequence (heavy, medium, light aircraft ordering) to maximize expected throughput, and the second stage assigns specific flights once uncertainties (e.g. exact arrival times) resolve. This approach improved upon both First-Come-First-Served (FCFS) and deterministic optimization to show that proactively sequencing by aircraft type can hedge against variability and yield higher realized arrival rates. The benefit of such analytical optimization indicates that at high demand levels the stochastic planner kept throughput higher than naive methods. Multi-objective optimization has also been introduced to account for trade-offs in runway operations. \cite{yin2021delay} formulated a multi-objective evolutionary algorithm to jointly maximize runway throughput (arrivals/departures) and minimize both delays and emissions. In their case study for Shanghai Pudong airport (i.e, two-runway system), they generated Pareto-optimal schedules balancing these objectives under separation and timing constraints. Their results indicated that the minimum-delay scheduling was often the best compromise, achieving near-maximal throughput with significantly lower emissions. This highlights that analytic models can incorporate environmental or workload considerations alongside capacity.

Analytical models tend to yield more generalizable insights and are computationally efficient. They are well-suited for strategic evaluations, such as estimating how much capacity a new runway should add under ideal conditions. However, they may overestimate capacity if real-world factors (e.g. variable human controller performance or suboptimal sequencing) are not captured or underestimated. This suggests that, in complex scenarios (i.e., varying uncertainties), simulation becomes necessary to capture interactions, whereas analytical formulas require excessive simplifications. As a summary, analytic methods provide essential tools for estimating and understanding runway arrival capacity, from simple to complex optimization formulation, but they are usually complemented by simulation or empirical calibration to ensure realism.

\subsection{Empirical Studies}
Empirical research on runway capacity involves using observed data and case studies to infer capacity limits and influencing factors. The most straightforward approach is to look at the historical peak traffic throughput under similar congested conditions and treat that as a estimated capacity. Empirical observations are often used to validate models, as mentioned in \cite{barrer2005analyzing}, the statistical analysis of the peak runway throughput during high demand period can yield capacity estimates, which also forms the basis of the reference tables in \cite{faa15050605}. However, the major purpose of empirical studies is to understand the influencing factors. 

Several studies have focused on identifying the contributive parameters to runway capacity. A review paper categorizes the influencing factors into five groups, (1) operations/procedures; (2) human factors; (3) infrastructure/geometry; (4) aircraft performance; (5) external factors. Notably, the investigation shows that operational/procedural factors (e.g., separation standards, sequencing techniques, percentage of arrivals in mix) were cited by the majority (i.e., about 52\%) of sources as key drivers of capacity \citep{putra2017review}. This suggests that better near-terminal procedures (e.g., optimizing approach sequences, reducing wake separations, better exit taxiway usage) can significantly boost arrival capacity. Similarly, \cite{farhadi2014runway} examined runway capacity at Doha International Airport under different scheduling approaches and configurations. Their study tested a heuristic scheduling algorithm against FCFS, taking into account local runway layout and fleet mix, and demonstrated that more efficient sequencing could increase the arrival throughput while keeping delays reasonable.

Other empirical works have leveraged machine learning and data science \citep{murcca2018predicting, Herrema2019, de2018data}. These studies often use real world meteorological (i.e., convective weather) and operational (i.e., radar tracks) data to model the interactions between these factors to capacity. These studies also frequently highlight uncertainty and variability in the real world. For example, analysis of empirical inter-arrival times at major U.S. airports has shown that capacity drops in poor visibility not only due to procedural increases in separation but also due to more variability in spacing. Field data has also revealed the impact of controller behavior (e.g. reaction times in issuing clearances, or sequencing strategies) which can cause actual throughput to be lower than theoretical capacity \citep{simaiakis2010impact, lehouillier2016measuring, murcca2018predicting}. 

In summary, empirical studies conduct data analysis and context to capacity estimation under varying operation conditions. By analyzing real operations, researchers and practitioners can identify leverage points. For example, reducing Runway Occupancy Time (ROT) by adding rapid-exit taxiways or improving pilot adherence to exit instructions can raise arrival capacity by a quantifiable amount. 

\subsection{Insights to Policy and Planning}
Accurate capacity estimation is integral to strategic airport planning and policy-making. At the policy level, the literature often frames capacity management as a choice among three broad strategies, (i) adding infrastructure; (ii) enhancing operations; (iii) managing demand. \cite{jacquillat2018roadmap} emphasizes that airport demand–capacity management requires a \textit{holistic roadmap} combining these approaches. This means decision-makers must decide whether to invest in new runways/taxiways (supply expansion), implement procedural or technological improvements to boost capacity (supply optimization), or control demand via slots and scheduling policies to mitigate congestion (demand management). The trade-offs are significant as building a new runway can dramatically increase capacity but involves additional cost and long lead times, whereas refining operations (e.g. better sequencing, new wake separation rules) can incrementally increase arrival rates at lower cost and faster implementation. Demand management (e.g., slot controls or congestion pricing) doesn’t increase capacity per se, but can align demand to available capacity, reducing delays at the cost of limiting flights. Many policy frameworks use declared (practical) capacity as a control variable. These declared values often come from a series of simulation, empirical analysis, and judgment, informed by studies like those reviewed above.

Long-term planning for airports relies heavily on capacity estimation to decide when and how to expand. Forecasts of future demand are compared to current practical capacity to identify when shortfalls will arise. However, numerous studies highlight the deep uncertainty in long-term demand forecasts. For example, demand forecast errors for U.S. airports have reached +210\% or –34\% over 15 years, and even 5-year projections have shown large deviations \citep{solvoll2020forecasting}. Recent studies \citep{xiao2013demand, li5132629joint} highlight the need for flexible uncertainty-aware runway planning, showing that under high demand volatility and uncertainty, traditional rules (e.g., expanding at 80\% utilization) may be suboptimal, and dynamic strategies like real options and shock-based models offer more effective investment timing and sizing.

Policy frameworks translate these insights into practice. The FAA, Eurocontrol, and ACRP provide standardized guidelines, such as the Eurocontrol ACAM Manual \citep{o2016airport}, which offer cost-benefit methods for assessing expansion needs. Practical capacity, typically associated with an average delay threshold (e.g., 4 minutes), forms the basis of declared airport arrival rates. When actual demand nears or exceeds this rate, planners are prompted to consider expansion or demand management. Accordingly, regulatory tools like IATA slot coordination and FAA initiatives are used to enforce throughput limits aligned with empirical capacity to avoid gridlock.

\subsection{Summary}
The literature on runway capacity estimation reveals a clear evolution from traditional deterministic models \citep{harris1973models, swedish1981upgraded} to more complex stochastic and simulation-based approaches that better reflect operational realities \citep{solak2018stochastic, liu2020stochastic, ikli2021aircraft}. As established, Arrival Manager (AMAN) systems are crucial for scheduling incoming aircraft \citep{hasevoets2010aman, de2014supporting}, but their effectiveness is constrained by runway capacity. While discrete-event and Monte Carlo simulations have become standard for capturing the probabilistic nature of airport operations \citep{ikli2021aircraft, attar2025simulation}, a significant gap persists in the literature. Many existing studies either use proprietary tools, which limits accessibility \citep{jurczyk2023simulation}, or analyze operations under idealized conditions. Specifically, they often oversimplify the profound impacts of human-induced variability and aeronautical communication uncertainties, which are critical drivers of inefficiency and delay in real-world settings \citep{putra2017review, jacquillat2018roadmap}. Furthermore, few studies have rigorously modeled the complex merging dynamics of converging arrival streams, such as a straight-in flow and a downwind flow, which are common in busy terminal areas.

Our research directly addresses these shortcomings by building upon the established foundation of stochastic simulation while introducing a higher degree of operational fidelity. We develop an open-source, Monte Carlo-based discrete-event simulation framework specifically designed to analyze a realistic merging scenario involving a straight-in stream and a downwind stream making a continuous turn-to-final. This approach allows for a granular investigation into how runway throughput is affected by explicitly modeled uncertainties, including pilot response delays and communication disruptions, which past research has often overlooked. Moreover, we extend our analysis beyond current operational paradigms by introducing a comparative case study featuring an automated air traffic control model based on inverse optimal control. By evaluating both a conventional and an automated system under the same strenuous uncertainty conditions, our work not only quantifies the vulnerabilities in today's voice-based procedures but also demonstrates the tangible efficiency and reliability gains achievable through future automation. This research fills a critical void by providing a comprehensive methodology to assess runway capacity in complex, uncertain environments and offers actionable insights for integrating autonomous systems and enhancing the resilience of air traffic management.

\section{Uncertainty Modeling and Performance Evaluation \label{sec: methodologies}}
Multiple sources of uncertainties are considered in our arrival simulation studies under various densities. The major components are, (i) the communication signal uncertainties, also referred as communication reliability; (ii) the pilot response delays, the differences between the time the pilot receives the signal to the pilot takes actions; (iii) the uncertain pilot-ATC communication time of a single complete transaction. Lastly, we discuss the evaluation metrics to measure the runway performance during final-approach arrival operations near the terminal.

\subsection{Communication Uncertainty Modeling}
Reliable communication, defined as the accurate transfer of information between the sender and receiver, is fundamental to current ATM services. The enhancement of communication, navigation, surveillance and air traffic management (CNS/ATM) has long been regarded as one of the key objectives by related authorities \citep{wgc1}. Communication reliability is the probability that the system performs correctly under defined conditions over time \citep{villemeur1992reliability, ahmad2017reliability}, which is impacted by both intended and unintended factors \citep{dave2022cyber, ukwandu2022cyber, hu2025reinforcement}. In the field of aviation, the communication reliability is assessed with the Required Communication Performance (RCP) concept defined by ICAO  \citep{ICAO9869}, or the equivalent Required Link Performance (RLP) framework by the Joint Authorities for Rulemaking of Unmanned Systems (JARUS) \citep{jarus4required} to address system-level communication performance for increasing future autonomous operations. RCP and RLP share similar metrics definitions, the transaction time, availability, continuity, and integrity, as indicators of communication performance that are necessary for safe and efficient operations in performance-based airspace. As in our previous work \citep{pang2026reliability}, this paper continues to use the RCP terminology. RCP is defined by four key parameters,
\begin{itemize}
    \item Transaction Time ($\tau_{msg}$): Maximum time span required to complete a communication transaction.
    \item Availability ($P_A$): Probability that the communication service can be initiated when needed.
    \item Continuity: Probability of completing a transaction within the specified time once service is available.
    \item Integrity: Probability of transactions completed within the transaction time without detected error.
\end{itemize}

A communication transaction involves human interaction, such as the issuing of clearances or instructions, combined with technical transmission times \citep{sollenberger2003effect}. The RCP framework groups required performance levels into discrete \textit{RCP types} (e.g., RCP10, RCP60, RCP120, RCP240, and RCP400), each tailored to different operational scenarios and separation requirements \citep{ICAO9869}. For instance, RCP10 supports precise, rapid interventions such as horizontal-separation interventions within 10 seconds, while RCP240 and RCP400 address longer-range communications typical of oceanic airspace \citep{PARKSATVOICE}. Regulatory authorities use these RCP types to assess whether aircraft communication capabilities meet operational requirements for a given airspace or procedure, enabling performance-based approvals rather than relying solely on equipment carriage \citep{ICAO9869, PARKSATVOICE}. We include this taxonomy as background context: the discrete RCP categories themselves are not used as inputs to our simulation. Instead, the simulation parameterizes the underlying performance metrics (notably the availability $P_A$ and the transaction time $\tau_{msg}$) and sweeps them continuously, so that any specific RCP type can be mapped onto a corresponding $(P_A, \tau_{msg})$ operating point in our reliability contours. This makes the results applicable across the full RCP spectrum without committing to a single category.

\begin{figure}
    \centering
    \includegraphics[width=0.85\textwidth]
    {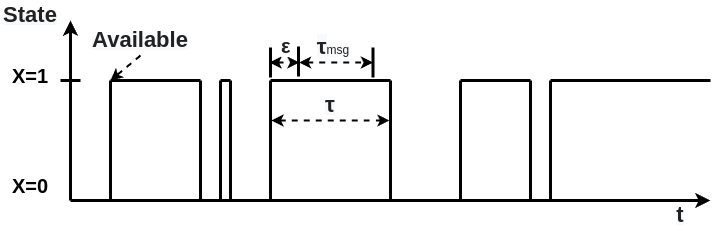}
    \caption{The illustration of communication signal reliability \citep{pang2026reliability}. }
    \label{fig: squarewave}
\end{figure}

RCP performance metrics, particularly Availability ($P_A$) and Continuity, are modeled using a Continuous-Time Markov Chain (CTMC), where the communication link alternates between \textit{on} and \textit{off} states. The full CTMC formulation and the derivation of these metrics, including the notation $P_A$ adopted in this paper, were developed and established in our previous work \citep{pang2026reliability}. Considering the following key metrics,
\begin{itemize}
    \item \textit{Availability} ($P_A$) represents the long-term proportion of time the communication service remains operational, calculated based on the expected durations of these states. $P_A$ is the primary communication-uncertainty parameter swept in our experiments.
    \item \textit{Continuity} indicates the probability that, once communication service becomes available, it stays continuously operational for the entire duration required to complete a transaction.
\end{itemize}

\Cref{fig: squarewave} is a simple illustration of the communication signal alternating between the \textit{on} and \textit{off} states, generated under the CTMC framework. In \Cref{fig: squarewave}, $\tau$ is the time since the system most recently became available, $\tau_{msg}$ is the minimum time duration required to finish the transmission of message in the communication system, while $\varepsilon$ represents the one way system latency in general. The detailed formulation and mathematical representations of these metrics is provided in Section 3 of \citep{pang2026reliability}, along with the newly developed communication reliability metric, \textit{Communicability}. We note that the specific $P_A$ values used in this paper ($P_A \in \{0.5, 0.6, 0.7, 0.8, 0.9, 1.0\}$) are adopted as an exploratory sensitivity sweep rather than being calibrated to a specific RCP category or link-budget measurement. The goal is to characterize the \textit{shape} of the reliability surface across the operationally plausible range of availability, from the idealized case ($P_A = 1.0$) down to severely degraded-link conditions. Mapping specific RCP categories or link budgets (voice VHF, CPDLC/AeroMACS, oceanic vs.\ terminal airspace) to empirically measured $P_A$ values is airspace- and technology-dependent and is left as future work.

\subsection{Pilot Response Time ($\eta$) Modeling}
\begin{figure}
    \centering
    \includegraphics[width=0.85\textwidth]
    {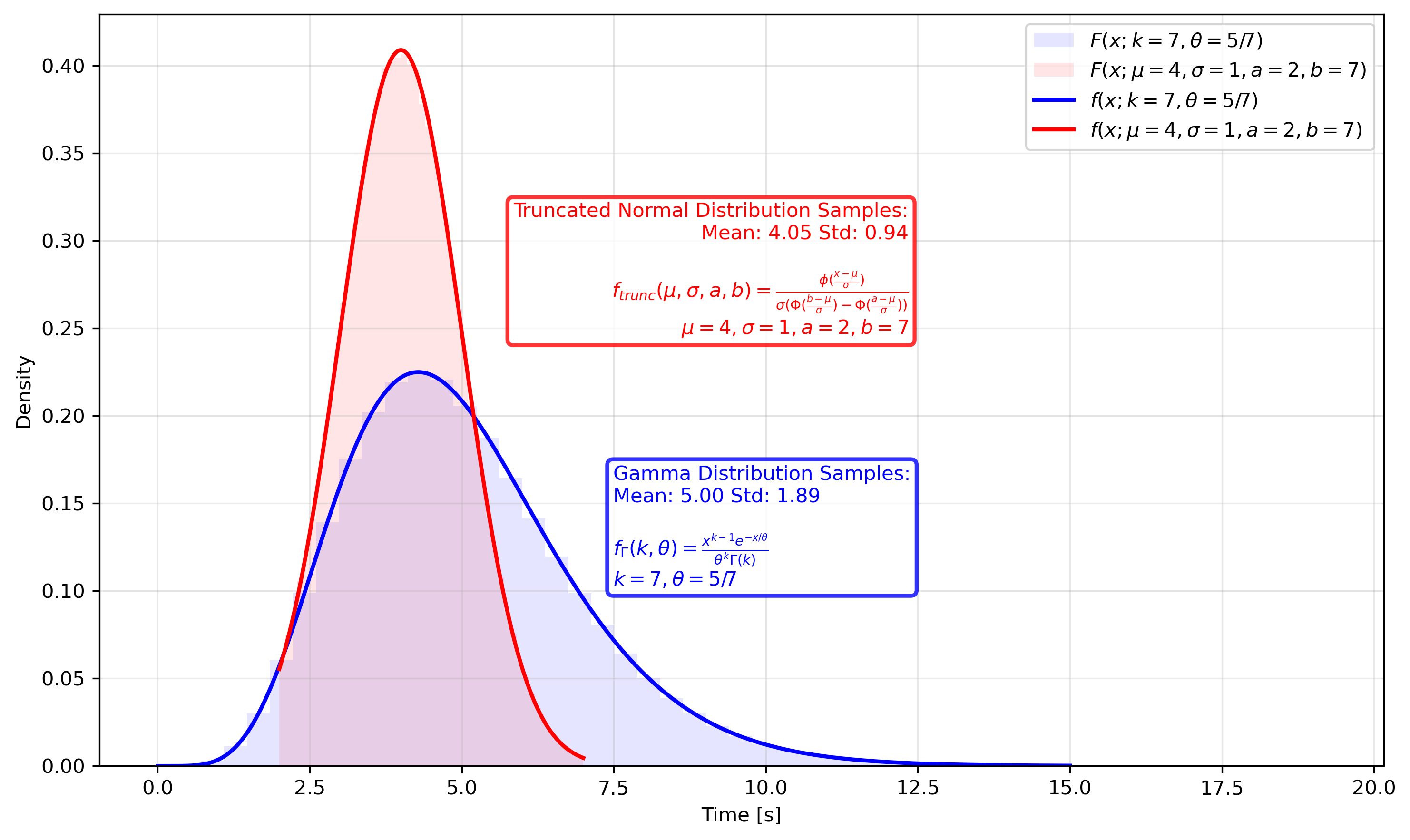}
    \caption{The pilot response model and message transaction time model. We adopt Gamma distribution of pilot response model, and truncated normal distribution in pilot-ATC message transaction time model.}
    \label{fig: uncertainty-models}
\end{figure}

The pilot response time has been studied and considered in various studies, such as the investigation of the benefit of utilizing satellite-based Automatic Dependent Surveillance-Broadcast (ADS-B) in Traffic Alert and Collision Avoidance System (TCAS) II as alternative to the onboard radar-based positioning \citep{romli2008impact}, and the studies on noise footprint, time/fuel efficiency, and arrival procedure design of Continuous Descent Approach (CDA) \citep{ta2001mitigating, clarke2006development, ren2007modeling, cao2011evaluation}. Pilot response time is typically referred as the time taken for the pilot to take physical action after receiving a certain instruction or clearance. As shown in \cite{kuchar1995probabilistic}, the pilot response delay can be modeled as a Gamma distribution parameterized by the shape parameter $k$ and scale parameter $\theta$ are used to control the desired mean value of the samples. This work provides a complete study of the effect of mean pilot response time on the risk of incident, and concludes that a two seconds increase (from 5 seconds to 7 seconds) on pilot response delay can increase the safety risk of TCAS II by a factor of 25 (from $2\times10^{-5}$ to $5\times10^{-4}$). Also, as pointed out in \citep{romli2008impact}, a valid assumption is that the pilot will response within 5 seconds after receiving a given command or clearance of traffic advisories. As a conclusion, we use the same assumption of 5 seconds mean pilot response time in our simulation. \Cref{fig: uncertainty-models} shows the pilot response distribution in blue.

\subsection{Message Transaction Time ($\tau_{msg}$) Modeling}
Effective pilot–ATC voice communication is critical for maintaining safety and operational efficiency in the National Airspace System, yet it is frequently challenged by a variety of cognitive, procedural, and technical factors \citep{pang2026voice}. Both field and simulation studies have shown that communication errors often stem from high message complexity, suboptimal timing, and pilot working memory limitations \citep{morrow1993influence, prinzo2002improving}. In particular, longer ATC messages, typically those with four or more commands, are significantly more prone to misunderstanding, incomplete readbacks, and clarification requests. These issues are exacerbated when controllers deliver messages too quickly or with insufficient spacing between transmissions, which interferes with pilots' ability to process and retain information \citep{morrow1993influence}. Conversely, segmenting complex instructions into shorter, well-timed messages has been found to reduce cognitive load and improve communication accuracy, albeit at the cost of additional transmission turns \citep{prinzo2002improving}. Timing studies indicate that short ATC messages generally last between 2 and 4 seconds, while longer messages extend to approximately 5 to 7 seconds \citep{morrow1993influence, prinzo2002improving}.

This distribution reflects operational and cognitive constraints inherent in radio-based communication and forms the empirical basis for modeling message duration using a truncated normal distribution. Specifically, a distribution defined by parameters $\mu=4, \sigma=1$, lower bound $a=2$, and upper bound $b=7$ is a reasonable and empirically grounded choice for modeling the duration of pilot-ATC message transaction $\tau_{msg}$ in voice communication, while the mean value of 4 seconds and standard deviation of 1 second covers the typical variability of message length. It captures the typical variability in message transmission time while excluding implausibly short or excessively long durations. Such modeling supports probabilistic analysis in human–automation systems and has been adopted in recent frameworks for evaluating pilot response latency and alerting system performance \citep{chryssanthacopoulos2011collision}. \Cref{fig: uncertainty-models} shows the samples from the message transaction time model in red.

\subsection{Performance Evaluation Metrics \label{subsec: metrics}}
We adopt three key metrics in evaluating the performance of final arrival procedures, (i) the runway throughput, which directly gauges the runway capacity and efficiency; (ii) the immediate turn fraction, which is the proportion of arriving aircraft that can turn from the downwind leg to final approach without any downwind holding delays; and (iii) the downwind holding delays, represents the straight-in congestion induced turn-to-base delays. 

\textit{Throughput} is defined as the number of aircraft that can be safely handled on a runway or runway system in a given period, typically expressed per hour \citep{simpson2016airport, tee2018modelling}. Higher throughput indicates that the runway configuration, separation minima, and traffic mix allow more flights to cycle through per unit time. In reliability engineering, tracking throughput under varying uncertainty (i.e., stochastic arrivals or intermittent control availability) reveals how robust a particular configuration is, while a steep drop in throughput under small perturbations signals poor resilience, whereas configurations that sustain high throughput despite uncertainty demonstrate superior reliability. Throughput is a primary indicator in runway capacity studies, which gauges the runway's capacity and efficiency, reflecting how many arrivals can be safely managed \citep{chen2024research}.

\textit{Immediate turn fraction} represents the percentage of arrivals that can transition from the downwind leg to final approach without any holding delays, preferable zero delay beyond their scheduled arrival time. A high immediate turn fraction signifies that most aircraft receive clearance for final approach as soon as they arrive in sequence (i.e., no downwind holding needed), denoting a smoothly functioning and resilient operation, whereas a low fraction means many aircraft require holding patterns (extended downwind legs) due to spacing adjustments or congestion. Incorporating such a metric is important because holding patterns are a standard tactic to manage excess demand, and tracking the number or duration of holds offers insight into operational efficiency and buffer use \citep{chen2024research}. Immediate turn fraction directly captures system reliability where higher fraction indicates strong resilience to uncertainty where more aircraft proceed uninterrupted.

\textit{Average downwind holding delay} captures the congestion-induced waiting time for arrivals in the downwind leg, and quantifies how much extra airborne time traffic must consume when demand approaches or exceeds capacity. In the simulation at Changi airport, the authors also record the average holding duration of arrival flights once runway capacity is reached, and shows that, once the runways have reached the maximum capacity, the average holding duration of airborne aircraft increases exponentially with additional flight movements \citep{tee2018modelling}. Another similar study highlights delay as a primary performance indicator, noting that many scheduling studies aim to minimize the total, average or weighted delay of all flights (e.g., FCFS inefficiencies and Constrained Position Shifting approaches) \citep{ng2022impact}. By focusing on downwind holding delays, one isolates the induced cost (i.e., fuel burn, controller workload, passenger inconvenience) of managing excess demand in the approach stream.

Together, these metrics provide a holistic view of runway behavior and resilience in high-density arrival scenarios. By incorporating various sources of uncertainty into the simulation, robust configurations can be identified that capture and maintain runway performance resilience under real-world variability.

\section{Discrete Event Simulation \label{sec: case1}}
This simulation models two distinct aircraft arrival flows converging toward a runway threshold, the straight-in flow and the downwind flow. This setup reflects operations in congested terminal areas, where controllers rely on vectoring along the downwind-to-final leg to manage spacing and maintain throughput \citep{favennec2009point}. Arrivals are modeled as modified Possion processes to capture the stochastic nature of high-demand operations, with minimum separation enforced according to wake vortex safety standards \citep{bolender2000evaluation}. A Discrete Event Simulation (DES) framework is used to represent and evaluate the performance of merging these flows, based on the following assumptions: (i) the simulation runs for a fixed duration of one hour, mirroring the hourly-capacity convention adopted by FAA AC 150/5060-5 \citep{faa15050605}. The metrics are computed only over events that complete within this one-hour window: the runway throughput counts all aircraft (both straight-in and successfully merged downwind) that cross the runway threshold within the hour; the downwind immediate-turn fraction is computed as the ratio of immediate-turn merges to the total number of downwind aircraft that enter the scenario within the hour; and the average downwind holding time is averaged only over those downwind aircraft that successfully merge before the cutoff, while aircraft still in the holding queue at the end of the hour are not included in the holding-time average. This handling preserves the conventional hourly-throughput definition but, as discussed in \Cref{subsec: case1-mc} and \Cref{fig: slope}, artificially depresses the estimated holding time at very high downwind demand because long-delay aircraft are truncated by the horizon -- a finite-horizon limitation we explicitly revisit in \Cref{sec: conclusion}; (ii) all aircraft are assumed to perform a constant-bank coordinated turn along a circular arc with a $3^\circ$ curvature, resulting in a fixed turning time of 60 seconds; and (iii) all aircraft follow the same speed profile for simplification. A Monte Carlo-based DES with importance sampling is employed to efficiently explore both typical and rare-edge scenarios, including severe communication failures and pilot delays. The mathematical formulation and merging logic are detailed in the following subsections.

\subsection{Generation of Arrival Flows}


The downwind and straight-in arrival flows are modeled using the shifted Possion processes with enforced minimum inter-arrival separations. Possion process with rate $\Lambda$ is adopted to describe the number of occurrence in a fixed time interval, which is used to model the arrival of aircraft waiting to merge in the near-terminal airspace. However, in the arrival flow generation process, we need to know when each arrival enters the vectoring scenario. This leads to the inter-arrival times between Possion process, which follows the exponential distribution with mean of $1/\Lambda$. That is, if the occurrence of arrivals $N(t)$ follow,
\begin{equation}
    N(t) \sim \text{Possion}(\Lambda t)
\end{equation}
Then, the inter-arrival times between each events follows,
\begin{equation}
    \Delta t \sim \text{Exp}(\Lambda) 
\end{equation}

The arrival generation process is thus summarized as adding multiple generated sampled inter-arrival times from the exponential distribution, and stop under the total simulation time $T_{max}$ is reached. Moreover, to make sure the generated arrival aircraft is well separated when entering the airspace of interest, and runway occupancy time of landing aircraft is properly considered in the simulation, we modify the standard exponential arrival to enforce the minimum separation time $T_{sep}$ to satisfy the IFR single-runway separation requirements \citep{odoni1987flow, jacquillat2018roadmap}, which defines two criteria for safe consecutive arrival operations, (a) the airborne final approach separation (i.e., \Cref{tab: separation}) $\mu$ must be satisfied ; (b) the leading aircraft must be clear of the runway before trailing aircraft touches down (i.e., the runway occupancy time $T_{\nu}$).

\begin{table}[]
\centering
\begin{tabular}{c|cccc}
\hline
Arrival-Arrival (nm)   & \multicolumn{4}{c}{Leading Aircraft}                                                       \\ \hline
Trailing Aircraft & \multicolumn{1}{c|}{Small} & \multicolumn{1}{c|}{Large} & \multicolumn{1}{c|}{B757} & Heavy \\ \hline
Small             & \multicolumn{1}{c|}{2.5}   & \multicolumn{1}{c|}{4}     & \multicolumn{1}{c|}{5}    & 6     \\ \hline
Large             & \multicolumn{1}{c|}{2.5}   & \multicolumn{1}{c|}{2.5}   & \multicolumn{1}{c|}{4}    & 5     \\ \hline
B757              & \multicolumn{1}{c|}{2.5}   & \multicolumn{1}{c|}{2.5}   & \multicolumn{1}{c|}{4}    & 5     \\ \hline
Heavy             & \multicolumn{1}{c|}{2.5}   & \multicolumn{1}{c|}{2.5}   & \multicolumn{1}{c|}{4}    & 4     \\ \hline
\end{tabular}
\caption{IFR Airborne separation requirements on single runway final approach for consecutive arrivals (in nautical miles).}
\label{tab: separation}
\end{table}

\begin{figure}
    \centering
    \includegraphics[width=0.5\textwidth]
    {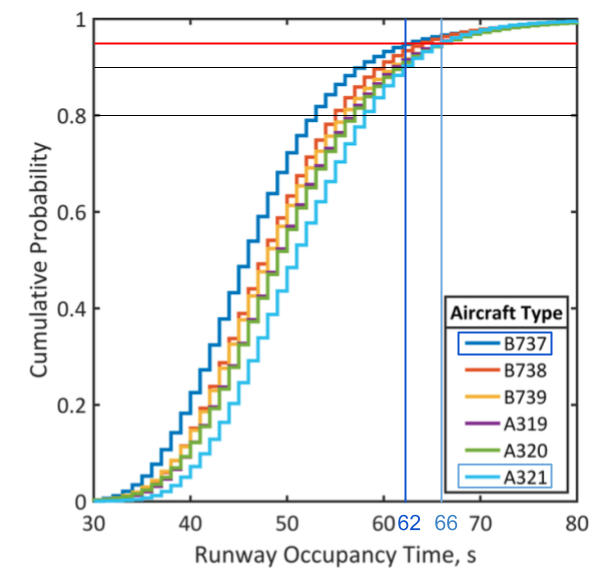}
    \caption{Cumulative distribution of runway occupancy time for common \textit{large} weight class \citep{meijers2019data}. From the figure, we show that the $95\%$ interval of runway occupancy time ($T_\nu$) for the arrival aircraft are between 62 seconds (for B737) to 66 seconds (for A321).}
    \label{fig: occupancy}
\end{figure}

In this sense, the $T_{sep} = \max(T_\mu, T_\nu)$ is the maximum value between the aircraft separation in seconds at runway threshold ($T_\mu$) and the runway occupancy time of the previous landed aircraft ($T_\nu$), where $T_\mu$ is simplified as $\mu/V_{ref}$ and assumed to be 64 seconds. This corresponding to the 2.5 nautical mile final spacing separation with the assumed final approach speed ($V_{ref}$) of 140 knots (i.e., 1.23 times the stall speed of A321 at 80\% maximum landing weight \citep{salgueiro2025aircraft}).

Finally, the arrival simulation modifies the stochastic nature of Possion arrivals while enforcing a separation regulation used in safety-critical systems,
\begin{equation}
    \Delta t_i = T_\text{sep} + X_{i}, \quad X_i \sim \text{Exp}(\Lambda)    
\end{equation}

Specifically, \textit{Straight-in} arrivals approach directly toward the runway threshold without the need for Base turns. The inter-arrival times of straight-in arrivals are represented as a shifted exponential distributions as,
\begin{equation}
H_i = T_{sep} + X_{StraightIn, i}, \quad X_{StraightIn, i} \sim \text{Exp}(\Lambda_{StraightIn}),
\end{equation}
resulting in the exponential distribution parameter:
\begin{equation}
\beta_{StraightIn} = \frac{1}{T_{sep} + \frac{1}{\Lambda_{StraightIn}}}, \quad \mathbb{E}[H_i] = T_{sep} + \frac{1}{\Lambda_{StraightIn}}
\end{equation}

Similarly, \textit{Downwind} arrivals initially approach along a downwind leg and merge into the straight-in flow by executing a constant bank angle Base turn. Their inter-arrival times are thus defined as,
\begin{equation}
D_j = T_{sep} + X_{Downwind, j}, \quad X_{Downwind, j} \sim \text{Exp}(\Lambda_{Downwind})
\end{equation}
with parameterization:
\begin{equation}
\beta_{Downwind} = \frac{1}{T_{sep} + \frac{1}{\Lambda_{Downwind}}}, \quad \mathbb{E}[D_j] = T_{sep} + \frac{1}{\Lambda_{Downwind}}
\label{eq: beta-dw}
\end{equation}

It is worth distinguishing the two rate parameters that appear in the formulation. $\Lambda$ is the rate of the underlying \textit{stochastic} component, i.e., the Poisson rate of the exponential inter-arrival gaps drawn before the minimum-separation shift is applied; it controls the variability of the arrival process. $\beta = 1/\mathbb{E}[H]$, in contrast, is the \textit{effective} arrival rate of the resulting shifted-Poisson process after the separation minimum $T_{sep}$ has been enforced, and is therefore the quantity that directly determines the realized aircraft flow. Because $T_{sep}$ caps the inter-arrival time from below, $\beta$ is bounded above by $1/T_{sep}$ even as $\Lambda \to \infty$, whereas $\Lambda$ has no such operational ceiling. For this reason, all sensitivity results in \Cref{subsec: case1-mc} are reported as functions of $\beta_{StraightIn}$ and $\beta_{Downwind}$ rather than $\Lambda$: $\beta$ is directly comparable to operational arrival rates (in aircraft/hour) and to the analytical throughput bound derived in \Cref{app: throughput-verification}. By adjusting the rate parameters $\Lambda_{StraightIn}$ and $\Lambda_{Downwind}$ of the exponential components, we can generate flows of aircraft at various effective densities $\beta_{StraightIn}$ and $\beta_{Downwind}$.

However, when sampling $\Lambda$, we employ a log-uniform sampling mechanism rather than a uniform sampling to ensure proper coverage across multiple orders of magnitude. This is mathematically implemented by first sampling $\Lambda_{StraightIn}$ and $\Lambda_{Downwind}$ uniformly in logarithmic space and then transforming back to linear space through exponentiation. Specifically, for a given range $[\Lambda_{\min}, \Lambda_{\max}]$, we generate samples using the transformation $\exp(U)$, where $U$ is uniformly distributed over $[\ln(\Lambda_{min}), \ln(\Lambda_{max})]$. This approach ensures that each order of magnitude has equal probability mass. This is particularly important for air traffic simulations where $\Lambda_{StraightIn}$ and $\Lambda_{Downwind}$ can span several orders of magnitude (e.g., from 1 to 3000), and we need to ensure that our simulation adequately explores both low and high arrival rate scenarios without bias towards any particular scale. The log-uniform distribution is therefore a more appropriate choice as it maintains scale invariance and provides equal sampling probability across different orders of magnitude, which is crucial for capturing the full range of possible scenarios in the simulation.

\subsection{Downwind Merging Logic}

\begin{figure}[htbp]
\centering
\resizebox{0.85\textwidth}{!}{
\begin{tikzpicture}[
    node distance=0.6cm and 0.5cm, 
    >=Stealth,
    box/.style={
        rectangle, 
        draw, 
        rounded corners, 
        minimum width=8cm, 
        minimum height=1.2cm, 
        align=center, 
        text width=7.5cm,
        fill=blue!10
    },
    decision/.style={
        diamond, 
        draw, 
        aspect=1.8, 
        minimum width=6cm, 
        minimum height=1cm, 
        align=center, 
        text width=5.5cm,
        fill=orange!20
    },
    arrow/.style={->, thick, shorten >=2pt, shorten <=2pt}, 
    startstop/.style={
        ellipse, 
        draw, 
        minimum width=6cm, 
        minimum height=1cm, 
        align=center,
        fill=green!20
    }
]

\node[startstop] (start) {Start: Downwind aircraft arrives at initial time $t_{0}$};

\node[box, below=of start] (signal) {Check continuous signal availability for duration $\tau_{msg}$ from current time $t$};

\node[decision, below=of signal] (signalcheck) {Is signal available?};

\node[box, right=of signalcheck, xshift=4cm] (increment_t) {Evaluation reset: \\$t \leftarrow t + dt$};

\node[box, below=of signalcheck, yshift=-0.2cm] (propose) {Compute proposed merge time: \\ $\widehat{t}_{merge} = t + \varepsilon + \eta + T_{Base}$};

\node[decision, below=of propose, yshift=-0.2cm] (gap1) {Initial Gap Verification: \\ Is $\widehat{t}_{merge}$ clear of traffic by $\pm T_{sep}$?};

\node[box, right=of gap1, xshift=4cm] (increment_fine) {Hold and re-evaluate: \\ $t \leftarrow t + dt$};

\node[box, below=of gap1, yshift=-0.2cm] (commit) {Pilot commits. Calculate: \\ $t_{turn} = t + \varepsilon + \eta$};

\node[box, below=of commit] (actualmerge) {Calculate actual merge time: \\ $t_{merge} = t_{turn} + T_{Base}$};

\node[decision, below=of actualmerge, yshift=-0.2cm] (gap2) {Final Gap Verification: \\ Is $t_{merge}$ still clear of traffic by $\pm T_{sep}$?};

\node[box, right=of gap2, xshift=4cm] (reset) {Evaluation reset: \\ $t \leftarrow t_{turn}$};

\node[box, below=of gap2, yshift=-0.2cm] (schedule) {Merge aircraft into schedule at $t_{merge}$. Record holding time.};

\node[startstop, below=of schedule] (stop) {End: Aircraft merged};

\draw[arrow] (start) -- (signal);
\draw[arrow] (signal) -- (signalcheck);
\draw[arrow] (signalcheck) -- node[right, pos=0.4] {Yes} (propose);

\draw[arrow, shorten >=6pt, shorten <=6pt] (signalcheck.east) -- node[above] {No} (increment_t.west);
\draw[arrow] (increment_t.east) -- ++(2.5cm,0) |- (signal.east);

\draw[arrow] (propose) -- (gap1);
\draw[arrow] (gap1) -- node[right, pos=0.4] {Yes} (commit);

\draw[arrow, shorten >=6pt, shorten <=6pt] (gap1.east) -- node[above] {No} (increment_fine.west);
\draw[arrow] (increment_fine.east) -- ++(2.5cm,0) |- (signal.east);

\draw[arrow] (commit) -- (actualmerge);
\draw[arrow] (actualmerge) -- (gap2);
\draw[arrow] (gap2) -- node[right, pos=0.4] {Yes} (schedule);

\draw[arrow, shorten >=6pt, shorten <=6pt] (gap2.east) -- node[above] {No} (reset.west);
\draw[arrow] (reset.east) -- ++(2.5cm,0) |- (signal.east);

\draw[arrow] (schedule) -- (stop);

\end{tikzpicture}
}
\caption{Flowchart showing the downwind merging logic. The process begins when a downwind aircraft is abeam the runway threshold, at which time it evaluates the gap between successive arrivals already in the straight-in traffic stream. For each evaluation, the system first checks if the communication signal is continuously available for the duration of message transmission time $\tau_{msg}$. If the signal is available, then the simulation computes a potential merging time by summing the system latency $\varepsilon$, the pilot response time $\eta$, and the time required to complete the standard constant bank angle turn (i.e., 60 seconds). The system perform two step verification process, (i) It first checks if the proposed merge time will properly maintain the minimum separation time requirement $T_{sep}$ with both the preceding and following aircraft in the straight-in stream, as well as considering the system latency and pilot response delay; (ii) The second verification is performed just before the actual merge to ensure the gap is still available, taking into account any changes in the straight-in flows that may have occurred during the turn. If both verification are passed, the downwind aircraft is inserted into the schedule at the merged position in the straight-in queue, and the corresponding holding time in the downwind leg is recorded. If the gap is unavailable anytime during the process, the downwind aircraft will re-evaluate starting from the point of commitment. The simulation continues continues until either a valid merging opportunity is found, or the maximum simulation time is reached.}
\label{fig: flowchart_compact}
\end{figure}
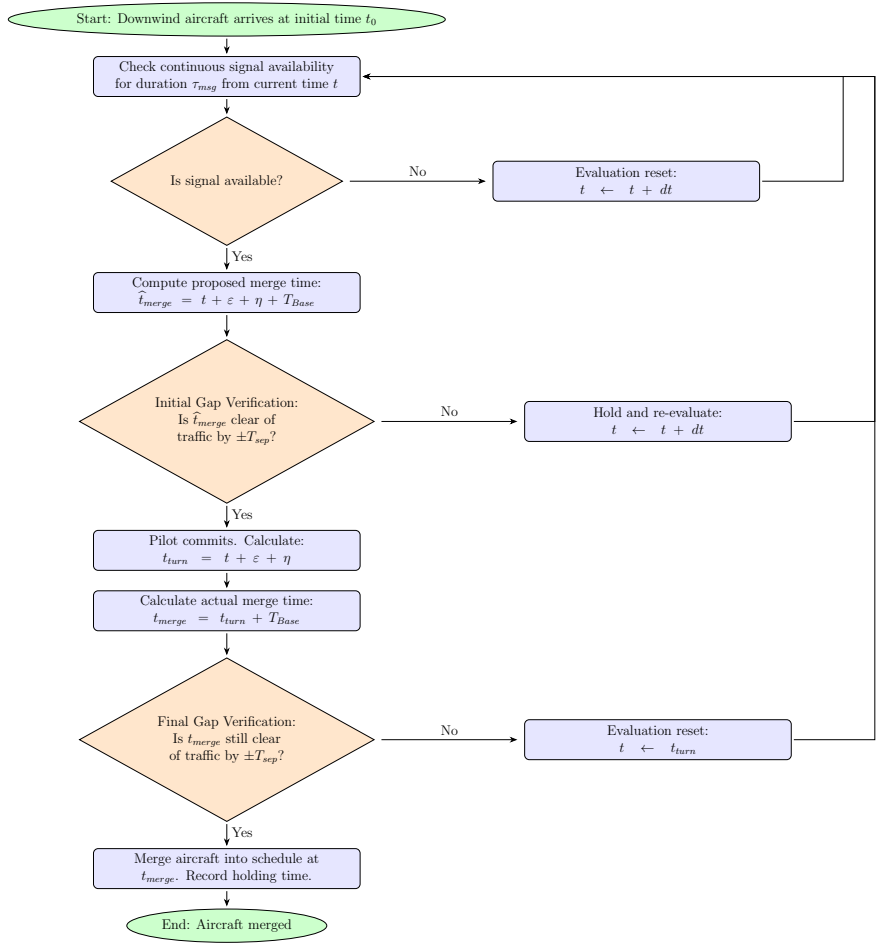

The downwind merging problem is complex because it requires continuously determination of the precise time at which the aircraft from the downwind leg can be integrated into the straight-in arrival flow. This process commences at the moment each downwind aircraft passes abeam of the final approach fix, an required evaluation of the temporal gaps between consecutive arrivals already scheduled in the straight-in traffic stream. A schematic diagram of the merging process from two arrival flows is depicted in \Cref{fig: flowchart} and the flowchart detailing the decision-making steps is shown in \Cref{fig: flowchart_compact}. 

At each evaluation timestep $t$, the system first assesses whether the communication link remains continuously available for at least the duration required to transmit a complete message $\tau_{msg}$. If the communication system is continuously available, the proposed merge time $\widehat{t}_{merge}$ is then computed by incorporating four explicit temporal segments,

\begin{equation}
    \widehat{t}_{merge} = t + \varepsilon + \eta + T_{Base}
\end{equation}

\noindent where $t$ is the current decision time, starting from the downwind aircraft’s initial arrival. $\varepsilon$ represents the one-way communication system latency. $\eta$ denotes the pilot response delay. $T_{Base}$ is the fixed duration (i.e., 60 seconds) required for the aircraft to complete a standard constant bank-angle turn onto final approach from Base.

Following this definition, a two-step verification process is developed to confirm the feasibility of final merging decision. The first verification confirms that the computed potential merge time $\widehat{t}_{merge}$ meets the minimum separation requirement $T_{sep}$ with respect to both the preceding ($S_k$) and following ($S_{k+1}$) scheduled straight-in aircraft in the queue as,
\begin{equation}
    S_{k} + T_{sep} \leq \widehat{t}_{merge} \leq S_{k+1} - T_{sep}
\end{equation}

Upon successful verification of the minimum separation requirement, the pilot commits to initiating the turn at time $t_{turn}$, which is the summation of the current time, random communication latency parameter, and random pilot response delay,
\begin{equation}
    t_{turn} = t + \varepsilon + \eta
\end{equation}

Subsequently, another verification immediately precedes the aircraft's physical merging onto the final approach trajectory, precisely at the actual merge time,
\begin{equation}
    t_{merge} = t_{turn} + T_{Base}
\end{equation}

This step reconfirms that the chosen gap still remains sufficiently large for downwind merging, satisfying the minimum separation constraint at the actual merge time,
\begin{equation}
    S_{k} + T_{sep} \leq t_{merge} \leq S_{k+1} - T_{sep}
\end{equation}

If both verification steps are successful, the downwind aircraft merges into the straight-in queue at the determined position, and the associated holding duration in the downwind leg is recorded accordingly,
\begin{equation}
    t_{hold} = t_{turn} - t_{entry}
\end{equation}

Should either verification fail at any point, the downwind aircraft resets its evaluation starting from the commitment point ($t_{turn}$), incrementally reassessing potential merging opportunities until either a suitable gap is identified or the simulation reaches its predefined maximum duration.

\subsection{Monte Carlo Analysis \label{subsec: case1-mc}}
In the previous sections, we discuss the the modeling of various uncertainty sources, provide the formulation of the discrete event simulation in the final approach merging scenario, as well as the key metrics of runway capacity estimation. Here, we show the high-fidelity contour plots under varying arrival rates from two streams. Specifically, we conduct two sets of experiments to simulation two scenarios, (i) voice-based VHF pilot-controller communication with voice-message length of $\tau_{msg} \sim \mathcal{TN}(\mu, \sigma; a, b)$, which represents the current legacy operational mode; (ii) AeroMACS/CPDLC-based digital datalink communication between remote pilot aircraft systems (RPAS) and the ground station with $\tau_{msg}$ smaller than the simulation time interval (i.e., dt=0.1s), which represents the digital-first communication environment anticipated for highly automated future ATM operations, where datalink is expected to be the primary mode of pilot-controller communication. Because ongoing CNS/ATM modernization programs such as FAA NextGen and SESAR explicitly identify digital datalink as the intended primary channel for future high-density and highly automated operations \citep{federal2013nextgen}, the RPAS/CPDLC configuration is the scenario of primary forward-looking interest in this study; the voice-based results are retained as a present-day baseline so that the efficiency and reliability gains of the transition to digital communication can be quantified under identical traffic and uncertainty conditions. Under these two configurations of the discrete event simulation framework, we conduct extensive Monte Carlo simulations with sampled values and evaluate the runway performance on the metrics defined above.

\subsubsection{Runway Throughput}
\begin{figure}
    \centering
    \begin{minipage}[t]{0.48\textwidth}
        \centering
        \includegraphics[trim={0 0 347.52bp 0},clip,width=\textwidth]{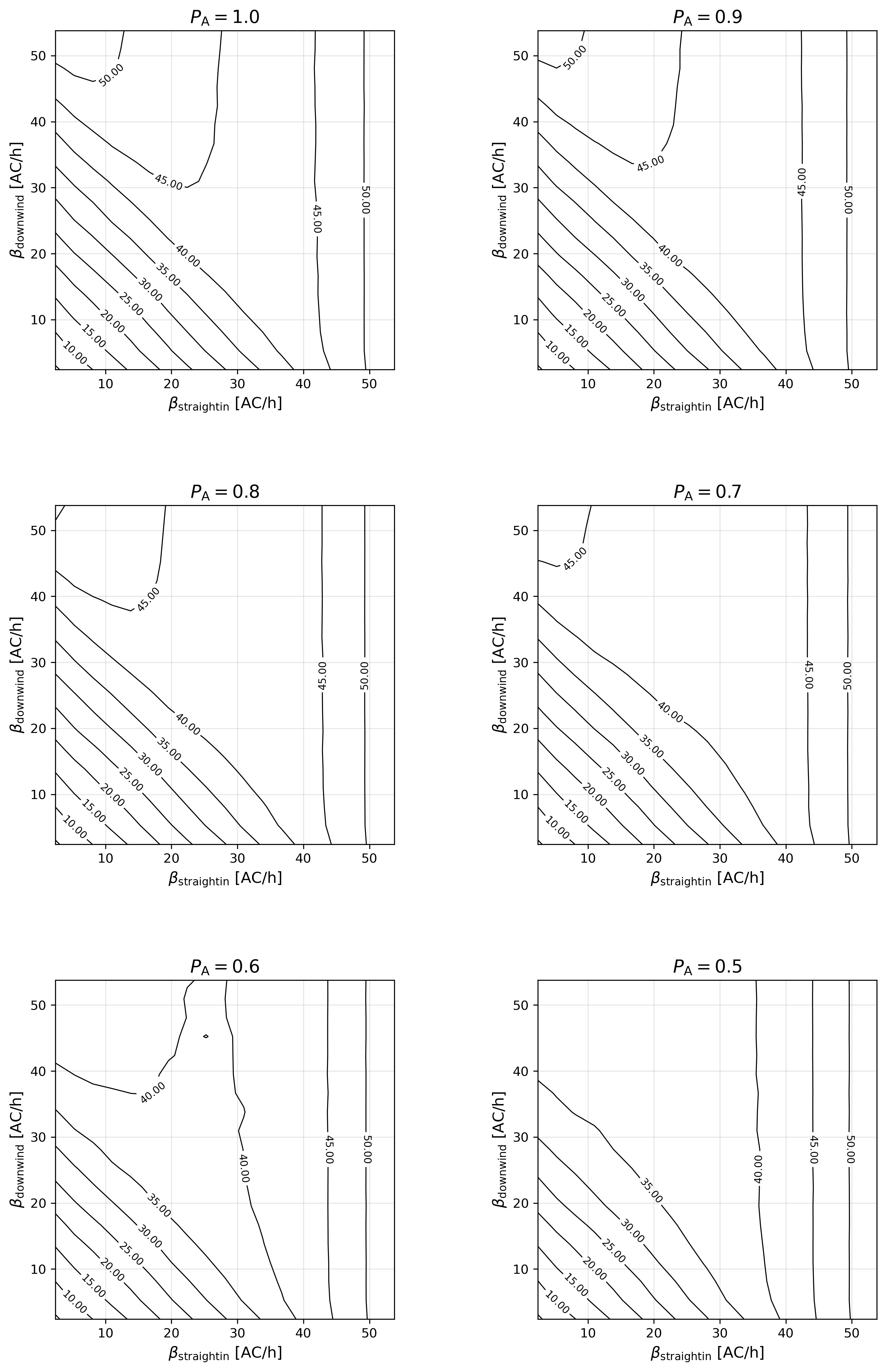}
    \end{minipage}
    \hfill
    \begin{minipage}[t]{0.48\textwidth}
        \centering
        \includegraphics[trim={0 0 347.52bp 0},clip,width=\textwidth]{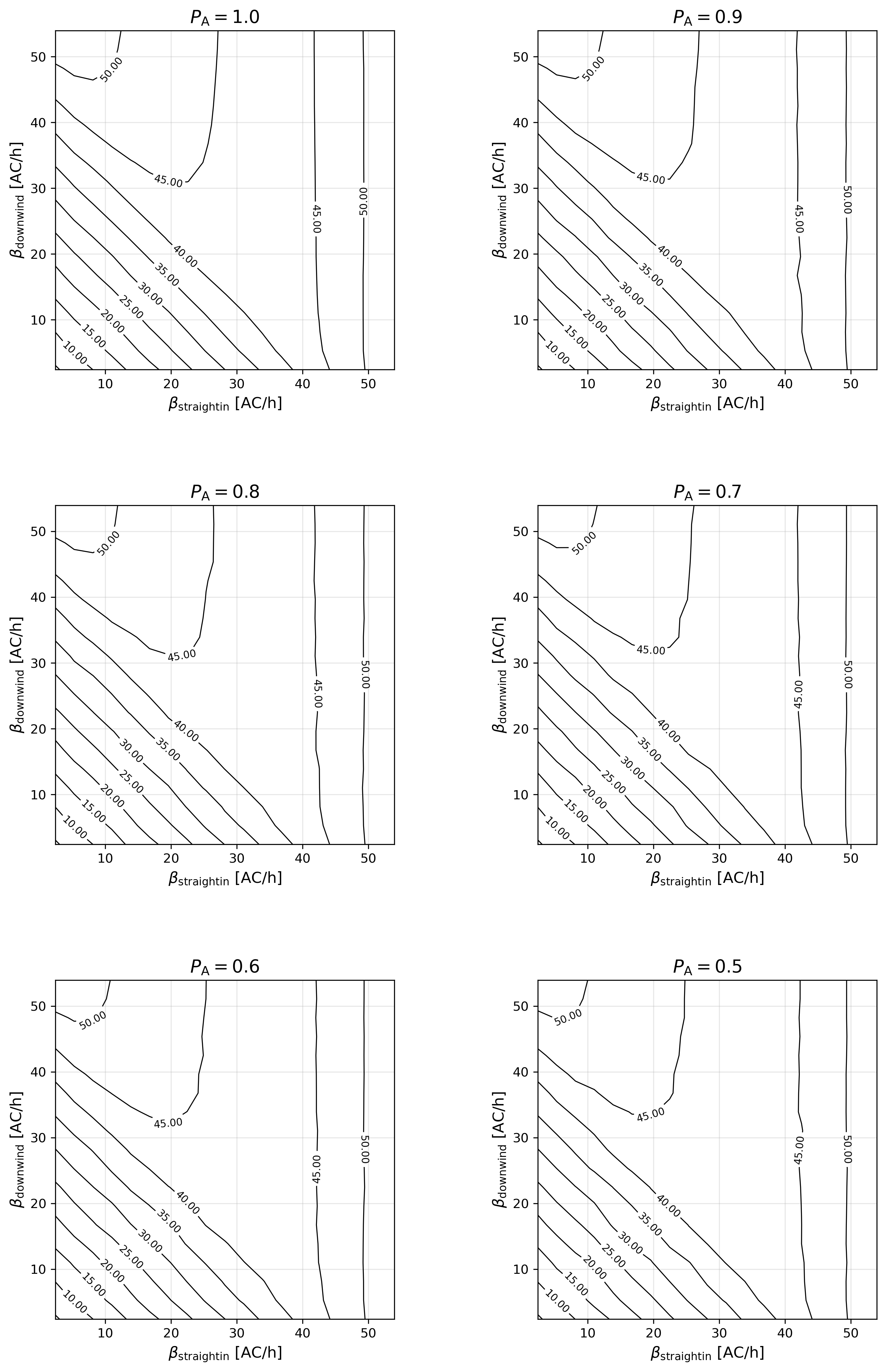}
    \end{minipage}
    \caption{Runway throughput under varying downwind and straight-in arrival rates at selected levels of aeronautical communication availability $P_A \in \{1.0, 0.8, 0.6\}$. Left column: legacy voice-based VHF pilot-controller communications with truncated-normal message transaction time. Right column: future RPAS/CPDLC digital datalink communications with near-instantaneous message transaction time. The complementary panels at $P_A \in \{0.9, 0.7, 0.5\}$ are provided in \Cref{fig: throughput-appendix}.}
    \label{fig: throughput-main}
\end{figure}

\Cref{fig: throughput-main} presents the contour plots of arrival runway throughput after collecting sufficient Monte Carlo sampled results, comparing the legacy voice-based and future RPAS/CPDLC digital datalink configurations side-by-side at $P_A \in \{1.0, 0.8, 0.6\}$. The complementary sweep at $P_A \in \{0.9, 0.7, 0.5\}$ is provided in \Cref{fig: throughput-appendix} of \Cref{app: extended-contours}. The figure illustrates the relationship between runway throughput and the arrival rates from two independent streams.

For moderate arrival rates, the throughput increases nearly linearly as the sum of the two arrival streams, confirming the expected additive effect under unconstrained conditions. However, this linear trend does not persist indefinitely. As either arrival rate $\beta$ approaches approximately 40 aircraft per hour (for those with a higher $P_A$), the throughput curves exhibit pronounced nonlinearities, which shows as curved envelopes in the contour plots. This departure from linearity signifies the impact of merging constraints, where runway throughput becomes dominated by the need to maintain adequate gaps for safe downwind integration. Specifically, when the straight-in arrival rate $\beta_{straightin}$ is low and $\beta_{downwind}$ is high, the system still supports increased throughput, but only to the extent that straight-in arrivals remain sparse enough to allow sufficient merging opportunities. Conversely, if $\beta_{straightin}$ is higher enough, runway throughput again becomes increasingly dominated by straight-in arrivals alone, as evidenced by the near-vertical contour lines in both panels of \Cref{fig: throughput-main}, indicating that few downwind aircraft can successfully merge. 


A critical insight from these figures is the significant influence of communication signal uncertainty on system performance. Under otherwise identical settings (pilot response, latency, etc.), increasing communication uncertainty (i.e., decreased $P_A$) leads to a marked reduction in estimated throughput. This effect is further exacerbated in the presence of longer message transaction times $\tau_{msg}$, as demonstrated by the larger throughput degradation observed in the voice-based communication scenario compared to the RPAS case. The RPAS case is notably more robust on arrival performance degradation, as its infinitesimal transaction time minimizes the impact of communication reliability on merging opportunities and overall throughput.

\begin{figure}
\centering
    \begin{subfigure}[t]{0.95\textwidth}
        \includegraphics[width=\textwidth]{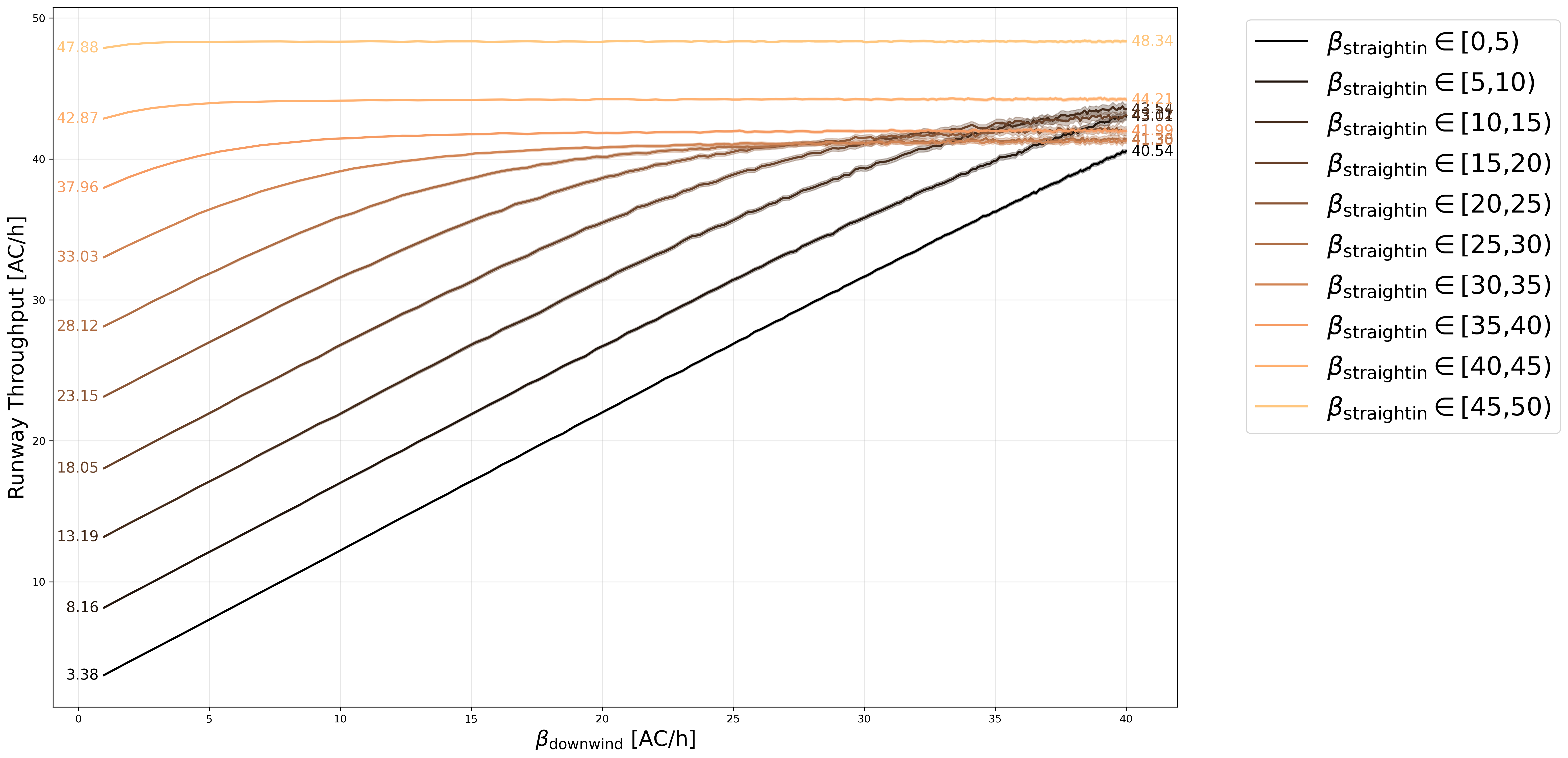}
        \caption{Runway throughput with voice-based communication under grouped straight-in arrival rates.}
        \label{fig: throughput-curve-voice}
    \end{subfigure}
    \begin{subfigure}[t]{0.95\textwidth}
        \centering
        \includegraphics[width=\textwidth]
        {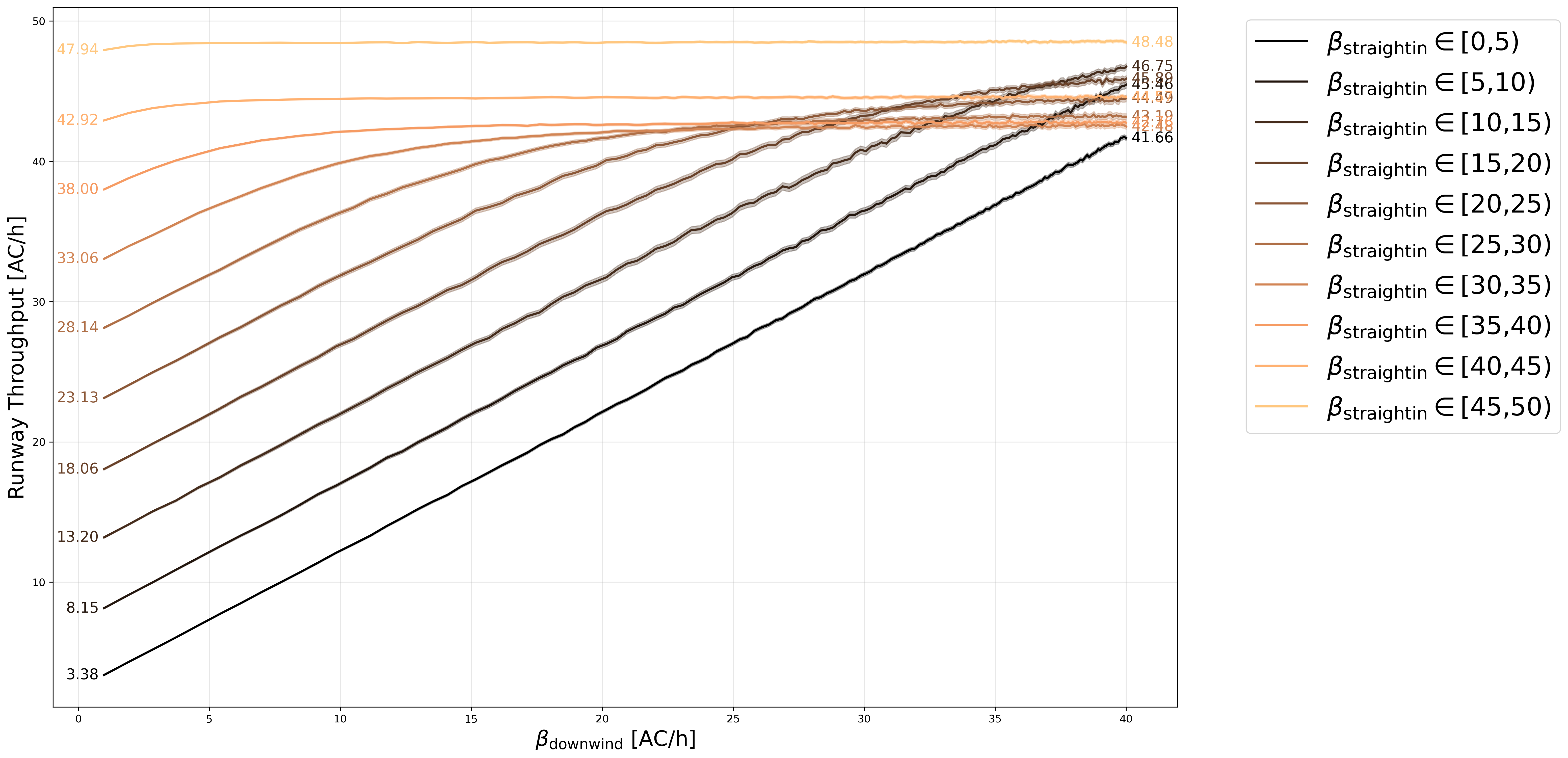}
        \caption{Runway throughput with RPAS under grouped straight-in arrival rates.}
        \label{fig: throughput-curve-rpas}
    \end{subfigure}
\caption{Impact of message transaction time on runway throughput.}
\label{fig: case1-throughput-curve}
\end{figure}

\Cref{fig: case1-throughput-curve} further investigates the behavior or runway throughput under varying arrival rates from two stream, especially the arrivals from the downwind leg. 
Increasing straight-in rates generally boost the runway throughput, until it saturates due to merging constraints, which demonstrates the nonlinear saturation behavior validated by the analytical study. Minimal difference on throughput between these two figures for low $\beta_{downwind}$ but the voice-based communication shows that for high $\beta_{downwind}$, the throughput under voice communication is systematically lower and shows increased curvature and variability in the contour structure. This finding underscores the compounding effect of both high traffic demand and communication delays, which jointly amplify congestion and restrict merging capacity.

These results have important implications for the design and management of terminal area operations. The agreement between analytical and simulated results not only validates the theoretical approach but also quantifies the performance losses attributable to operational imperfections, such as delayed or unreliable communication. As arrival rates approach system capacity, particularly in future high-density or mixed-fleet environments, accounting for realistic transaction dynamics and communication uncertainty becomes essential for accurate capacity estimation and robust operational planning.


Notably, to rigorously consolidate these observed behaviors, we supplement our simulation results with analytical throughput analysis in \Cref{app: throughput-verification}. We show that theoretical estimations closely match the simulation contours, confirming the nonlinear saturation effects and providing confidence in the simulated merging models.

\subsubsection{Downwind Immediate Turn Percentage}
\begin{figure}
    \centering
    \begin{minipage}[t]{0.48\textwidth}
        \centering
        \includegraphics[trim={0 0 347.52bp 0},clip,width=\textwidth]{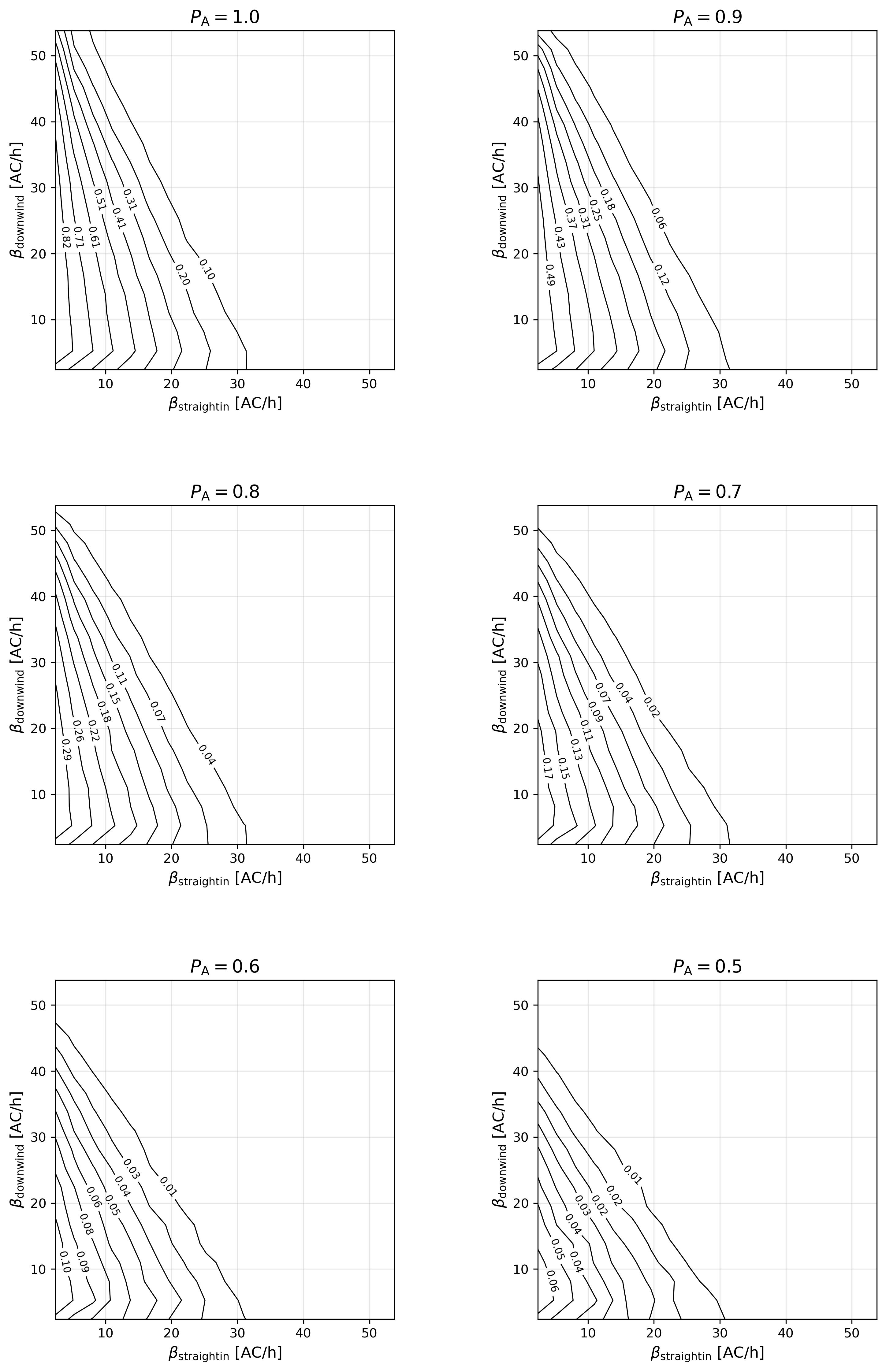}
    \end{minipage}
    \hfill
    \begin{minipage}[t]{0.48\textwidth}
        \centering
        \includegraphics[trim={0 0 347.52bp 0},clip,width=\textwidth]{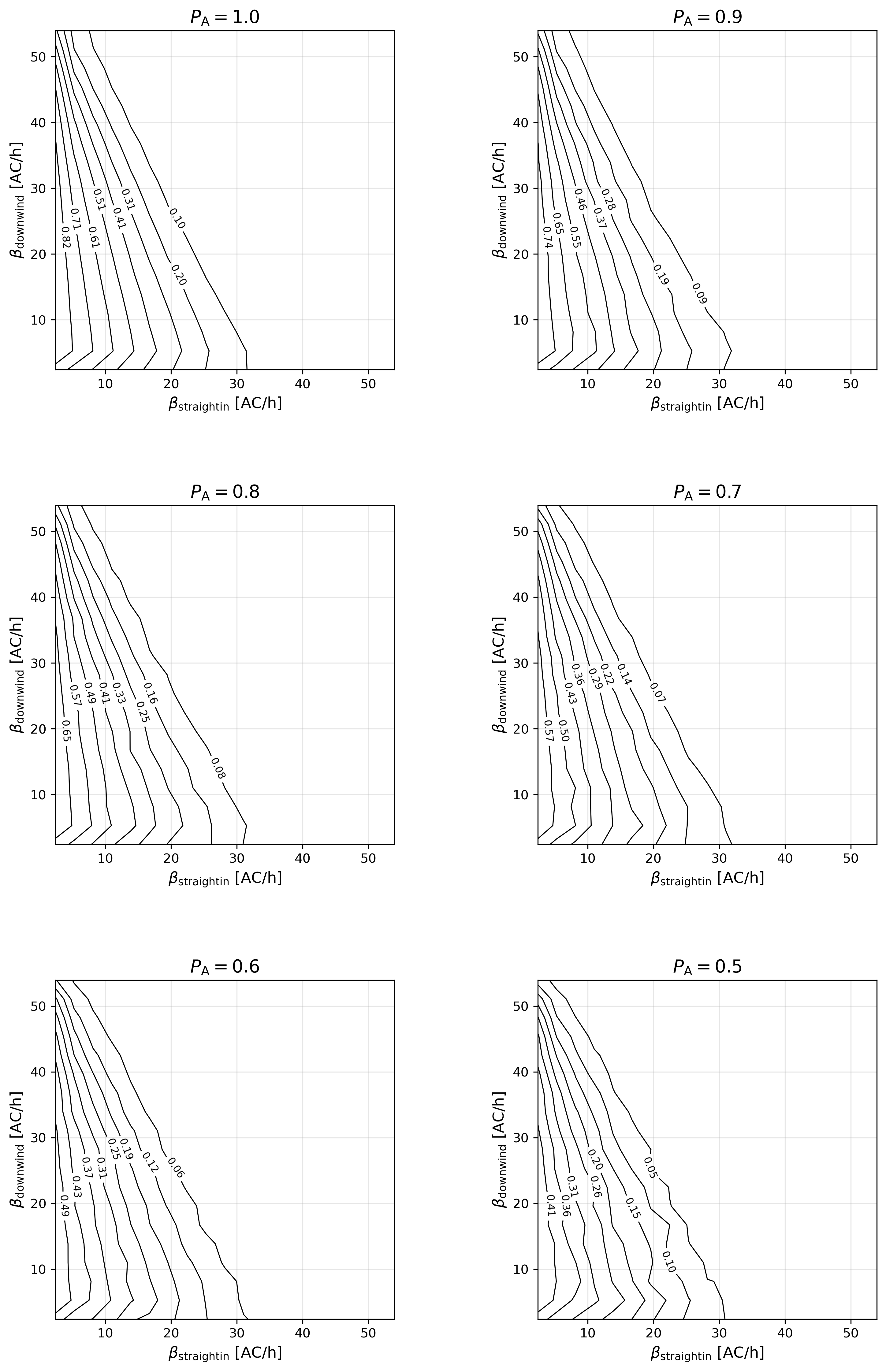}
    \end{minipage}
    \caption{Percentage of immediate turns in downwind arrivals under varying downwind and straight-in arrival rates at $P_A \in \{1.0, 0.8, 0.6\}$. Left column: legacy voice-based VHF pilot-controller radio communications. Right column: future RPAS/CPDLC digital datalink communications. The complementary panels at $P_A \in \{0.9, 0.7, 0.5\}$ are provided in \Cref{fig: immediate-fraction-appendix}.}
    \label{fig: immediate-fraction-main}
\end{figure}

As mentioned, the second runway capacity evaluation metric is the percentage of immediate turns for downwind arrivals upon receiving the turn advisory from the controller. The contour plots in \Cref{fig: immediate-fraction-main} illustrate how the immediate turn fraction varies across the joint parameter space of straight-in and downwind arrival rates ($\beta_{straightin}$ and $\beta_{downwind}$) for the legacy voice-based and future RPAS/CPDLC datalink configurations, shown side-by-side at $P_A \in \{1.0, 0.8, 0.6\}$. The complementary panels at $P_A \in \{0.9, 0.7, 0.5\}$ are provided in \Cref{fig: immediate-fraction-appendix} of \Cref{app: extended-contours}. Each subplot represents a fixed value of $P_A$.

It is obvious that as $P_A$ decreases, the area in the ($\beta_{straightin}$ and $\beta_{downwind}$) space where the immediate turn fraction remains high contracts sharply. For high $P_A$, moderate to high traffic levels can still support relatively efficient merging. As communication becomes less reliable, only very low traffic scenarios maintain a substantial immediate turn fraction. Moreover, The plots also reveal that higher straight-in arrival rates severely restrict the opportunities for downwind aircraft to merge immediately. This effect is evident in the way the contours shift and compress leftward as $\beta_{straightin}$ increases, indicating that straight-in arrivals dominate runway access and reduce available gaps for merging. Furthermore, comparing the RPAS (i.e., minimal message transaction time) to the voice-based vectoring scenario (i.e., longer message transaction time), the RPAS contours consistently show larger feasible regions for immediate merging. In contrast, the voice-based scenario displays more rapid performance degradation as communication becomes less reliable, highlighting the vulnerability of legacy systems to communication-induced delays.

\begin{figure}
\centering
    \begin{subfigure}[t]{0.85\textwidth}
        \includegraphics[width=\textwidth]{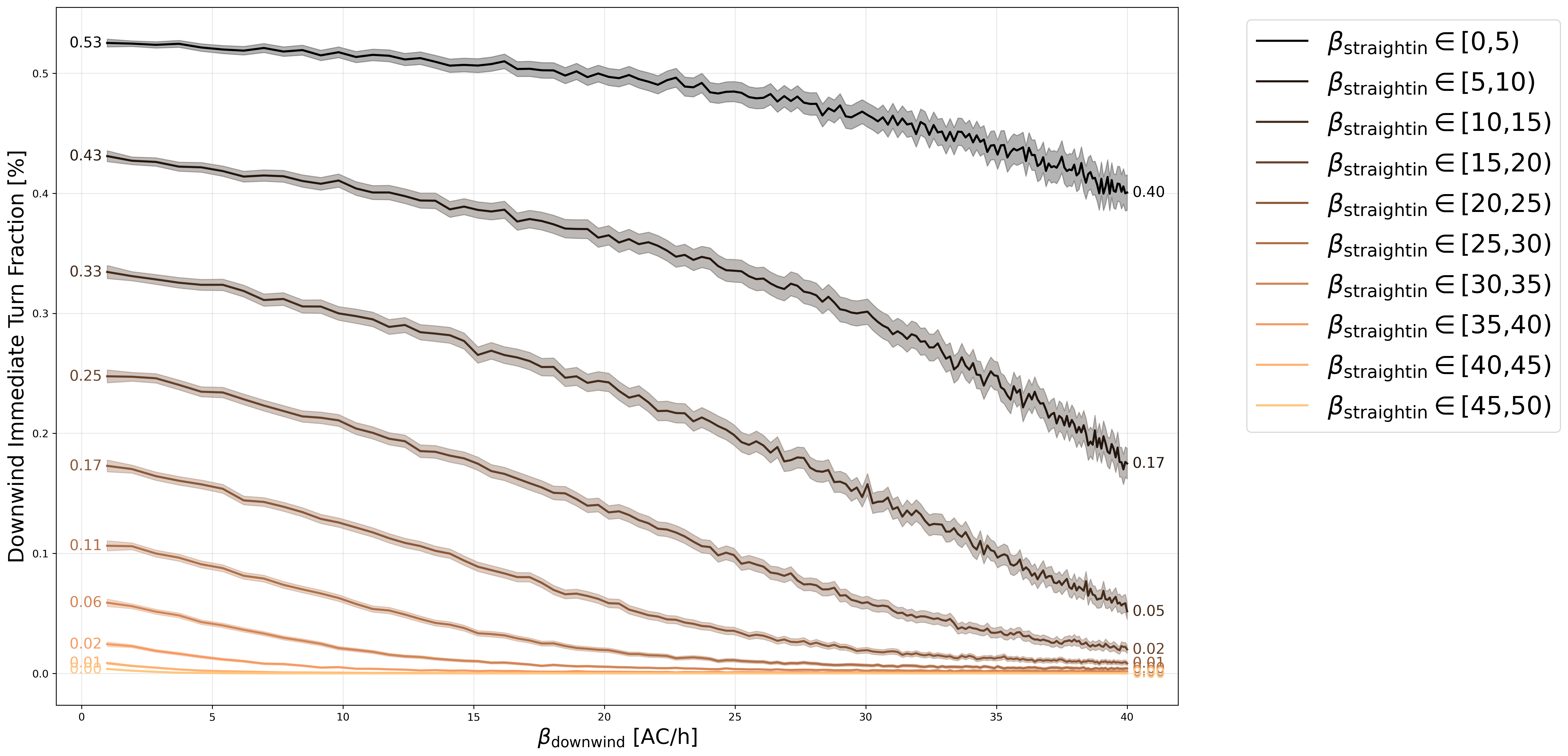}
        \caption{Downwind immediate turn fraction with voice-based communication.}
        \label{fig: immediate-fraction-curve-transaction}
    \end{subfigure}
    \begin{subfigure}[t]{0.85\textwidth}
        \centering
        \includegraphics[width=\textwidth]
        {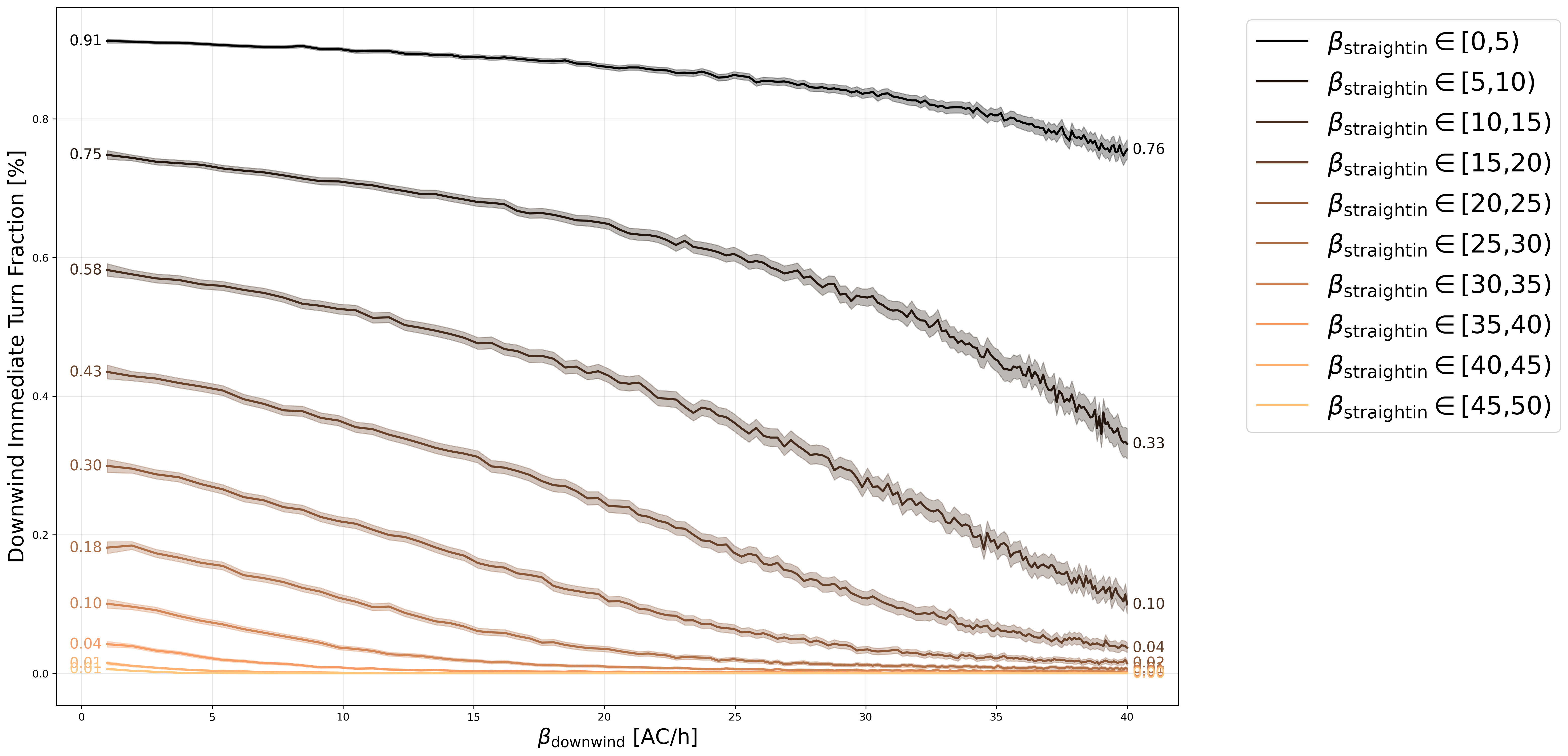}
        \caption{Downwind immediate turn fraction with RPAS.}
        \label{fig: immediate-fraction-curve}
    \end{subfigure}
\caption{Impact of message transaction time on percentage of immediate turns in downwind arrivals.}
\label{fig: fraction-curve}
\end{figure}

Turning to the curve plots in \Cref{fig: fraction-curve}, these present the immediate turn fraction as a function of $\beta_{downwind}$, with each curve corresponding to a range of straight-in arrival rates. In both communication scenarios, the immediate turn fraction decreases monotonically with increasing downwind arrival rate, reflecting the rising likelihood of merge conflicts as traffic demand grows. However, the RPAS scenario consistently supports higher immediate turn fractions across all traffic levels, especially at moderate to high arrival rates. The decline in performance is more gradual, and substantial immediate merging remains feasible even at elevated traffic levels. In contrast, the voice-based scenario exhibits sharply reduced immediate turn fractions as either arrival stream intensifies, with the majority of downwind arrivals requiring holding at high traffic densities or under unreliable communication conditions. The complement of the immediate turn fraction represents the average holding fraction for downwind arrivals. As communication uncertainty increase, a larger proportion of downwind traffic must enter holding, especially in dense traffic scenarios.

In summary, these figures underscore the critical role of communication uncertainty in determining system efficiency. While RPAS architectures are robust even under challenging traffic and uncertainty, voice-based systems are more susceptible to performance degradation due to message transaction times.

\subsubsection{Averaged Downwind Holding Time}
\begin{figure}
    \centering
    \begin{minipage}[t]{0.48\textwidth}
        \centering
        \includegraphics[trim={0 0 347.52bp 0},clip,width=\textwidth]{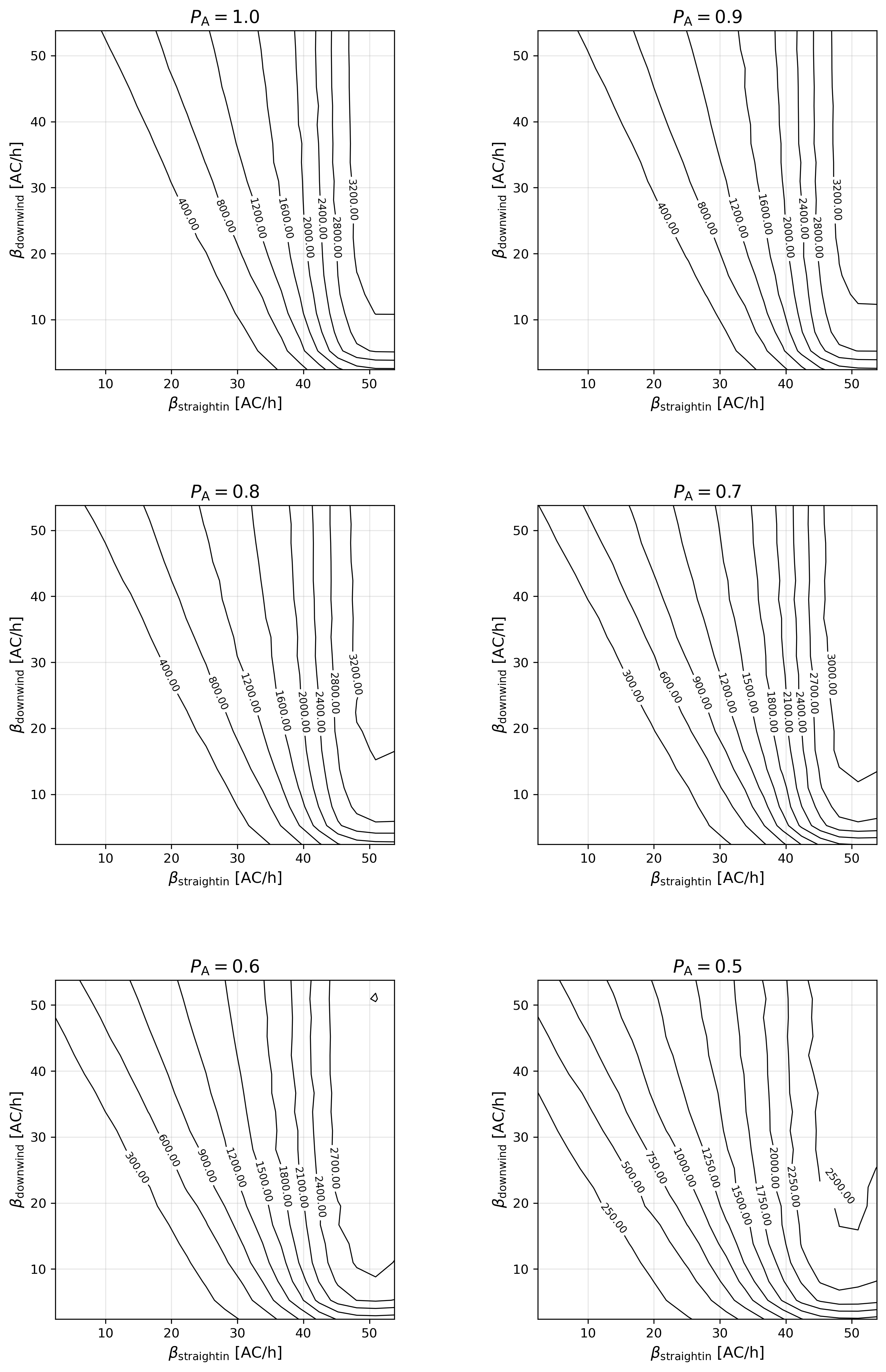}
    \end{minipage}
    \hfill
    \begin{minipage}[t]{0.48\textwidth}
        \centering
        \includegraphics[trim={0 0 347.52bp 0},clip,width=\textwidth]{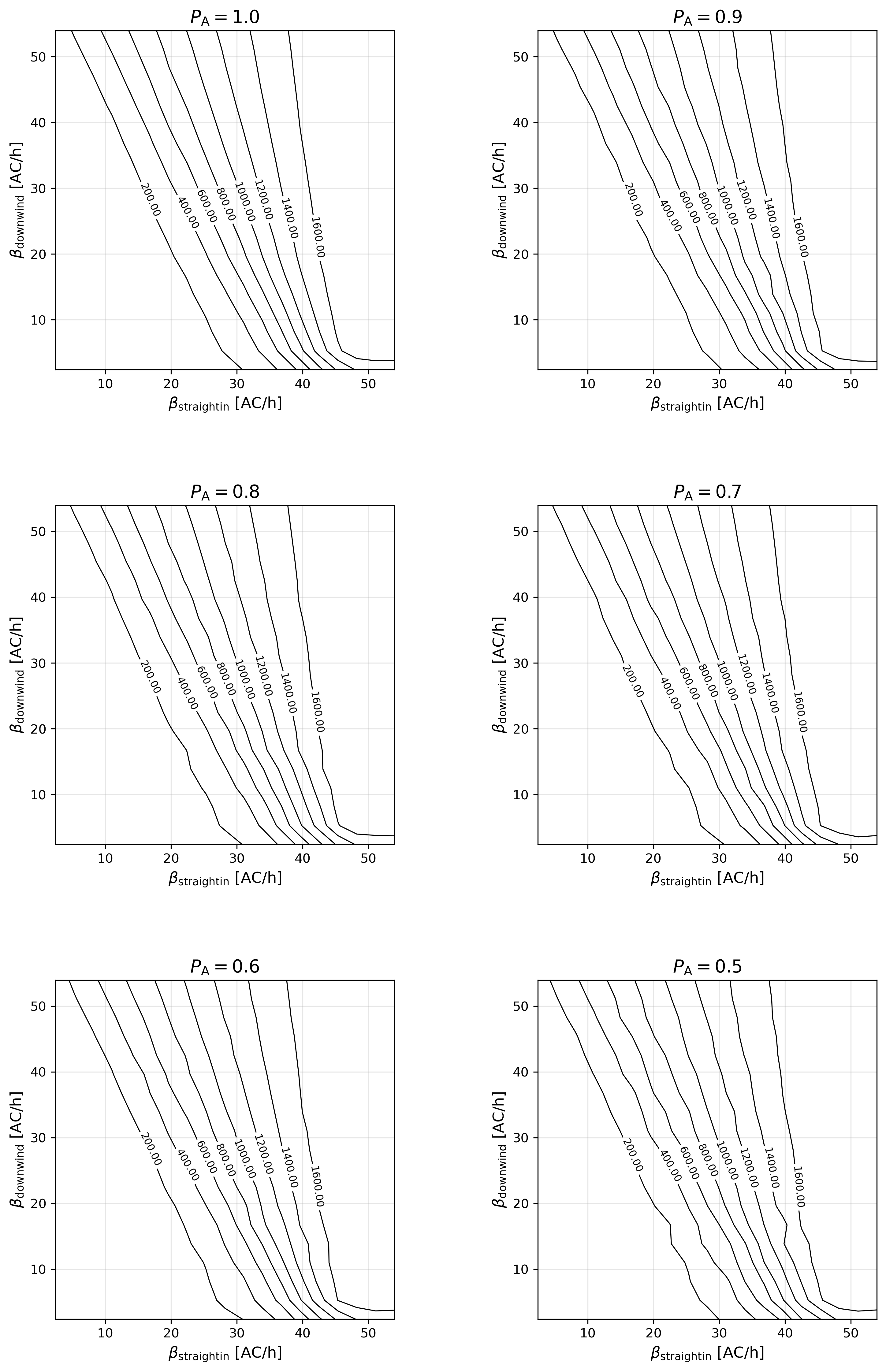}
    \end{minipage}
    \caption{Average holding time of downwind arrivals under varying downwind and straight-in arrival rates at $P_A \in \{1.0, 0.8, 0.6\}$. Left column: legacy voice-based VHF communications. Right column: future RPAS/CPDLC digital datalink communications. The complementary panels at $P_A \in \{0.9, 0.7, 0.5\}$ are provided in \Cref{fig: avg-holding-appendix}.}
    \label{fig: avg-holding-main}
\end{figure}

The contour plots in \Cref{fig: avg-holding-main} illustrate how the average holding time experienced by downwind arrivals varies as a function of both straight-in and downwind arrival rates ($\beta_{straightin}$, $\beta_{downwind}$) for the legacy voice-based and future RPAS/CPDLC datalink configurations, shown side-by-side at $P_A \in \{1.0, 0.8, 0.6\}$. The complementary panels at $P_A \in \{0.9, 0.7, 0.5\}$ are provided in \Cref{fig: avg-holding-appendix} of \Cref{app: extended-contours}.

Not surprisingly, average holding times increase significantly with increased traffic density from both straight-in and downwind streams. This pattern is consistently observed across both communication scenarios and all acceptance probabilities. Also, reduced communication reliability (lower $P_A$) exacerbates holding delays, especially at higher arrival rates. Contours shift toward lower traffic densities with reduced reliability, indicating increased sensitivity to merging delays. The voice-based scenario, characterized by substantial message transaction times, exhibits significantly higher holding times across the parameter space than the RPAS scenario. This clearly demonstrates the operational penalty imposed by communication uncertainties in legacy systems.

\begin{figure}
\centering
    \begin{subfigure}[t]{0.85\textwidth}
        \includegraphics[width=\textwidth]
        {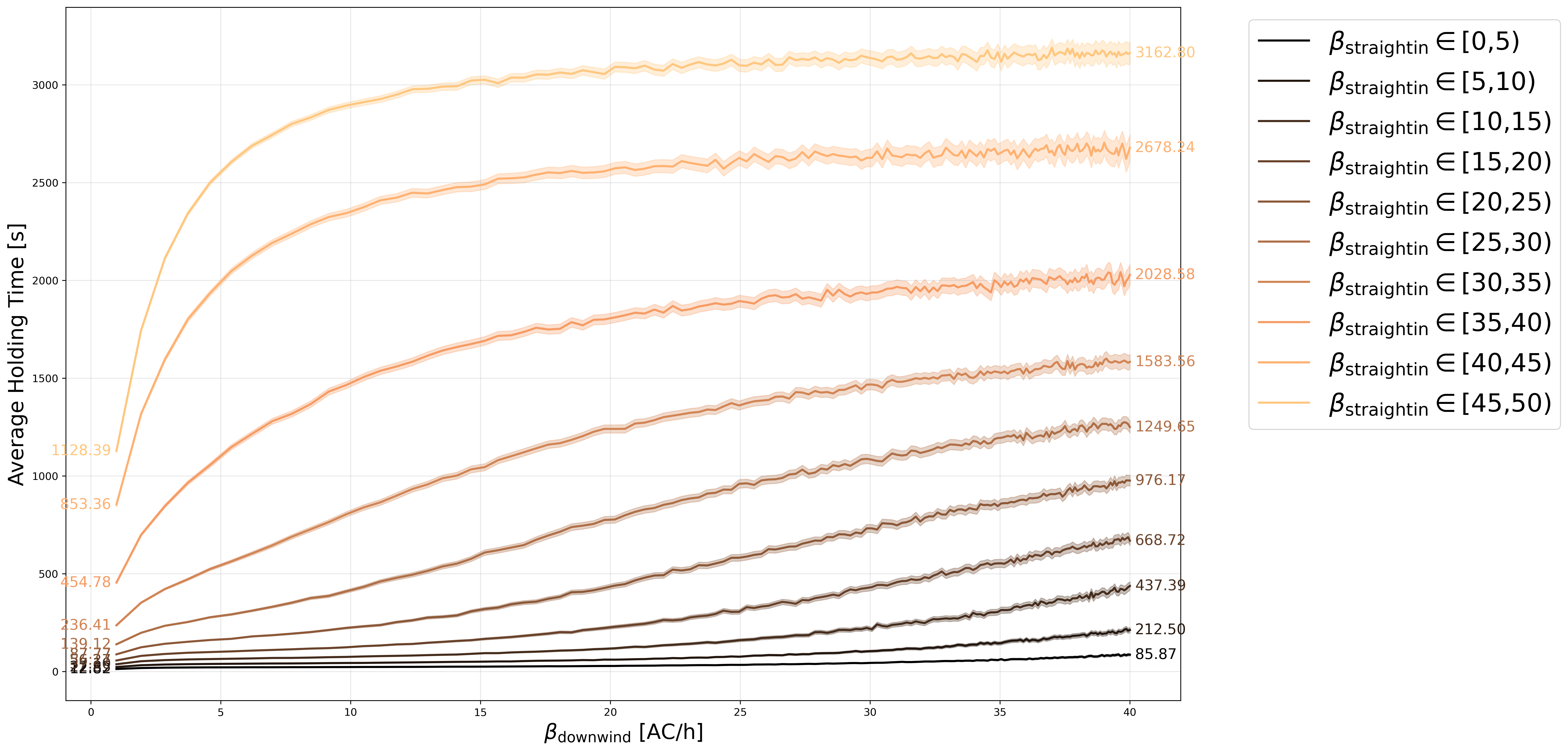}
        \caption{Downwind average holding time for VHF radio communication.}
        \label{fig: holding-curve-transaction}
    \end{subfigure}
    \begin{subfigure}[t]{0.85\textwidth}
        \centering
        \includegraphics[width=\textwidth]
        {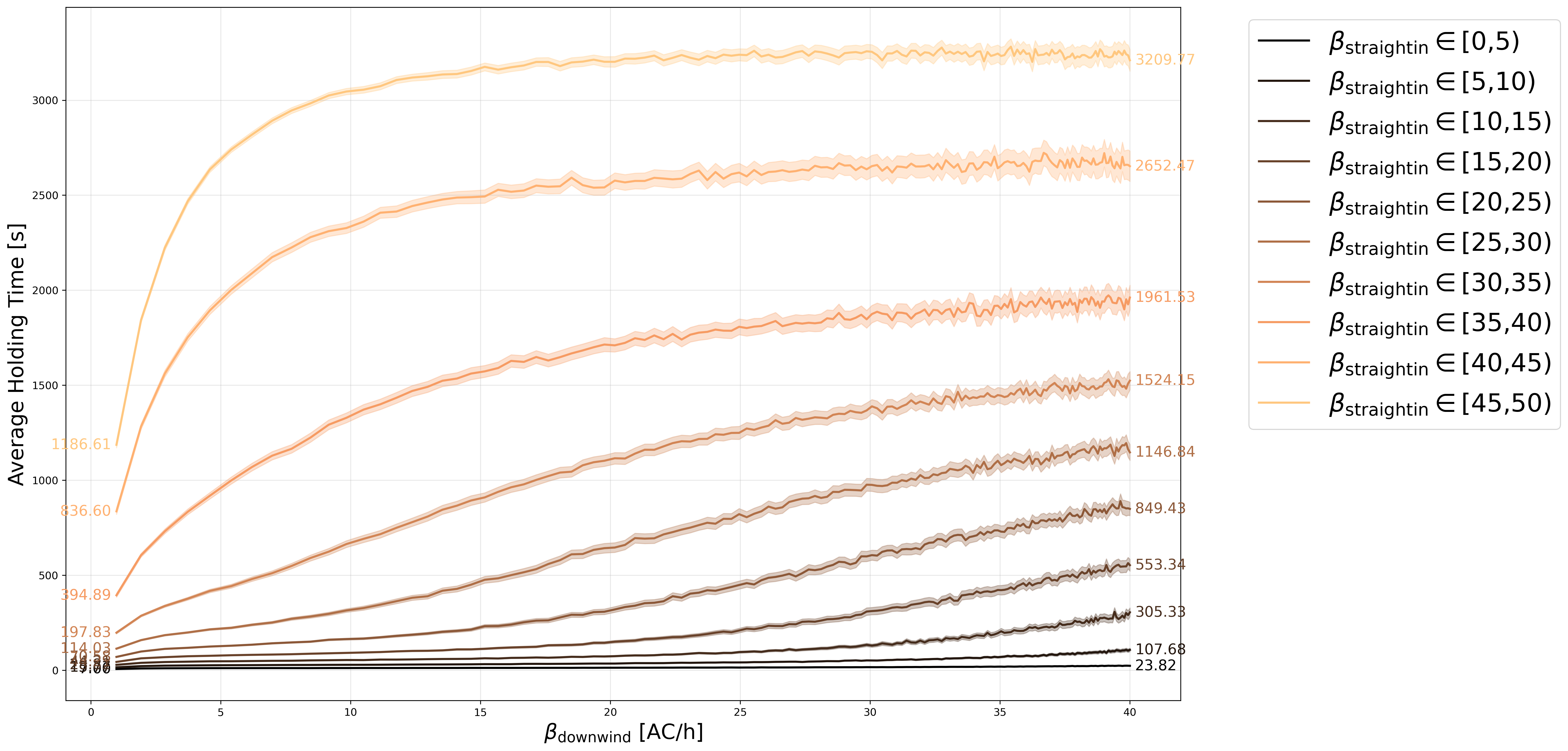}
        \caption{Downwind average holding time for RPAS.}
        \label{fig: holding-curve}
    \end{subfigure}
\caption{Impact of message transaction tome on average holding time of the downwind arrivals.}
\label{fig: holding-time-curve}
\end{figure}

As usual, we further investigate how the downwind holding times respond specifically to changes in downwind arrival rates for fixed intervals of straight-in arrivals in \Cref{fig: holding-curve-transaction} and \Cref{fig: holding-curve}. We noticed that for low-to-moderate downwind arrival rates, the average holding time exhibits an near exponential increase, as the downwind waiting time grow rapidly as merging opportunities become scarce. At higher downwind arrival rates and particularly for higher straight-in rates, the average holding time curves flatten significantly. This saturation occurs because, under these congested conditions, many downwind aircraft exceed the finite simulation horizon, \textit{artificially} limiting measurable delays and thus producing an underestimated steady-state holding time. Aircraft unable to merge within the simulation horizon artificially lower the observed holding time by not being accounted for fully, thus masking the true magnitude of delays. This is purely due to the limited simulation time. 

\begin{figure}
    \centering
    \includegraphics[width=0.75\textwidth]
    {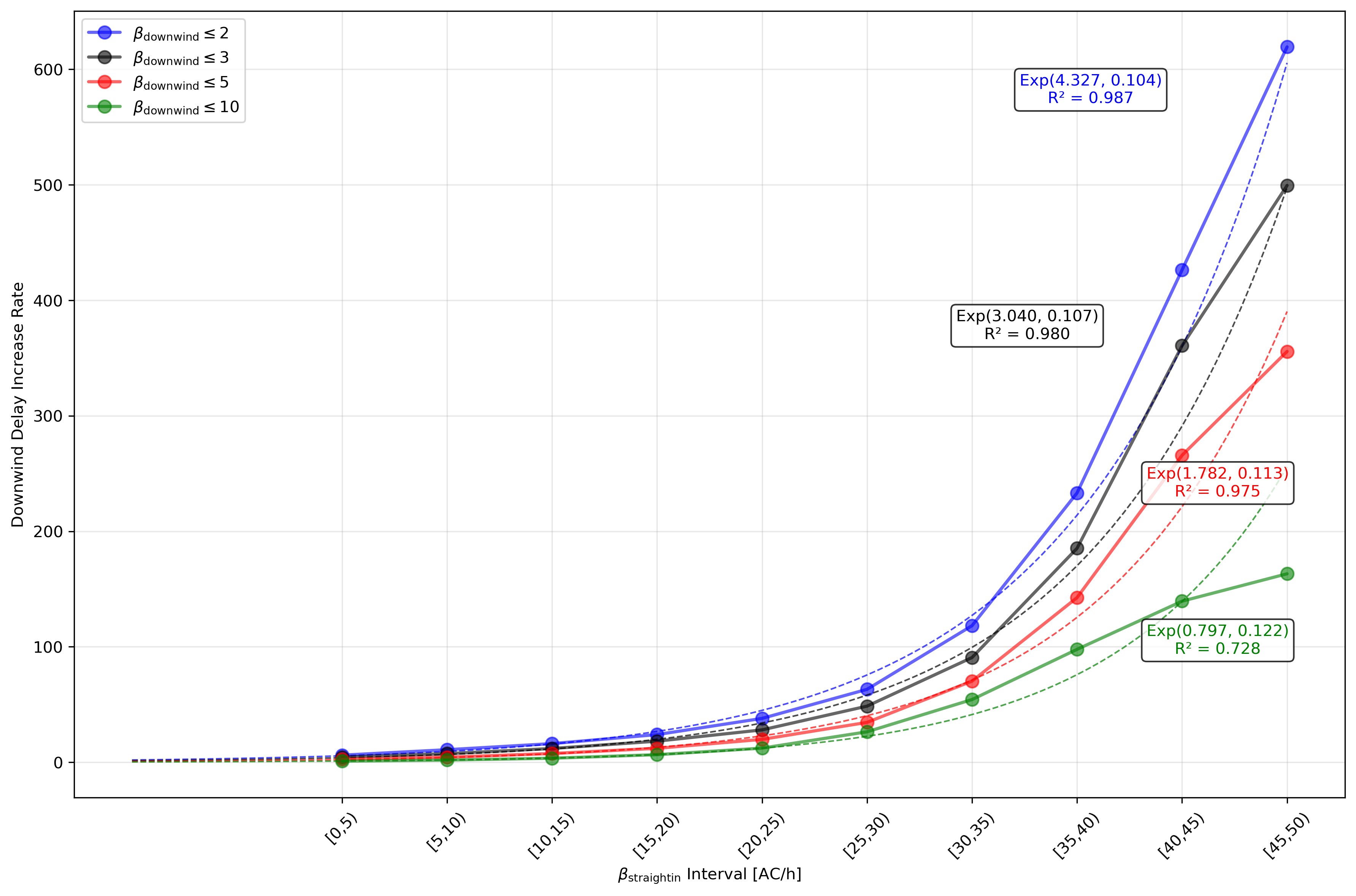}
    \caption{The rate of increase on average downwind holding time at different straight-in arrival rates. The exponential distribution is fit on each downwind rates. This also echoes the conclusion from \cite{tee2018modelling}. It is obvious that the downwind delay time saturated when increasing the downwind arrival rate, due to the finite time simulation setup in our case study.}
    \label{fig: slope}
\end{figure}

To further quantify and confirm this saturation effect, \Cref{fig: slope} provides the analysis of the rate of holding time increase and reveals a clear exponential growth at low-to-medium straight-in intervals (i.e., indicated by the $R^2$ values). However, as the straight-in rate increases, the rate-of-increase diminishes markedly (i.e., the goodness-of-fit significantly deteriorates) highlighting how finite simulation times distort true system behavior at high traffic densities.

This analysis suggests that ensuring sufficiently long simulation horizons at high-density traffic scenarios is essential to accurately capture true system delays. Explicitly adjust simulation length or implement analytical corrections to mitigate these effects are listed as a major future work.

\section{Automated Terminal Operation Simulation \label{sec: case2}}
This case study adopts an automated terminal vectoring planner in air traffic control to further study runway capacity under future automated vectoring systems using a more realistic scenario. Specifically, it implements an imitation learning approach based on maximum entropy inverse reinforcement learning (MaxEnt IRL) to understand the underlying control strategies from real-world trajectories. By learning a cost function directly from historical flight track data, the method identifies implicit human decision-making rules, which are subsequently applied within a search-based motion planner. This planning framework discretizes both state and control spaces and leverages the learned cost function to generate trajectories that are not only safe and compliant with established aeronautical separation regulations but also efficient in various traffic environments. Consequently, the planned trajectories effectively mimic expert human vectoring preferences, offering a robust and realistic choice as a benchmark for analyzing runway capacity and airspace efficiency under autonomous terminal operations.

\subsection{Inverse Optimal Planning \label{subsec: iop}}
We provide a brief introduction of the inverse optimal planning framework here \citep{tolstaya2019inverse}. This work aims to automate terminal-area air traffic control procedures through imitation learning from historical air traffic data, specifically FlightAware arrival recordings at Seattle--Tacoma International Airport (KSEA) on 11--13 January 2016, provided as GPS (WGS-84) waypoints at roughly 30-second intervals and converted to a local east-north-up Cartesian frame for training \citep{tolstaya2019inverse}. The aircraft trajectories are represented by state vectors $\mathbf{s}(t) = [x(t), y(t), z(t), \psi(t)]$, which include three-dimensional spatial coordinates and aircraft heading angles, as well as control inputs $\mathbf{u}(t) = [u_z(t), u_\psi(t)]$, describing the vertical climb/descent rates and horizontal turning rates, respectively. The path optimization thus have the following formulation,
\begin{equation}
\begin{aligned}
\min C(\varsigma) &= \int_{t_0}^{T} \left[1 + J(\mathbf{s}(t))\right]\|\dot{\mathbf{s}}(t)\|\ dt, \\
s.t., \quad \quad \\
    \mathbf{s}(t) &= (\cos(\psi), \sin(\psi), u_z(t), u_\psi(t)) \\
    \mathbf{s}(t_0) &= \mathbf{s}_0, \mathbf{s}(T) = \mathbf{s}_g \\
    \mathbf{u}(t) &\in \mathcal{U}
\end{aligned}
\end{equation}
The arrival trajectory planning problem is formulated as minimizing the total path cost $C(\varsigma)$ from an initial state $\mathbf{s}_0$ to the goal state $\mathbf{s}_g$ (i.e., the runway). The cost function represents two primary considerations, (i) minimizing the total traveled distance (equivalent to time); (ii) adhering to a learned cost function $J(\mathbf{s}(t))$, derived from historical arrival flight track data. This learned cost function encapsulates implicit human decision-making behaviors, operational preferences, and safety constraints. The state dynamics, $\mathbf{s}(t) = (\cos(\psi), \sin(\psi), u_z(t), u_\psi(t))$, capture the horizontal aircraft heading ($\psi$), the vertical motion rate ($u_z(t)$), and the horizontal turning rate ($u_\psi(t)$). Boundary conditions enforce that the trajectory begins at the initial state $s(t_0) = s_0$ and reaches exactly the goal state $s(T) = s_g$. Control inputs $u(t)$ are bounded by allowable operational constraints $u(t)\in U$. 

In the MaxEnt IRL formulation, we do not directly optimize the cost function $C(\varsigma)$. Instead, we optimize the parameters $\theta$ of the learned cost function by maximizing the likelihood of observed expert trajectories. Specifically, MaxEnt IRL assumes expert trajectories follow a probability distribution $P(\varsigma|\theta) = \frac{1}{Z(\theta)} e^{-C_\theta(\varsigma)}$, where the trajectory cost is parameterized as $C_\theta(\varsigma) = \int_{t_0}^{T}[1 + J_\theta(\mathbf{s}(t))]\|\dot{\mathbf{s}}(t)\|\,dt$, and $J_\theta(\mathbf{s}(t)) = \theta^\top f(\mathbf{s}(t))$ is a linear combination of features $f(\mathbf{s}(t))$. The parameters $\theta$ are estimated through gradient ascent by maximizing the log-likelihood of expert demonstrations. The gradient is computed as the difference between the empirical feature counts from observed expert trajectories and the expected feature counts from trajectories generated using the current cost function, 
\begin{equation}
    \nabla_\theta \mathcal{L}(\theta) = \mathbb{E}_{\varsigma\sim P_{\text{expert}}}[f(\varsigma)] - \mathbb{E}_{\varsigma\sim P_\theta}[f(\varsigma)]
\end{equation}
By iteratively updating $\theta$, the learned cost function $J_\theta(\mathbf{s})$ converges, capturing implicit human controllers' decision-making patterns, and subsequently enabling automated trajectory generation closely aligned with realistic ATC vectoring behaviors.

\subsection{Planning Performance Evaluation \label{subsec: case2-eval}}
\begin{figure}
    \centering
    \includegraphics[width=0.98\textwidth]
    {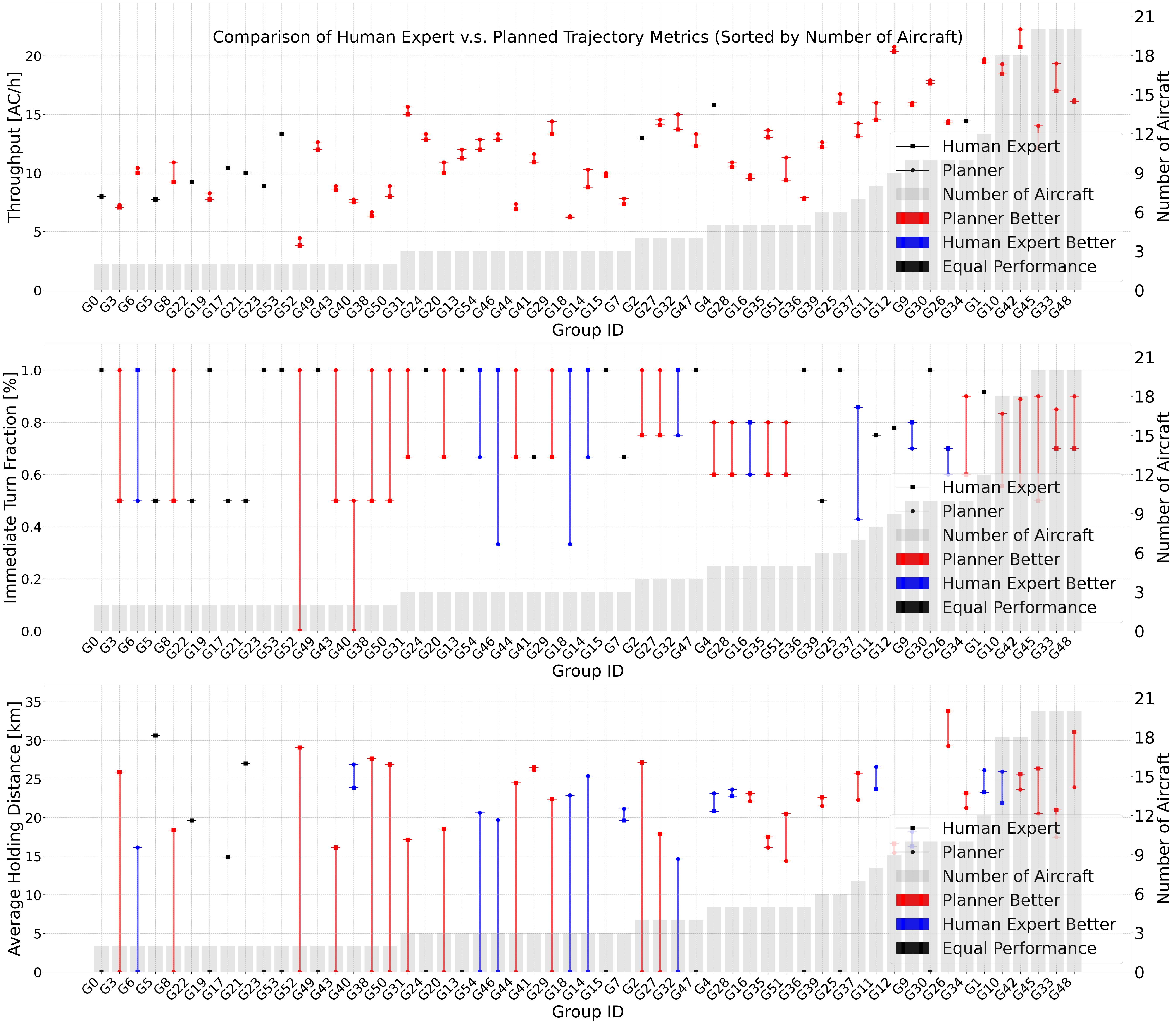}
    \caption{The evaluation of inverse optimal planning performance compared to the history trajectory (human experts). A total of 55 scenario groups are first sorted based on the number of arrivals aircraft in the scenario, then assessed using consistent performance metrics, including runway throughput, immediate downwind turn percentage, and average holding distance (used in place of holding time for this autonomous ATC planning task). Circle markers represent the performance of planned trajectories, while square markers denote the historical trajectories. Performance improvements by the planner over human experts are highlighted in red; degradations are shown in blue.}
    \label{fig: case2-group-eval}
\end{figure}

We evaluate the performance of the automated planner using the test scenarios provided by the original authors, employing the same metrics as used in the first case study. The results are shown in \Cref{fig: case2-group-eval}. The figure compares the automated planner trajectories against historical human expert flight tracks across three distinct performance metrics: throughput, immediate turn fraction, and average holding distance. Overall, the automated planner consistently demonstrates better performance, particularly in group IDs with lower density (i.e., left part of the plots). Notably, the automated planner achieves a guaranteed improvement in throughput compared to the historical trajectories, as indicated by the consistent positive performance margin across all evaluated groups.

\begin{figure}[htbp]
\centering
\begin{subfigure}[t]{0.19\textwidth}
    \includegraphics[width=\textwidth]{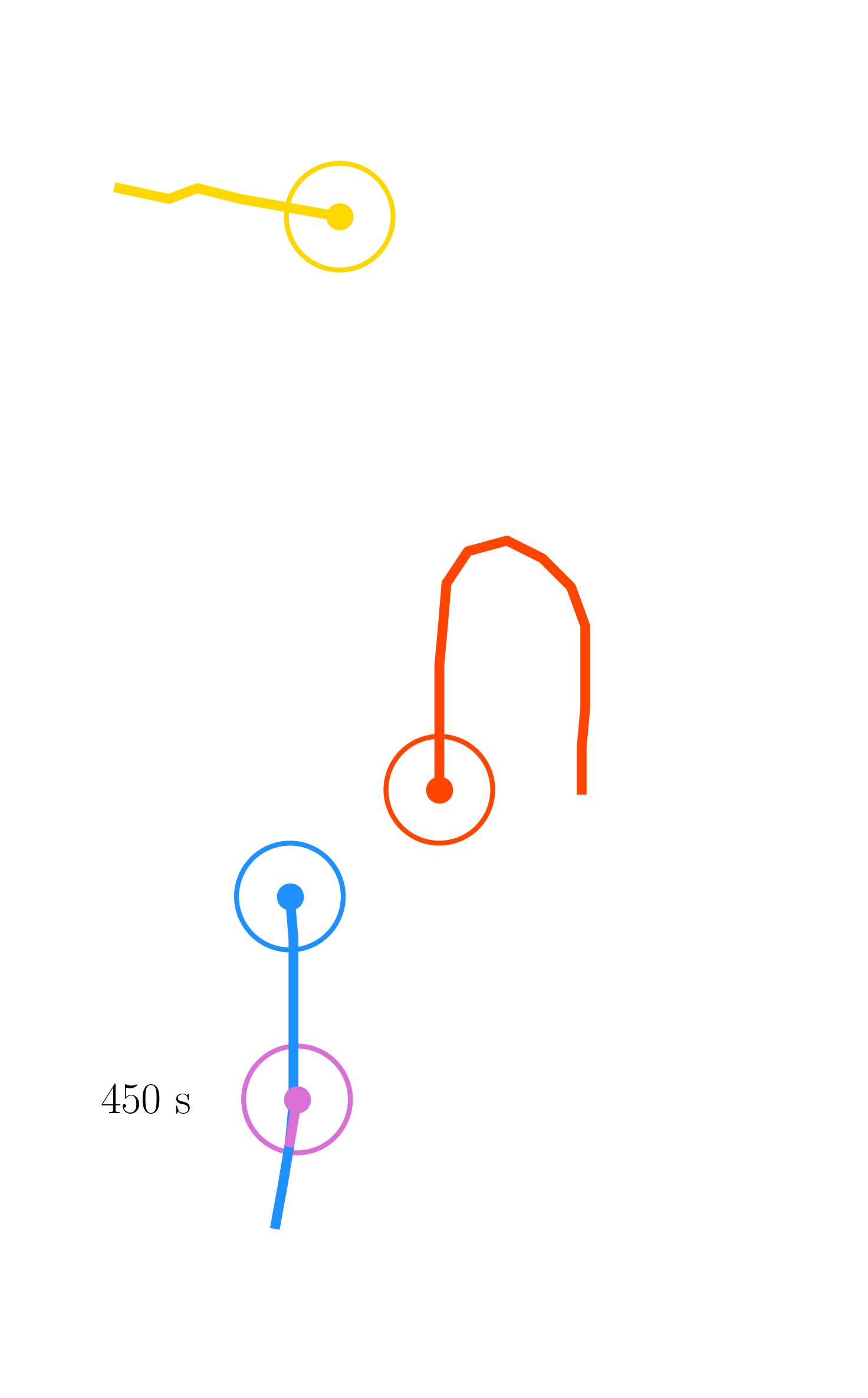}
\end{subfigure}
\hfill
\begin{subfigure}[t]{0.19\textwidth}
    \includegraphics[width=\textwidth]{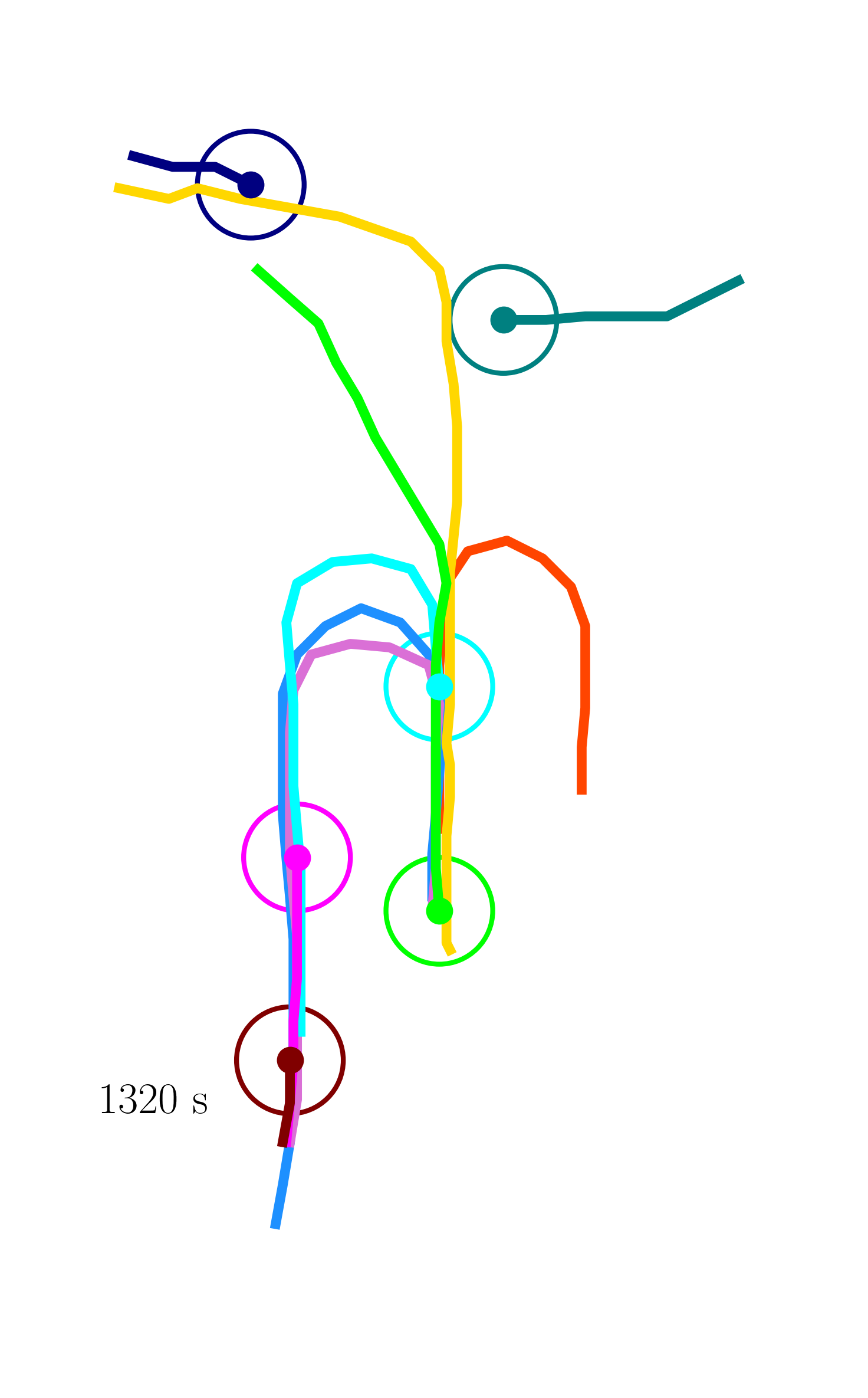}
\end{subfigure}
\hfill
\begin{subfigure}[t]{0.19\textwidth}
    \includegraphics[width=\textwidth]{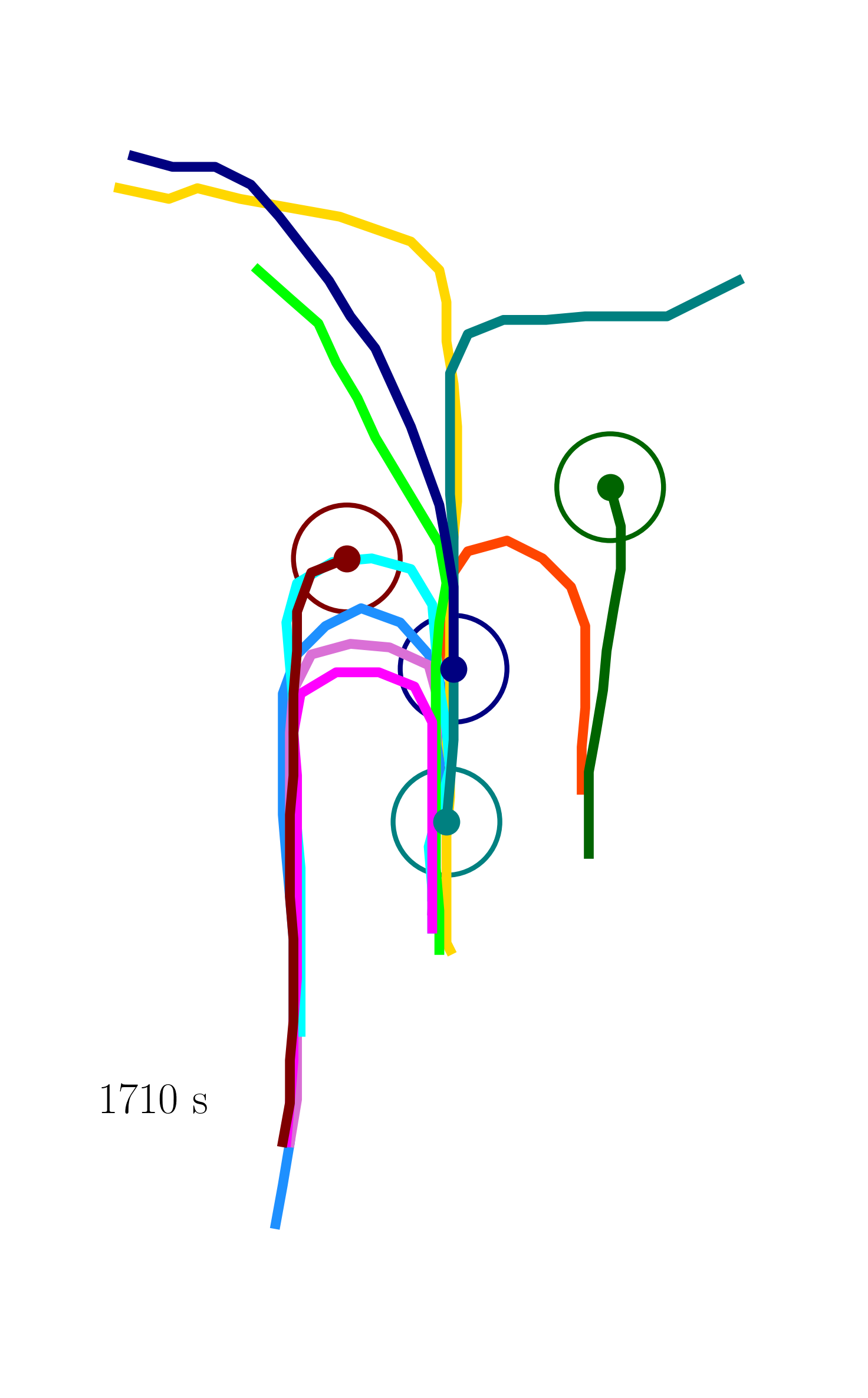}
\end{subfigure}
\hfill
\begin{subfigure}[t]{0.19\textwidth}
    \includegraphics[width=\textwidth]{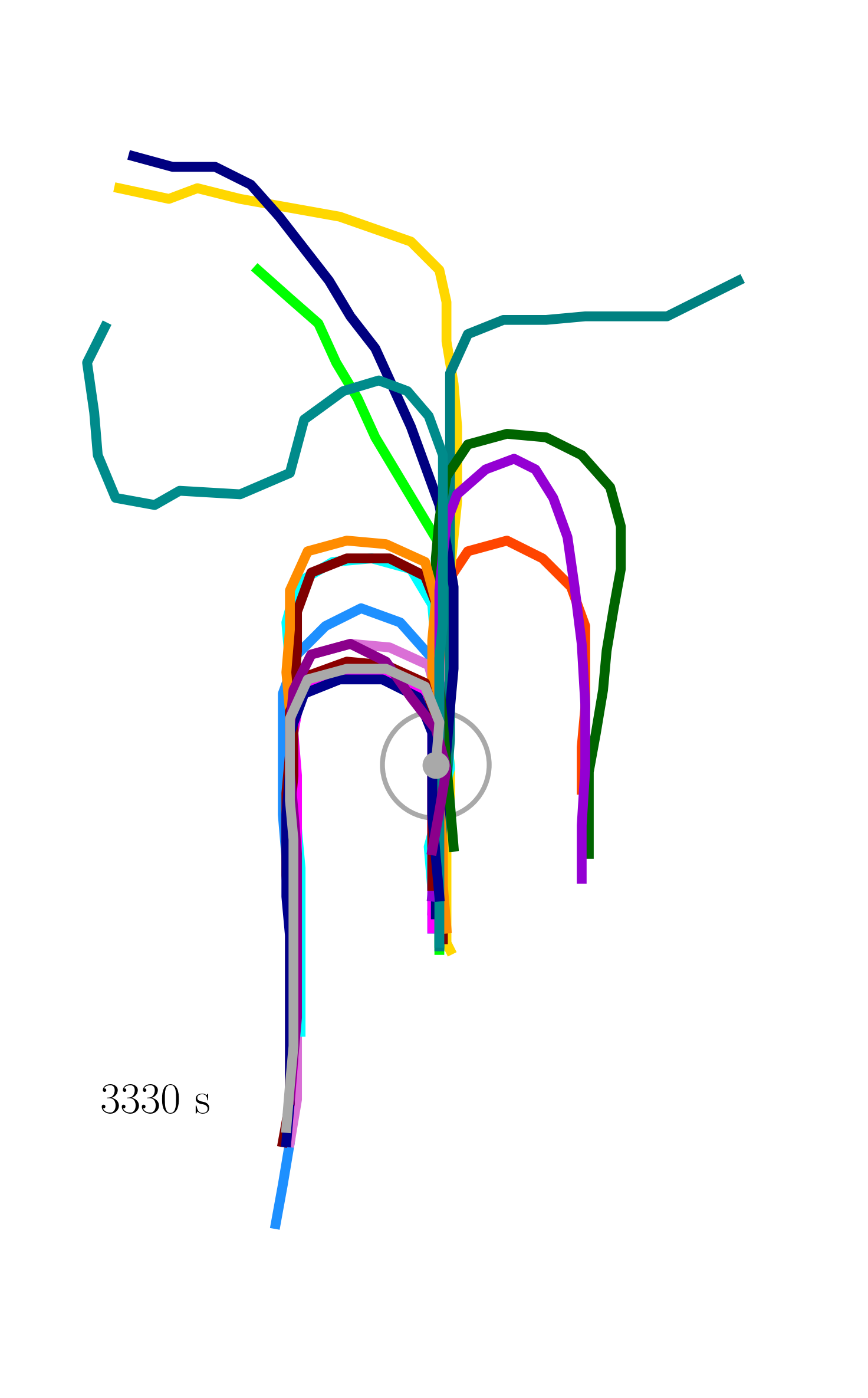}
\end{subfigure}
\\
\begin{subfigure}[t]{0.19\textwidth}
    \includegraphics[width=\textwidth]{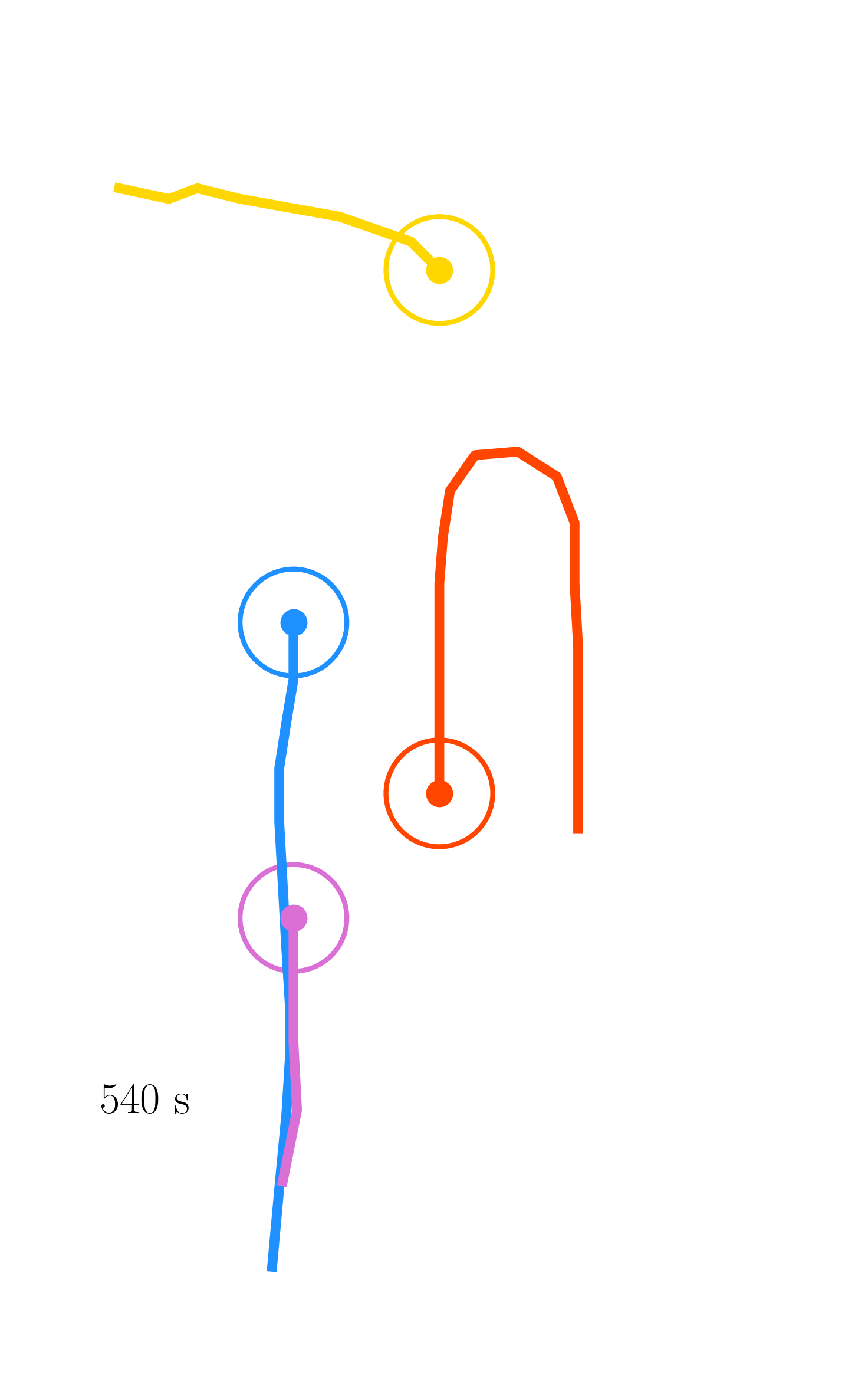}
\end{subfigure}
\hfill
\begin{subfigure}[t]{0.19\textwidth}
    \includegraphics[width=\textwidth]{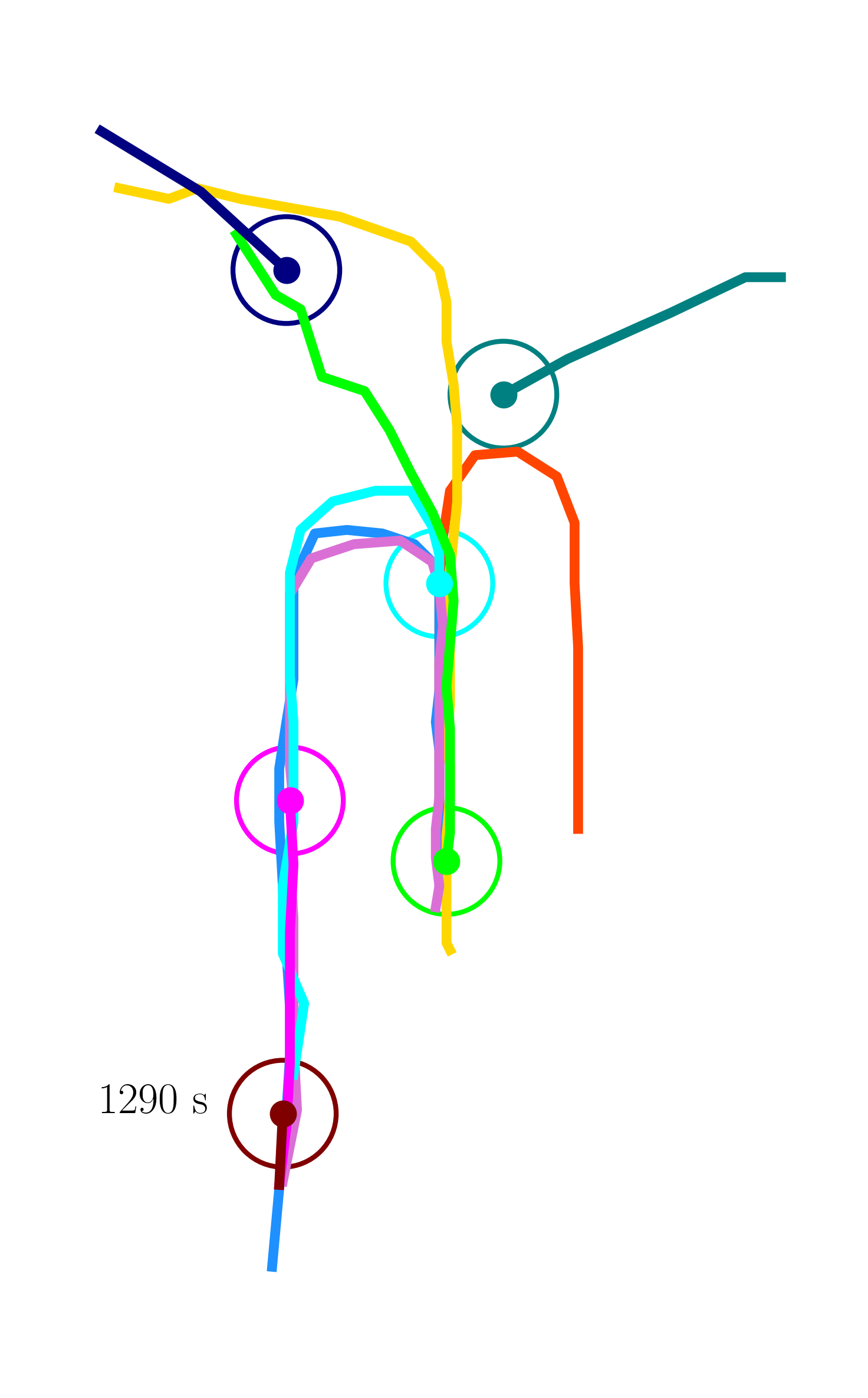}
\end{subfigure}
\hfill
\begin{subfigure}[t]{0.19\textwidth}
    \includegraphics[width=\textwidth]{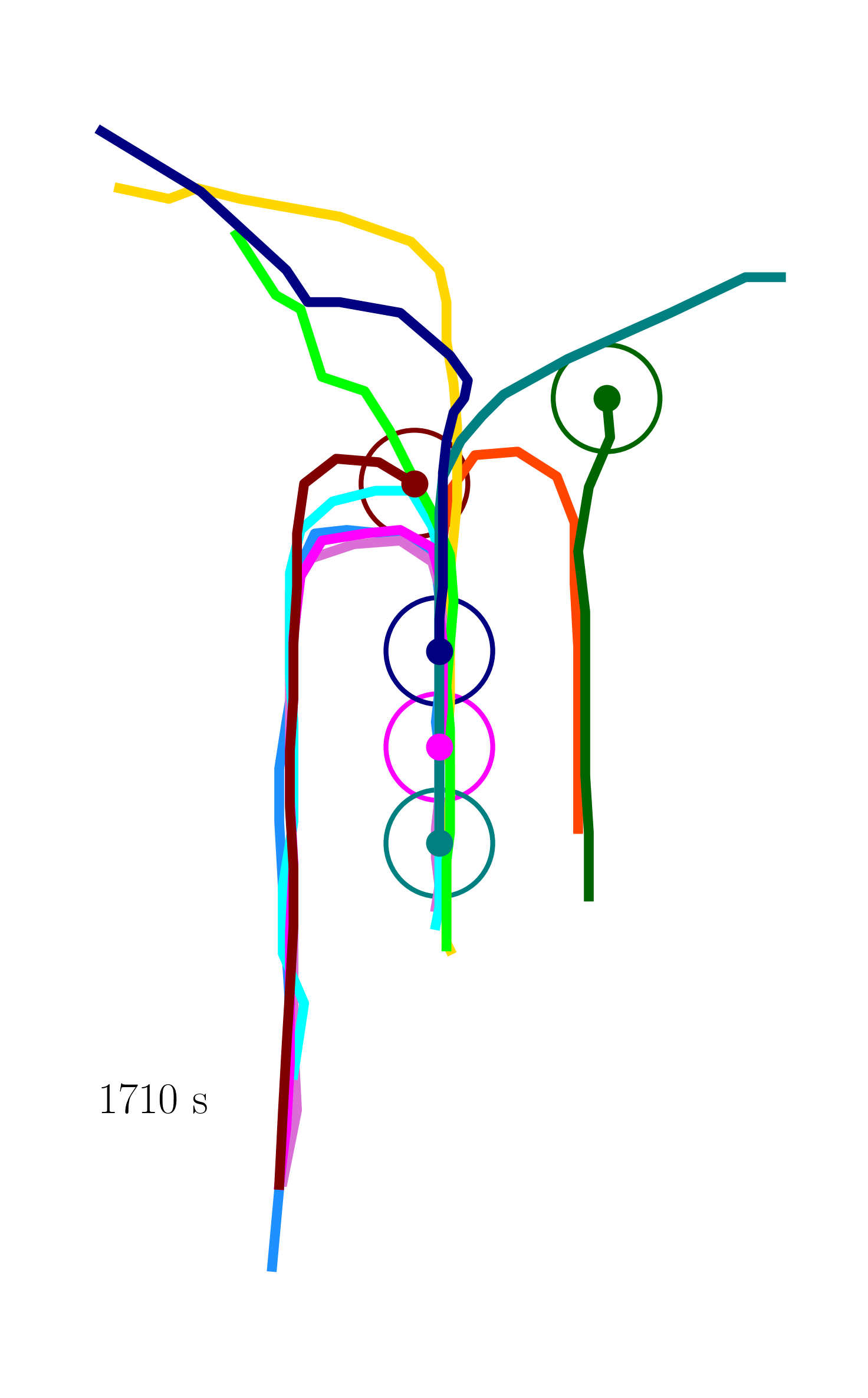}
\end{subfigure}
\hfill
\begin{subfigure}[t]{0.19\textwidth}
    \includegraphics[width=\textwidth]{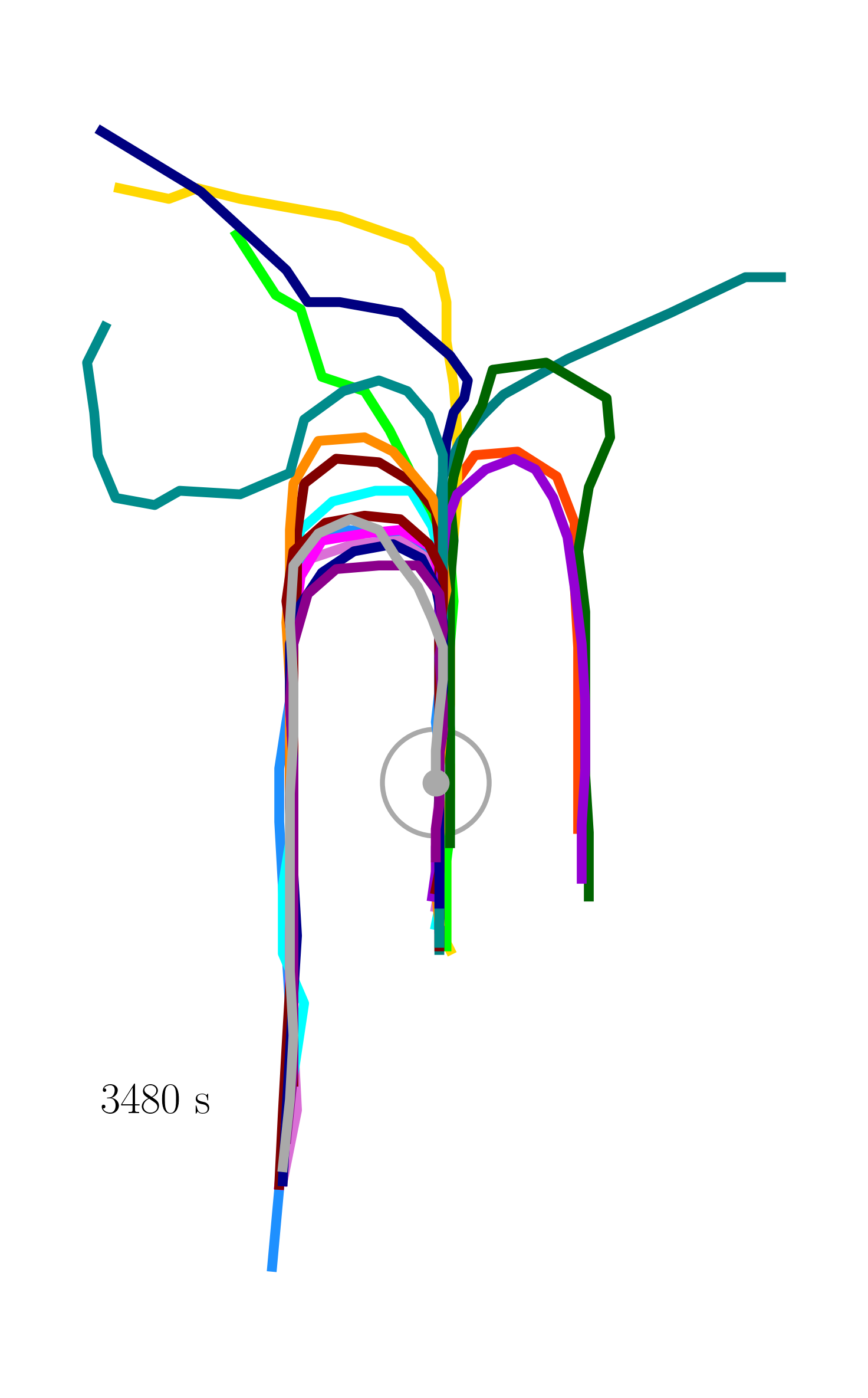}
\end{subfigure}
\caption{Visualization of planner trajectories (top row) and historical trajectories (bottom row) in the focused scenario.}
\label{fig: case2-planner-history-group10}
\end{figure}

Specifically, in \Cref{fig: case2-planner-history-group10}, we show the planned trajectories and the history trajectories of scenario group 10, to better compare the results. Four timestamps are visualized. The upper row finished planner slightly quicker than the lower row, indicating mild throughput increase. 

\subsection{High-Density Augmentation \label{subsec: augmentation}}

To replicate and extend the experiments conducted in the first case study, we focused on a representative arrival scenario (i.e., Group 10) and augmented it to simulate high-density terminal operations. The original scenario included arrival aircraft from three streams: west downwind, straight-in, and east downwind. To simplify the analysis and better control the augmentation process, we removed arrivals from the east downwind stream and retained only the west downwind and straight-in approaches. These two streams were selected due to their operational relevance and the clarity they provide in testing the planner’s behavior under increasing arrival volume.

\begin{figure}
    \centering
    \includegraphics[width=0.8\textwidth]
    {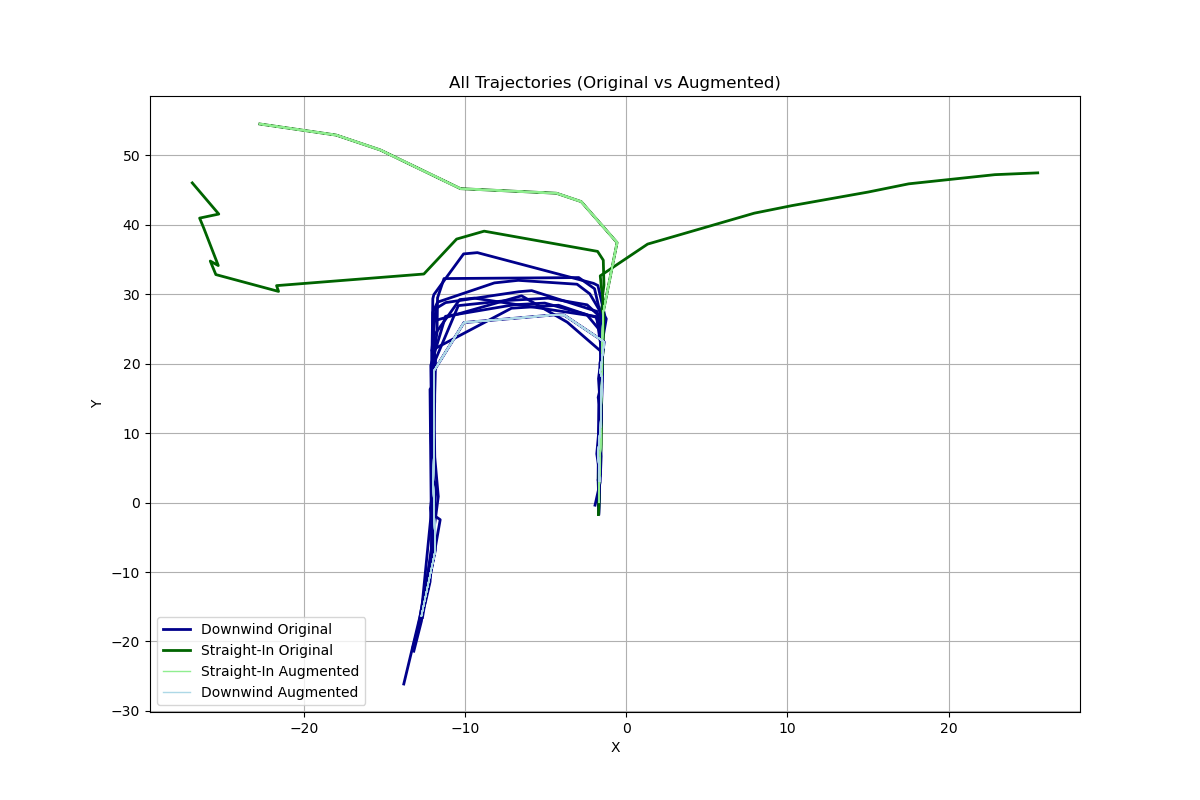}
    \caption{The visualization of the scenario considered (group 10) for augmentation, to simulate high-density arrival operations. One specific trajectory from the Downwind and Straight-In arrivals used as the referenced trajectory for augmentation are highlighted. The unit of $\mathsf{X}$ and $\mathsf{Y}$ are in kilometers.}
    \label{fig: case2-aug-traj}
\end{figure}

The augmentation process involved duplicating representative trajectories from the remaining two arrival streams. Specifically, one trajectory from each stream was chosen as a reference path, which was then used to generate additional arrivals. These augmented trajectories were introduced into the scenario with a carefully enforced minimum time separation from the existing aircraft, thereby ensuring safe and realistic initial conditions for the planner. As visualized in \Cref{fig: case2-aug-traj}, these reference trajectories are highlighted and serve as the baseline patterns for the synthetic arrivals added to the scenario.

\begin{figure}
\centering
    \begin{subfigure}[t]{0.85\textwidth}
        \includegraphics[width=\textwidth]
        {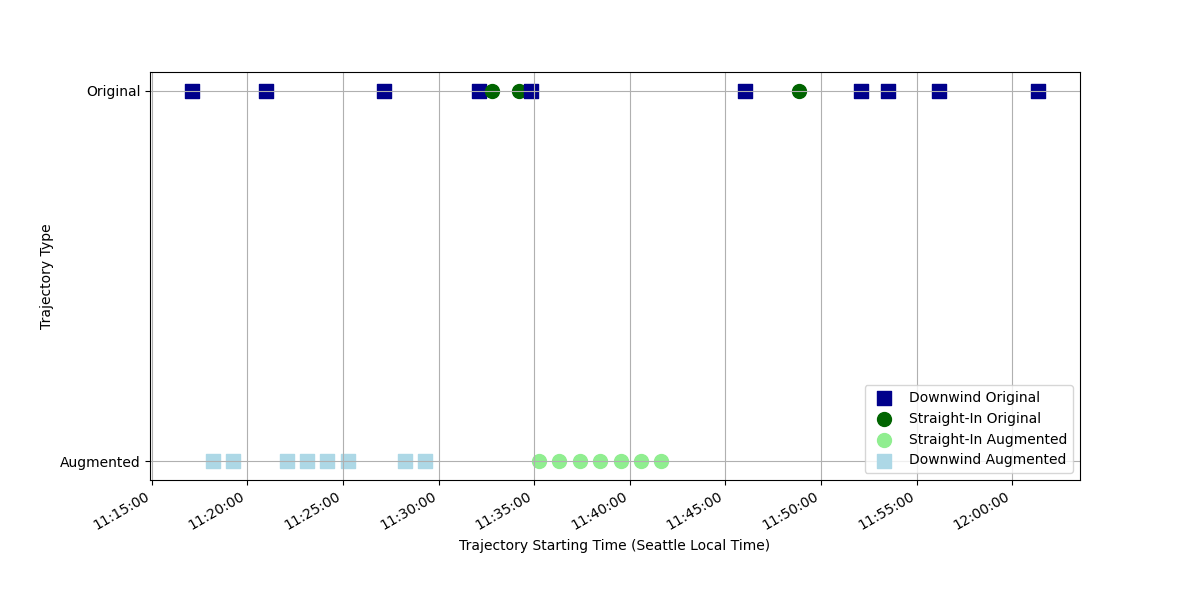}
        \caption{In this demonstration, 8 Downwind and 7 Straight-In arrivals are augmented.}
        \label{fig: start-d8s7}
    \end{subfigure}
    \begin{subfigure}[t]{0.85\textwidth}
        \includegraphics[width=\textwidth]
        {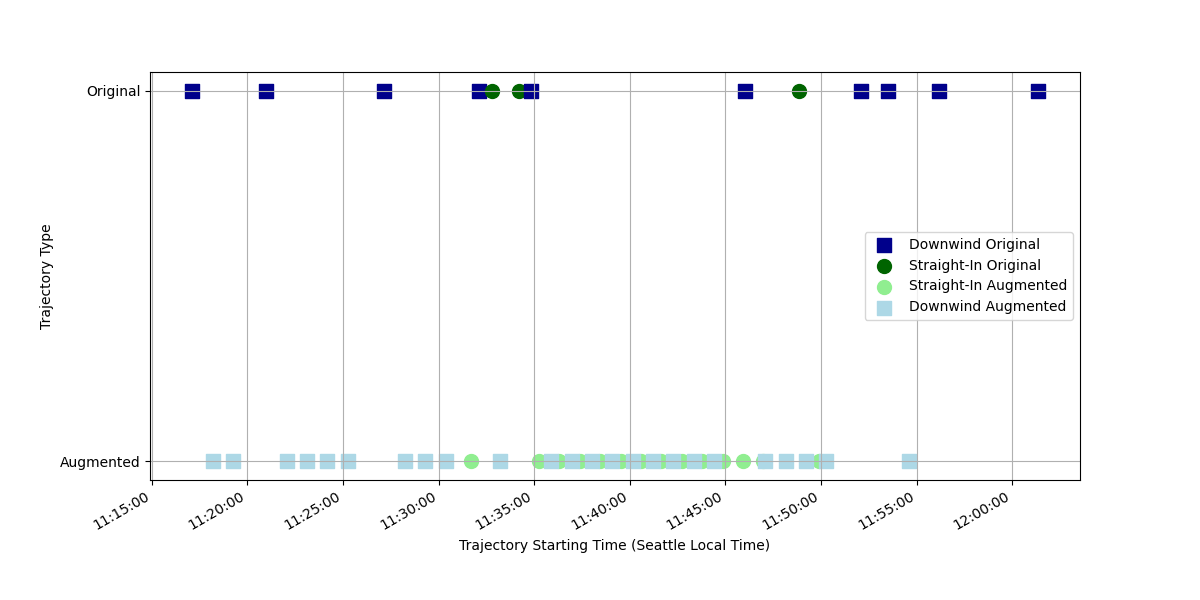}
        \caption{In this demonstration, 24 Downwind and 27 Straight-In arrivals are augmented.}
        \label{fig: start-d24s27}
    \end{subfigure}
\caption{The scenario entering time for the arrival aircraft to be considered by the planner. Two demonstrations are shown.}
\label{fig: case2-starttime}
\end{figure}

The resulting traffic patterns are illustrated in \Cref{fig: case2-starttime}, which shows the entry times of all aircraft into the terminal area for two different augmentation levels. In the first case (\Cref{fig: start-d8s7}), a moderate density is simulated by adding 8 downwind and 7 straight-in arrivals. In the second case (\Cref{fig: start-d24s27}), a significantly denser scenario is created with 24 downwind and 27 straight-in augmentations. In both plots, darker bars indicate the original aircraft, while lighter ones represent the newly added arrivals, offering a clear visual contrast between baseline and augmented data.

\begin{figure}
\centering
    \begin{subfigure}[t]{0.45\textwidth}
        \includegraphics[width=\textwidth]
{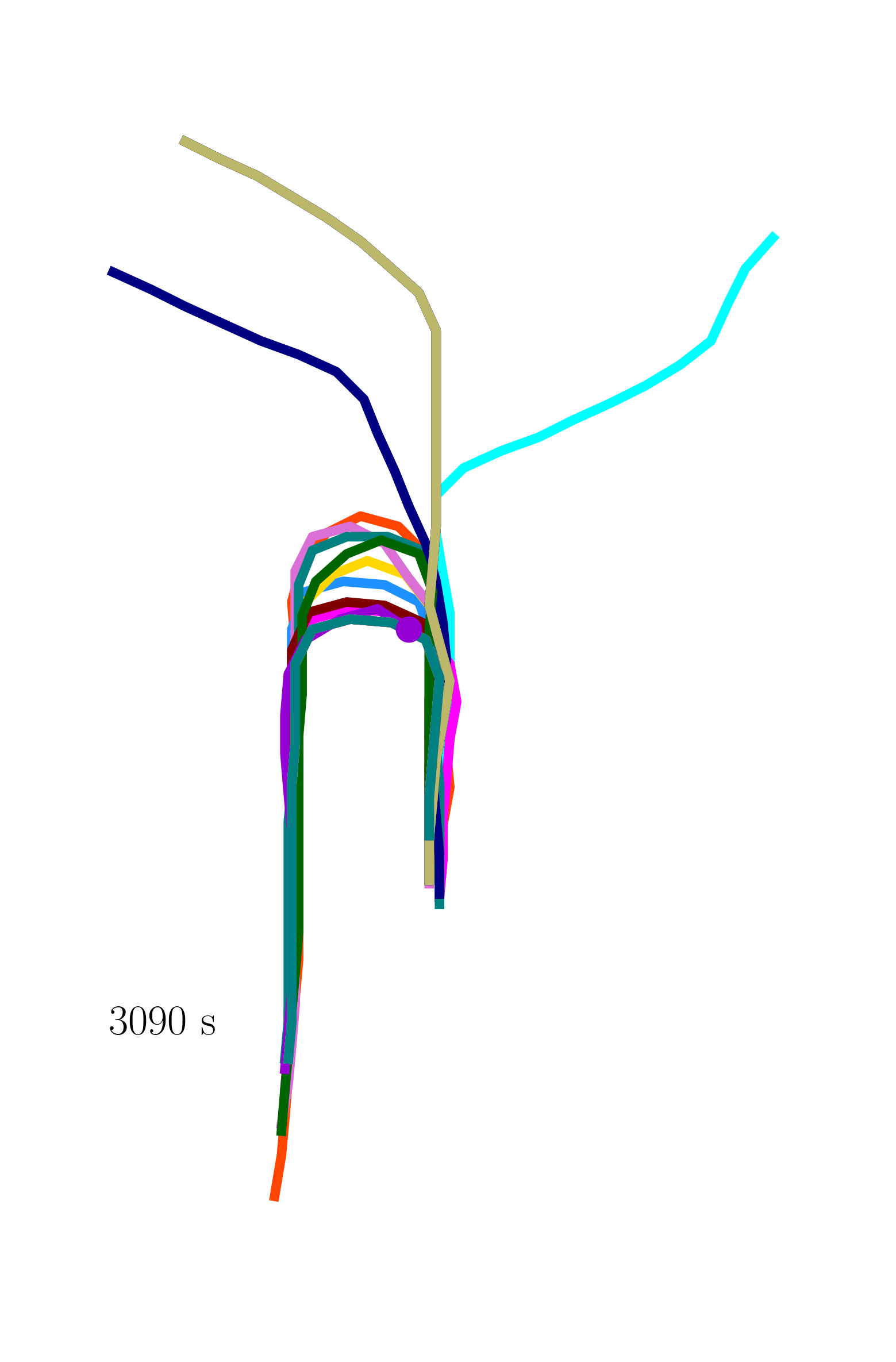}
        \caption{Planner output where 8 Downwind and 7 Straight-In arrivals are augmented.}
        \label{fig: traj-d8s7}
    \end{subfigure}
    \begin{subfigure}[t]{0.45\textwidth}
        \includegraphics[width=\textwidth]{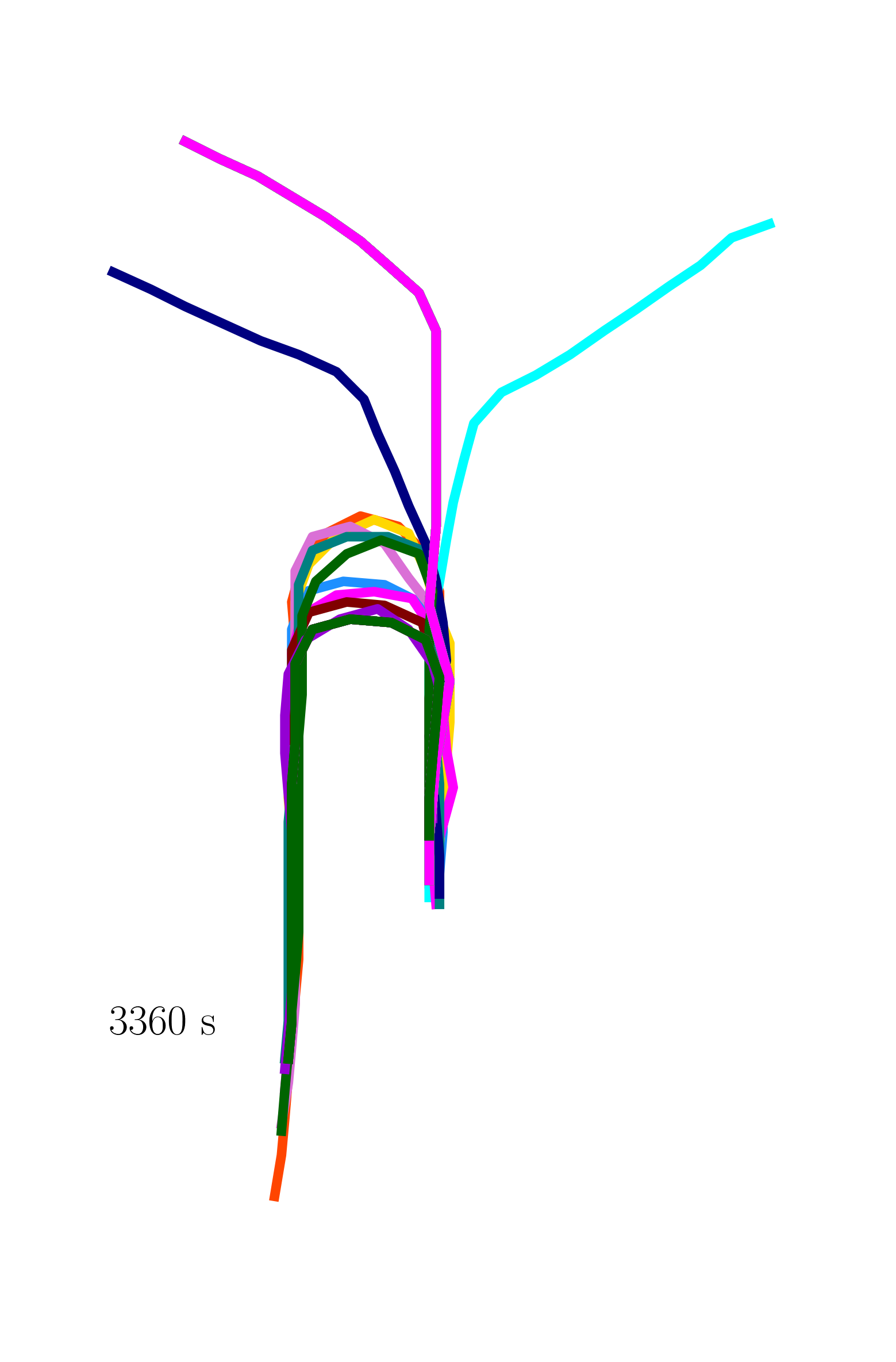}
        \caption{Planner output where 24 Downwind and 27 Straight-In arrivals are augmented.}
        \label{fig: traj-d24s27}
    \end{subfigure}
\caption{The planned trajectory for the augmented high-density arrival scenario. Only arrival trajectories to the same landing runway are considered (i.e., KSEA 16L). Downwind arrivals from the east are removed. Same color are applied for the same arrival aircraft. The different color at the same trajectory location are augmented trajectories overlaid.}
\label{fig: case2-traj}
\end{figure}

Finally, the sample output of the trajectory planner under these high-density conditions is shown in \Cref{fig: case2-traj}. The planned paths demonstrate the system’s ability to manage increased complexity while maintaining safe and conflict-free operations. Each subplot corresponds to one of the two augmentation levels and includes only arrivals landing on runway 16L at KSEA. Consistent color coding is applied to each aircraft's path, and overlapping colors indicate where augmented trajectories follow the same spatial routes. These results confirm the planner’s robustness and adaptability in managing increased arrival demand.

\subsection{Monte Carlo Analysis \label{subsec: case2-mc}}
In this second case study, we apply the inverse optimal planning framework to evaluate runway capacity metrics under the more realistic and future-oriented terminal airspace management setting. This case study reflects a more realistic and future-facing scenario, where terminal-area arrival trajectories are generated through an optimization-based imitation learning approach. Similar to the first case study, we present three contour plots, total runway throughput, downwind immediate turn fraction, average downwind holding distance, evaluated across different configurations of arrival demands and varying levels of communication signal availability. Unlike the first case study, we use downwind holding distance instead of time in seconds, as the inverse optimal planning algorithm is inherently distance-based. Finally, arrival aircraft are assumed to be RPAS conducting efficient automated final approach. Consistent with the RPAS/CPDLC configuration introduced in \Cref{subsec: case1-mc}, communication uncertainty in this case study is incorporated through the signal availability $P_A$ and the one-way latency $\varepsilon$ only; the message transaction time is effectively negligible ($\tau_{msg} < dt = 0.1$~s) because digital datalink messages are assumed to be transmitted within a single simulation tick. Operationally, this means that whenever the planner is invoked at a given time step, the link is checked for availability (a Bernoulli draw with probability $P_A$): if the link is available, the advisory is delivered after the latency $\varepsilon$, otherwise the planner waits one tick and re-checks. Transaction-time variability and continuity (which were the dominant communication-uncertainty drivers in the voice-based scenario of \Cref{sec: case1}) therefore contribute negligibly here, by design, since this case study targets the future automated regime in which datalink, rather than voice, is the primary mode of pilot-controller communication.

\begin{figure}
    \centering
    \includegraphics[width=0.95\textwidth]
    {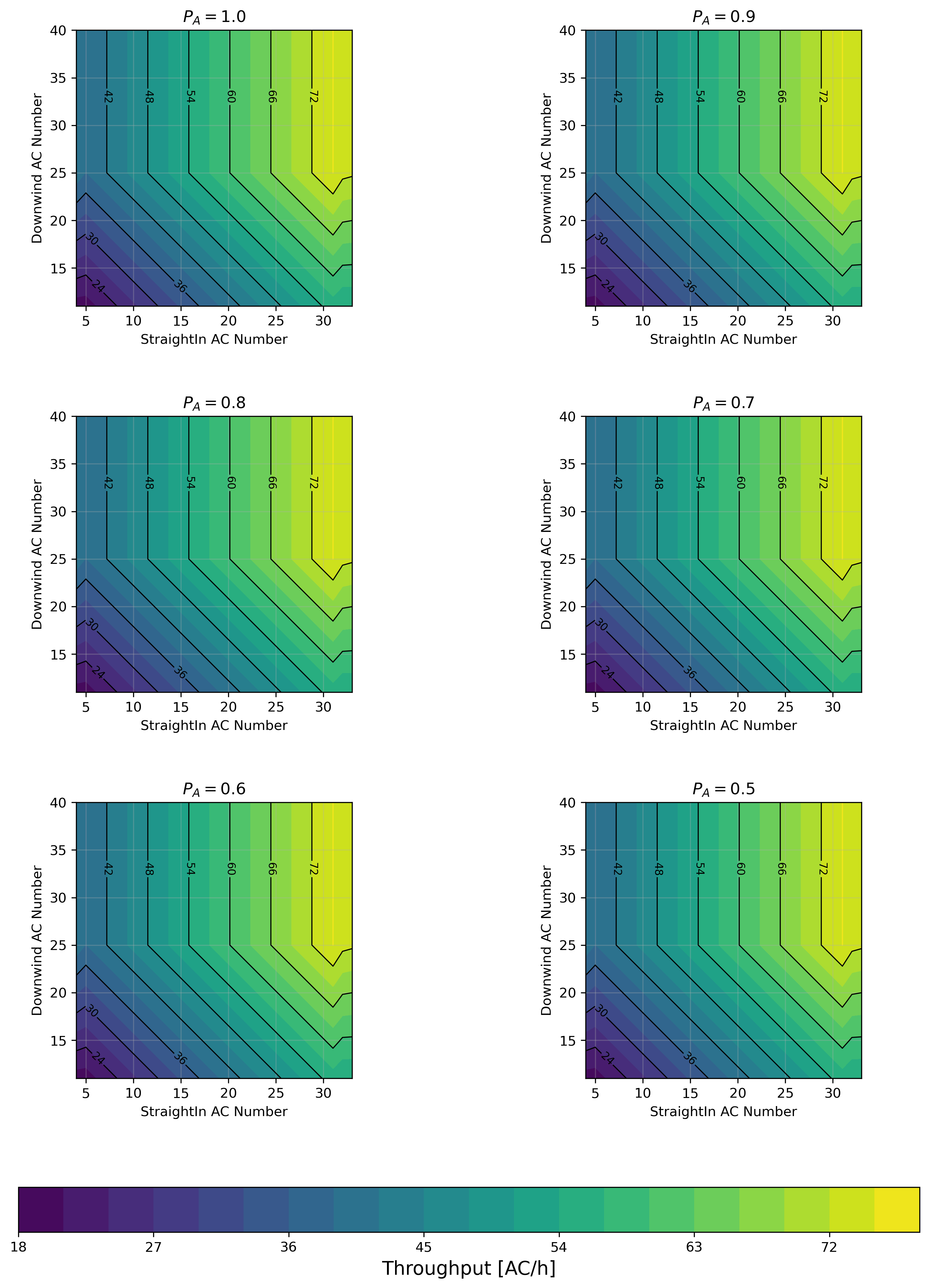}
    \caption{Runway throughput under varying downwind and straight-in arrival aircraft counts across different levels of aeronautical communication uncertainty. }
    \label{fig: case2-throughput}
\end{figure}

The runway throughput contours (\Cref{fig: case2-throughput}) show consistency across various communication signal availability levels, highlighting the advantage of RPAS with near-instantaneous message transactions and continuous trajectory re-evaluation after the turn advisory point. This robustness suggests that even under degraded signal conditions, the system can maintain efficient gap-filling behavior and high utilization of runway capacity.

\begin{figure}
    \centering
    \includegraphics[width=0.95\textwidth]
    {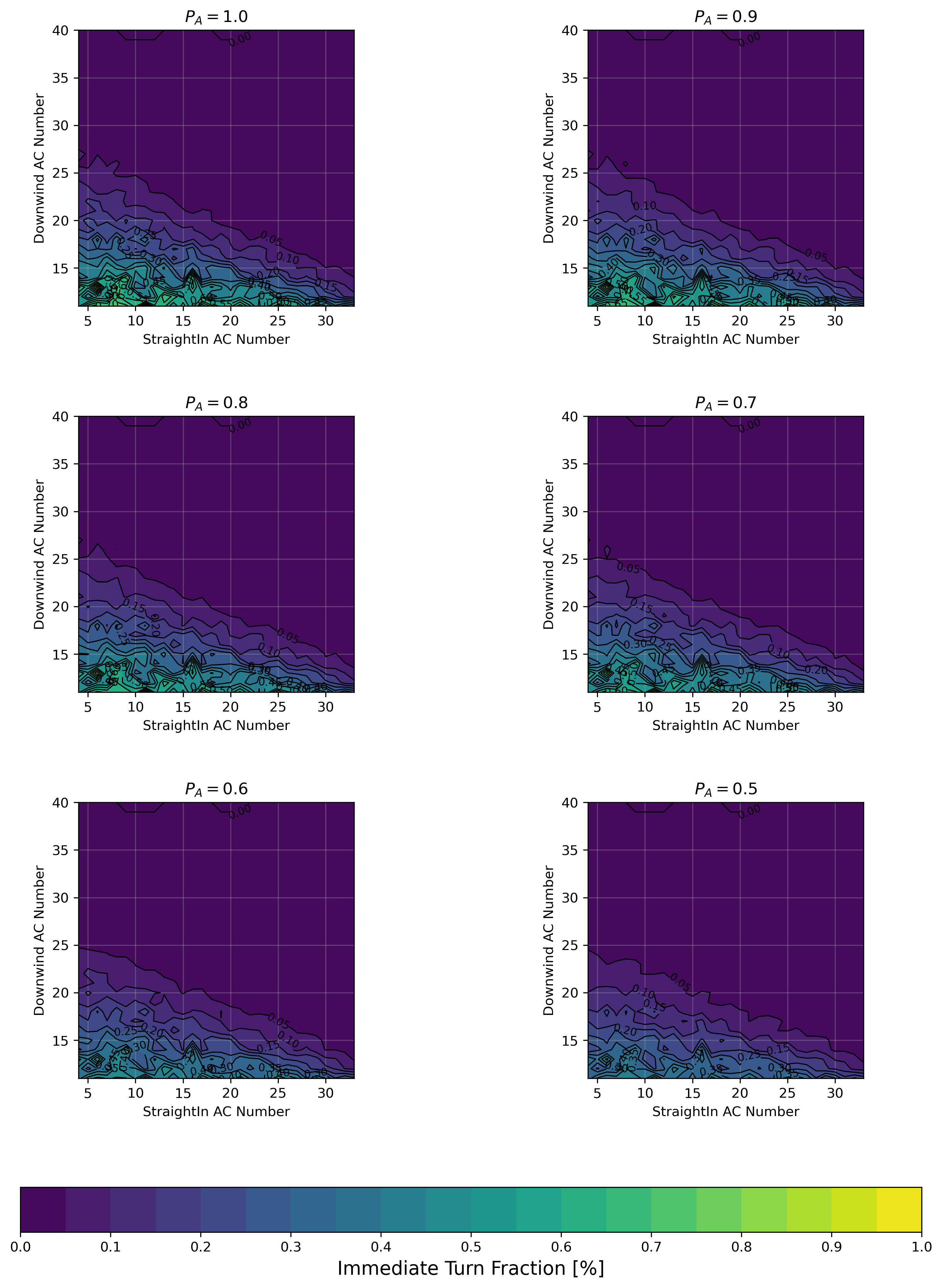}
    \caption{Percentage of immediate turns in downwind arrivals under varying number of downwind and straight-in arrivals across different levels of aeronautical communication uncertainty. }
    \label{fig: case2-immediate-turn}
\end{figure}

In contrast, the downwind immediate turn fraction contours (\Cref{fig: case2-immediate-turn}) demonstrates a strong dependency on signal availability $P_A$. As $P_A$ decreases, aircraft are less likely to receive timely turn instructions at the advisory point, leading to a reduction in immediate turn execution. This behavior is evident in the contour plots, where high immediate turn fractions concentrate in regions with low downwind traffic and high signal availability. As availability diminishes, the opportunity for timely turns drops off sharply, forcing more aircraft to continue on extended downwind legs.

\begin{figure}
    \centering
    \includegraphics[width=0.95\textwidth]
    {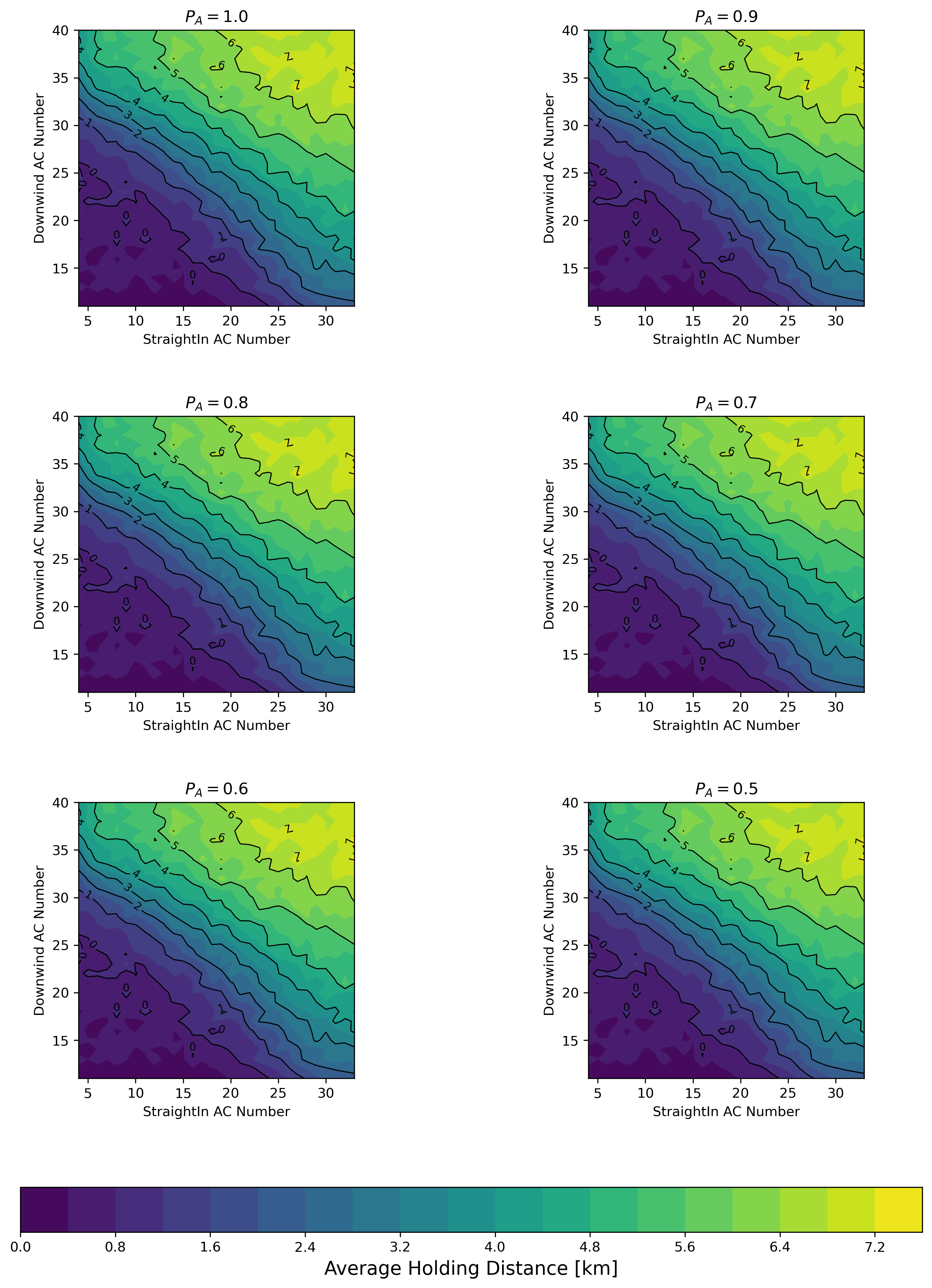}
    \caption{Downwind average holding distance in downwind arrivals under varying downwind and straight-in arrival aircraft numbers across different levels of aeronautical communication uncertainty.}
    \label{fig: case2-avg-holding}
\end{figure}

Consequently, as in \Cref{fig: case2-avg-holding}, the average downwind holding distance increases with reduced $P_A$, particularly in high-traffic scenarios. This increase reflects the growing number of aircraft that must delay their turns and extend their paths to safely merge onto final approach. However, despite these inefficiencies, the RPAS-based system retains relatively controlled delay growth due to its ability to continuously adapt plans in real time. These findings highlight the resilience under degraded communications.

Together, these results demonstrate that this inverse optimal planning approach enables efficient and automated terminal operations. By capturing the statistical patterns, safety margins, and implicit decision-making behaviors embedded in historical ATC data, the method adapts effectively to varying levels of signal reliability and traffic demand. While there remain opportunities to further enhance the automated vectoring planner, this framework offers a foundation and a glimpse into the potential of future automated air traffic management systems.


\section{Conclusions\label{sec: conclusion}}
This study set out to examine how aeronautical communication uncertainties and human induced uncertainties affect runway capacity in terminal arrival scenario, from low to high density. By focusing on a converging arrival flow (with a straight-in stream and a downwind stream merging via a continuous turn-to-final), we address the core objective of quantifying the impact of various level of uncertainties on arrival performance. The research contributes a modeling framework with necessary analysis that incorporate these real-world uncertainties into runway capacity estimation, providing insight into maintaining efficiency and safety under congested conditions.

In the first simulation, we develop a Monte Carlo-based framework to simulate an approach environment under current voice-based operations and future autonomous operations. This simulation generated two streams of arriving aircraft using modified Possion processes subject to minimum separation constraints, simulating traffic of various demand. A rule-based merging logic was implemented, including standard bank-angle turns for the downwind aircraft, dynamic gap assessments, and stochastic pilot reaction times to controller commands. Communication uncertainty was introduced as the probability of delayed or missed voice instructions, reflecting unreliable radio contact. Using extensive simulation runs, we evaluated metrics such as runway throughput, the downwind immediate turn rate, and average holding time on the downwind leg. The simulation reveals that under degraded communication reliability and longer pilot response times, runway throughput declines while holding delays grow, highlighting the vulnerability of traditional operations to these uncertainties. We also observed that as traffic demand increases, maintaining a high immediate turn rate becomes difficult when communications are unreliable, leading to more aircraft entering holding patterns to maintain safe spacing.

In the second scenario, we investigated an automated vectoring planner for the same merging scenario, representing a future highly-automated Air Traffic Management setting. This planner is based on an inverse optimal control approach (i.e., Auto-ATC) that learns vectoring decision-making process from historical flight trajectories. By augmenting recorded approach paths and embedding the same uncertainties (e.g., communication and human-induced) into the planning process, the automated system was tested in a similar environment to the first case study. The outcome from this second study shows that the automated approach can achieve efficient merging, demonstrating the potential benefits of automation in mitigating the impacts of human and communication uncertainties.

The two setup provide a comprehensive perspective on how aeronautical communication and human-induced uncertainties shape runway capacity, as they were found to notably reduce performance. In the voice-based legacy system, reduced radio communication reliability led to more frequent holding and a drop in immediate merge success, directly cutting into runway throughput. In contrast, the system simulating autonomous operations (i.e., RPAS) maintained higher throughput and merge efficiency, even under equivalent traffic loads and uncertainty levels. These findings underscore that incorporating realistic uncertainties is essential when estimating capacity, and failing to do so will overestimate what an airport can handle safely. By modeling a high-density arrival scenario with stochastic communication situations and human-in-the-loop delays, our study captured complex interactions that simpler models might overlook. This approach yielded reliability surfaces for runway capacity, mapping how throughput, immediate turn fraction, and holding delay vary across ranges of traffic demand and communication situation. The probabilistic characterization of runway capacity is valuable for operational decision-making, as it identifies the conditions under which runway capacity degrades and quantifies the benefits of improved communication or automation.

This simulation has made several modeling assumptions, each motivated by both tractability and the focal objective of isolating the effect of communication and human-induced uncertainties on runway capacity. We acknowledge that these assumptions also shape the generalizability of the reported results, as discussed below.

(i) A fixed inter-arrival separation of 64 seconds is imposed at the runway threshold, corresponding to a 2.5-nautical-mile final-approach spacing at a reference speed of 140 knots (1.23 times the stall speed of an A321 at 80\% maximum landing weight \citep{salgueiro2025aircraft}). This value is selected to represent a homogeneous medium-weight-class arrival stream (e.g., B737/A321) so as to separate the capacity impact of communication and human uncertainties from the additional variability introduced by wake-category pairings (see \Cref{tab: separation}). In practice, the required $T_{sep}$ depends on the leading-trailing weight class combination, and the actual stall speed scales with landing weight, both of which would further inflate inter-arrival variance. The reliability contours reported here should therefore be interpreted as an \textit{upper envelope} for a single-weight-class fleet; a mixed-fleet operation would reduce the usable throughput but leave the qualitative trends (monotonic degradation with reduced $P_A$, robustness of the RPAS/datalink case relative to voice) unchanged.

(ii) The simulation adopts a continuous $180^\circ$ constant-bank turn from downwind to final, rather than the rectangular (\textit{boxed}) pattern common in civil aviation. This geometry is chosen for three reasons: it yields a constant Base-turn duration (60 seconds) that cleanly separates the gap-selection decision from the trajectory generation; it is consistent with the inverse-optimal planner used in \Cref{sec: case2}, which operates in a continuous state-control space; and it enables the closed-form analytical throughput verification presented in \Cref{app: throughput-verification}. A rectangular pattern would introduce a discretionary Base leg and thus additional slack for gap matching, so our continuous-turn setup is comparatively conservative on the immediate-turn fraction. Although the continuous-turn-to-final approach has been used operationally in military aviation, its validity for commercial and general aviation remains under active investigation \citep{AOPA2016circpattern}; nonetheless, we expect the relative sensitivities of throughput and holding time to $P_A$, $\eta$, and $\tau_{msg}$ to carry over to more traditional approach geometries.

(iii) A finite simulation horizon of one hour is adopted, consistent with the hourly-capacity convention used by FAA AC 150/5060-5 \citep{faa15050605}. Aircraft that cannot merge within this horizon are excluded from the holding-time average, producing the saturation artifact visible in \Cref{fig: slope} at high downwind demand. Longer simulation windows or an analytical finite-horizon correction would sharpen the high-density delay estimates at the cost of significant additional compute, and a principled correction of this effect is critical to correctly understand the delay propagation behavior of downwind arrivals. We identify this as a priority for future work.

Further investigation into developing efficient and reliable automated terminal vectoring tools remains a significant and promising direction for future research. Current aviation systems, while robust, still rely heavily on human controllers to perform sequencing, spacing, and conflict resolution, particularly within terminal airspace. This reliance inherently limits the efficiency, consistency, and scalability of operations, especially under complex or high-density traffic scenarios. Automation can consistently maintain optimal aircraft spacing, improving runway throughput, and reducing airborne holding patterns that contribute to fuel inefficiency and increased emissions. Moreover, validating the safety and robustness of such automated systems through high-fidelity simulations and real-world flight trials is crucial. This includes extensive human-in-the-loop experiments to ensure seamless integration with existing operations and acceptance by air traffic controllers. Ultimately, a transition towards increasingly automated air traffic management systems offers immense potential for addressing the growing challenges of air traffic demand, environmental sustainability, and operational safety.

Despite these limitations, this work offers timely and practical insights for the future of air traffic management, particularly in the context of ongoing efforts to next Generation ATM initiatives and AAM integration \citep{federal2013nextgen}, which continue to reshape terminal operations through automation and enhanced airspace utilization. By examining the arrival scenario from low-density to high-density scenarios under multiple uncertainties, our simulation studies contribute to the development of future sequencing, spacing, and vectoring strategies that maintain high throughput while safely managing variability. The demonstrated resilience of automated vectoring under degraded communication conditions highlights the value of automated systems in mitigating human-induced delays and maximizing runway efficiency. These findings inform the design of procedures and decision-support tools to ensure safe, reliable, and scalable traffic flows in increasingly complex and congested airspace environments.

\section*{Acknowledgment}
This work was supported by the National Aeronautics and Space Administration (NASA) University Leadership Initiative (ULI) program under project “Autonomous Aerial Cargo Operations at Scale”, under grant No.80NSSC21M071 to the University of Texas at Austin. Any opinions, findings, conclusions, or recommendations expressed in this material are those of the authors and do not necessarily reflect the views of the project sponsor.

\bibliography{ref}

\clearpage
\appendix
\section{Theoretical Verification of Runway Throughput \label{app: throughput-verification}}

To assess the validity of the simulation, we derive the expected steady state characteristics of the arrival flow merging scheme under idealized conditions without communication latency or outages. We investigate the maximum throughput the straight-in flow may sustain without making accommodations to expand gaps for merging downwind aircraft. The number of aircraft that may be accommodated in the gap behind a leading straight-in aircraft, plus the leading aircraft itself, is as follows.

\begin{equation}
N_{gap}(X_{StraightIn}) = 1+\bigg\lfloor \frac{X_{StraightIn}}{T_{sep}} \bigg\rfloor
\end{equation}

The expected number of aircraft that may be accommodated by a gap in the straight-in flow is computed given the exponentially distribution gap size.

\begin{equation}
\begin{aligned}
 \mathbb{E}[N_{gap}] &= \int_0^\infty N_{gap}(X) p(X)dX\\
 &= \int_0^\infty\bigg(1+\bigg\lfloor \frac{X}{T_{sep}} \bigg\rfloor \bigg) \Lambda_{StraightIn}e^{-\Lambda_{StraightIn}X}dX\\
 &=\sum_{n=0}^\infty \int_{nT_{sep}}^\infty \Lambda_{StraightIn}e^{-\Lambda_{StraightIn}X} dX\\
 &=\sum_{n=0}^\infty e^{-n\Lambda_{StraightIn} T_{sep}}\\
 &=\frac{1}{1-e^{-\Lambda_{StraightIn} T_{sep}}}
\end{aligned}
\end{equation}

The maximum throughput rate is computed as the ratio of the expected aircraft per straight-in gap and the expected time between straight-in aircraft.

\begin{equation}
\begin{aligned}
\beta_{max} &=\frac{\mathbb{E}[N_{gap}]}{\mathbb{E}[H_I]}=\frac{1}{(1-e^{-\Lambda_{StraightIn}T_{sep}})(T_{sep}+\Lambda_{StraightIn}^{-1})}
 \end{aligned}
\end{equation}

From \Cref{eq: beta-dw}, we have,
\begin{equation}
\Lambda_{StraightIn} = \frac{1}{\frac{1}{\beta_{StraightIn}}-T_{sep}}
\end{equation}

That is,
\begin{equation}
\begin{aligned}
\beta_{max} 
&=\frac{\mathbb{E}[N_{gap}]}{\mathbb{E}[H_I]}\\
&=\frac{1}{(1-e^{-\Lambda_{StraightIn}T_{sep}})(T_{sep}+\Lambda_{StraightIn}^{-1})}\\
&=\frac{1}{(1-e^{-\frac{T_{sep}}{\frac{1}{\beta_{StraightIn}}-T_{sep}}})(T_{sep}+\frac{1}{\beta_{StraightIn}}-T_{sep})}\\
&=\frac{1}{(1-e^{-\frac{T_{sep}}{\frac{1}{\beta_{StraightIn}}-T_{sep}}})(\frac{1}{\beta_{StraightIn}})}\\
&=\frac{\beta_{StraightIn}}{1-e^{-\frac{T_{sep}}{\frac{1}{\beta_{StraightIn}}-T_{sep}}}}
\end{aligned}
\end{equation}

The total realized throughput given a downwind arrival rate is,
\begin{equation}
\begin{aligned}
\beta_{Throughput} 
&= \min(\beta_{max}, \quad \beta_{Downwind}+\beta_{StraightIn}) \\
&= \min(\frac{\beta_{StraightIn}}{1-e^{-\frac{T_{sep}}{\frac{1}{\beta_{StraightIn}}-T_{sep}}}}, \quad \beta_{Downwind}+\beta_{StraightIn})
\end{aligned}
\end{equation}

\begin{figure}
\centering
\includegraphics[width=0.75\textwidth]{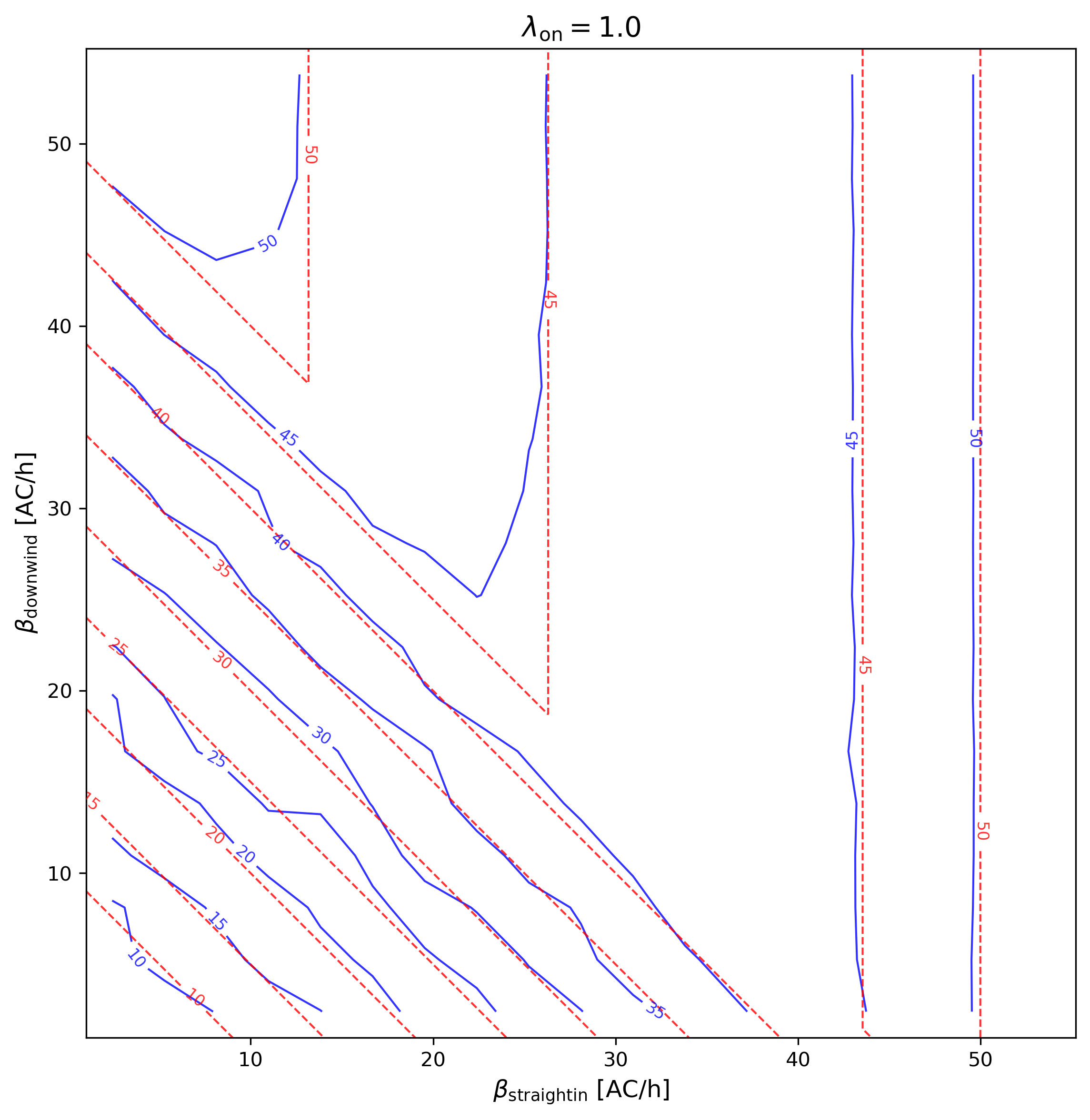}\\
\caption{The validation of the simulation results with the analytical contour lines. Red solid lines are the theoretical values, and blue dashed lines are the contour lines from simulated results. Note: Communication uncertainties (i.e., $\lambda_{on}=1$) or pilot-induced delays are not modeled into this simulation, as the primary focus is for theoretical validation.}
\label{fig:throughput-verification}
\end{figure}

This figure compares the results from theoretical throughput analysis and the sampled results in our simulation setup. Our discrete event simulation formulation can match the theoretical results with minimal differences. For the large discrepancies of the curvatures for the throughput at 45 [AC/h] and 50 [AC/h] contour lines, the averaging function in plotting the blue smoothed out the sharp corners.

\section{Extended Reliability Contours at $P_A \in \{0.9, 0.7, 0.5\}$ \label{app: extended-contours}}
This appendix provides the complementary reliability contour panels corresponding to the intermediate availability levels $P_A \in \{0.9, 0.7, 0.5\}$ for each of the three runway-capacity metrics considered in \Cref{subsec: case1-mc}. Main-text figures (\Cref{fig: throughput-main}, \Cref{fig: immediate-fraction-main}, and \Cref{fig: avg-holding-main}) show the corresponding panels at $P_A \in \{1.0, 0.8, 0.6\}$. Together, the main-text and appendix figures constitute the full $P_A \in \{0.5, 0.6, 0.7, 0.8, 0.9, 1.0\}$ sweep for each metric and each communication configuration.

\begin{figure}[H]
    \centering
    \begin{minipage}[t]{0.48\textwidth}
        \centering
        \includegraphics[trim={347.52bp 0 0 0},clip,width=\textwidth]{figs/beta_throughput_contour_lines_is_pr_transaction_time.png}
    \end{minipage}
    \hfill
    \begin{minipage}[t]{0.48\textwidth}
        \centering
        \includegraphics[trim={347.52bp 0 0 0},clip,width=\textwidth]{figs/beta_throughput_contour_lines_is_pr.png}
    \end{minipage}
    \caption{Runway throughput at $P_A \in \{0.9, 0.7, 0.5\}$, complementing \Cref{fig: throughput-main}. Left column: legacy voice-based VHF pilot-controller communications. Right column: future RPAS/CPDLC digital datalink communications.}
    \label{fig: throughput-appendix}
\end{figure}

\begin{figure}[H]
    \centering
    \begin{minipage}[t]{0.48\textwidth}
        \centering
        \includegraphics[trim={347.52bp 0 0 0},clip,width=\textwidth]{figs/beta_immediate_fraction_contour_lines_is_pr_transaction_time.png}
    \end{minipage}
    \hfill
    \begin{minipage}[t]{0.48\textwidth}
        \centering
        \includegraphics[trim={347.52bp 0 0 0},clip,width=\textwidth]{figs/beta_immediate_fraction_contour_lines_is_pr.png}
    \end{minipage}
    \caption{Percentage of immediate turns in downwind arrivals at $P_A \in \{0.9, 0.7, 0.5\}$, complementing \Cref{fig: immediate-fraction-main}. Left column: legacy voice-based VHF communications. Right column: future RPAS/CPDLC digital datalink communications.}
    \label{fig: immediate-fraction-appendix}
\end{figure}

\begin{figure}[H]
    \centering
    \begin{minipage}[t]{0.48\textwidth}
        \centering
        \includegraphics[trim={347.52bp 0 0 0},clip,width=\textwidth]{figs/beta_avg_holding_contour_lines_is_pr_transaction_time.png}
    \end{minipage}
    \hfill
    \begin{minipage}[t]{0.48\textwidth}
        \centering
        \includegraphics[trim={347.52bp 0 0 0},clip,width=\textwidth]{figs/beta_avg_holding_contour_lines_is_pr.png}
    \end{minipage}
    \caption{Average holding time of downwind arrivals at $P_A \in \{0.9, 0.7, 0.5\}$, complementing \Cref{fig: avg-holding-main}. Left column: legacy voice-based VHF communications. Right column: future RPAS/CPDLC digital datalink communications.}
    \label{fig: avg-holding-appendix}
\end{figure}

\end{document}